\documentclass[a4paper,journal,twocolumns,10pt]{IEEEtran}
\usepackage[ left=0.64in, right=0.64in, top=0.75in, bottom=1.1in]{geometry}
\usepackage{amsmath}
\usepackage{amssymb}
\usepackage{bm}
\usepackage{color}
\usepackage{graphicx}
\usepackage{textcomp}
\usepackage[linesnumbered,ruled,lined]{algorithm2e}
\usepackage{tabularx}
\usepackage[outdir=./]{epstopdf}
\usepackage{caption}
\usepackage{lipsum}
\usepackage{cite}
\usepackage{multirow}
\usepackage{subfig}
\usepackage{fancyhdr}
\usepackage{array}
\newcolumntype{L}[1]{>{\raggedright\let\newline\\\arraybackslash\hspace{0pt}}m{#1}}
\newcolumntype{C}[1]{>{\centering\let\newline\\\arraybackslash\hspace{0pt}}m{#1}}
\newcolumntype{R}[1]{>{\raggedleft\let\newline\\\arraybackslash\hspace{0pt}}m{#1}}

\SetKwInput{Input}{input}
\SetKwInput{Output}{output}
\usepackage{float}

\begin{document}
	
	\title{Rate-Splitting Multiple Access for Downlink Communication Systems:
		Bridging, Generalizing and Outperforming SDMA and NOMA
	}

	\author{Yijie Mao,
		 Bruno Clerckx, 
		 and Victor O.K. Li
		\thanks{This work is partially supported by the U.K. Engineering and Physical
			Sciences Research Council (EPSRC) under grant EP/N015312/1.}}%


\maketitle

\thispagestyle{empty}
\pagestyle{empty}

\begin{abstract}
	\par Space-Division Multiple Access (SDMA) utilizes linear precoding to separate users in the spatial domain and relies on \textit{fully} treating any residual multi-user interference as noise. Non-Orthogonal Multiple Access (NOMA) uses linearly precoded superposition coding with successive interference cancellation (SIC) to superpose users in the power domain and relies on user grouping and ordering to enforce some users to \textit{fully} decode and cancel interference created by other users. 
	
	\par In this paper, we argue that to efficiently cope with the high throughput, heterogeneity of Quality-of-Service (QoS), and massive connectivity requirements of future multi-antenna wireless networks, multiple access design needs to depart from those two extreme interference management strategies, namely \textit{fully} treat interference as noise (as in SDMA) and \textit{fully} decode interference (as in NOMA). 
	
	\par Considering a multiple-input single-output broadcast channel, we develop a novel multiple access framework, called Rate-Splitting Multiple Access (RSMA). RSMA is a more general and more powerful multiple access for downlink multi-antenna systems that contains SDMA and NOMA as special cases. RSMA relies on linearly precoded rate-splitting with SIC to decode part of the interference and treat the remaining part of the interference as noise. This capability of RSMA to \textit{partially} decode interference and \textit{partially} treat interference as noise enables to softly bridge the two extremes of fully decoding interference and treating interference as noise, and provide room for rate and QoS enhancements, and complexity reduction.
	
	\par The three multiple access schemes are compared and extensive numerical results show that RSMA provides a smooth transition between SDMA and NOMA and outperforms them both in a wide range of network loads (underloaded and overloaded regimes) and user deployments (with a diversity of channel directions, channel strengths and qualities of Channel State Information at the Transmitter). Moreover, RSMA provides rate and QoS enhancements over NOMA at a lower computational complexity for the transmit scheduler and the receivers (number of SIC layers).
	
\end{abstract}
\maketitle


\section{Introduction}
\par With the dramatic upsurge in the number of devices expected in 5G and beyond, wireless networks will be operated in a variety of regimes ranging from underloaded to overloaded (where the number of scheduled devices is smaller and larger than the number of transmit antennas at each access point, respectively). Moreover due to the heterogeneity of devices (high-end such as smartphones and low-end such as Internet-of-Things and Machine-Type Communications devices), deployments and applications in 5G and beyond, the transmitter will need to serve simultaneously users with different capabilities, deployments and qualities of Channel State Information at the Transmitter (CSIT). This massive connectivity problem together with the demands for high throughput and heterogeneity of Quality-of-Service (QoS) has recently spurred interests in re-thinking multiple access for the downlink of communication systems.

\par In this paper, we propose a new multiple access called Rate-Splitting Multiple Access (RSMA). In order to fully assess the novelty of the proposed multiple access paradigm and the design philosophy, we first review the state-of-the-art of two major multiple accesses, namely Non-Orthogonal Multiple Access (NOMA) \cite{NOMA2013YSaito}, also called Multi-User Superposition Transmission (MUST) in 3GPP LTE Rel-13 \cite{3gpp.36.859} and Space-Division Multiple Access (SDMA). We identify their benefits and limitations and make critical observations, before motivating the introduction of the novel and more powerful RSMA.

\subsection{SDMA and NOMA: The Extremes}
\par Contrary to Orthogonal Multiple Access (OMA) that schedules users or groups of users in orthogonal dimensions, e.g. time (TDMA), frequency (FDMA), NOMA superposes users in the same time-frequency resource via the power domain or the code domain, leading to the power-domain NOMA (e.g. \cite{NOMA2013YSaito}) and code-domain NOMA (e.g. sparse code multiple access (SCMA) \cite{HNikoSCMA2013}). Power-domain NOMA\footnote{In the sequel, power-domain NOMA will be referred to simply by NOMA.} relies on superposition coding (SC) at the transmitter and successive interference cancellation (SIC) at the receivers (denoted in short as SC--SIC) \cite{NOMA2013YSaito,NOMAsurvey2015,NOMAsurvey2017Ding,wshin2017RSNOMA}. Such a strategy is motivated by the well-known result that SC--SIC achieves the capacity region of the Single-Input Single-Output (SISO) (Gaussian) Broadcast Channel (BC) \cite{Tcover1972,tsefundamentalWC2005}. It is also well known that the capacity region of the SISO BC is larger than the rate region achieved by OMA (e.g. TDMA) when users experience a disparity of channel strengths \cite{tsefundamentalWC2005}. On the other hand, when users exhibit the same channel strengths, OMA based on TDMA is sufficient to achieve the capacity region \cite{tsefundamentalWC2005}. 

\par The \textit{benefit of single-antenna NOMA} using SC--SIC is therefore to be able, despite the presence of a single transmit antenna in a SISO BC, to cope with an overloaded regime in a spectrally efficient manner where multiple users experience potentially very different channel strengths/path losses (e.g. cell centre users and cell edge users) on the same time/frequency resource.   

\par The \textit{limitation of single-antenna NOMA} lies in its complexity as the number of users grows. Indeed for a $K$-user SISO BC, the strongest user needs to decode using SIC the $K-1$ messages of all co-scheduled users and therefore peel off $K-1$ layers before accessing its intended stream. Though SIC of a small number of layers should be feasible in practice\footnote{Recall that SU--MIMO in LTE Rel. 8 was designed with Minimum Mean Square Error--SIC (MMSE--SIC) in mind \cite{brunoMIMO2010}.}, the complexity and likelihood of error propagation  becomes quickly significant for a large number of users. This calls for ways to decrease the number of SIC layers at each user. One could divide users into small groups of users with disparate channels and apply SC--SIC in each group and schedule groups on orthogonal resources (using OMA), but that may lead to some performance loss and latency increase.

\par In nowadays wireless networks, access points are often equipped with more than one antenna. This spatial dimension opens the door to another well-known type of multiple access, namely SDMA. SDMA superposes users in the same time-frequency resource and separates user via a proper use of the spatial dimensions. Contrary to the SISO BC, the multi-antenna BC is non-degraded, i.e. users cannot be ordered based on their channel strengths in general settings. This is the reason why SC--SIC is not capacity achieving and the complex Dirty Paper Coding (DPC) is the only strategy that achieves the capacity region of the Multiple-Input Single-Output (MISO) (Gaussian) BC with perfect CSIT \cite{capacityRegion2006HW}. DPC, rather than performing interference cancellation at the receivers as in SC--SIC, can be viewed as a form of enhanced interference cancellation at the transmitter and relies on perfect CSIT to do so. Due to the high computational burden of DPC, linear precoding is often considered the most attractive alternative to simplify the transmitter design \cite{clerckx2013mimo}. Interestingly, in a MISO BC, Multi-User Linear Precoding (MU--LP), e.g. either in closed form or optimized using optimization methods, though suboptimal, is often very useful when users experience relatively similar channel strengths or long term Signal-to-Noise Ratio (SNR) and have semi-orthogonal to orthogonal channels \cite{ZFrateRegion2006}. SDMA is therefore commonly implemented using MU--LP. The linear precoders create different beams with each beam being allocated a fraction of the total transmit power. Hence, similarly to NOMA, SDMA can also be viewed as a superposition of users in the power-domain, though users are separated at the transmitter side by spatial beamformers rather than by the use of SIC at the receivers.

\par SDMA based on MU--LP is a well-established multiple access that is nowadays the basic principle behind numerous techniques in 4G and 5G such as Multi-user
Multiple-Input Multiple-Output (MU--MIMO), Coordinated MultiPoint (CoMP) coordinated beamforming, network MIMO, millimeter-wave MIMO and Massive MIMO. 

\par The \textit{benefit of SDMA} using MU--LP is therefore to reap all spatial multiplexing benefits of a MISO BC with perfect CSIT with a low precoder and receiver complexity. 

\par The \textit{limitations of SDMA} are threefold.

\par \textit{First}, it is suited to the underloaded regime and performance of MU--LP in the overloaded regime quickly drops as it requires more transmit antennas than users to be able to efficiently manage the multi-user interference. When the MISO BC becomes overloaded, the current and popular approach for the transmitter is to schedule group of users over orthogonal dimensions (e.g. time/frequency) and perform linear precoding in each group, which may increase latency and decrease QoS depending on the application. 

\par \textit{Second}, its performance is sensitive to the user channel orthogonality and strengths and requires the scheduler to pair semi-orthogonal users with similar channel strengths together. The complexity of the scheduler can quickly increase when an exhaustive search is performed, though low complexity (suboptimal) scheduling and user pairing algorithms exist \cite{clerckx2013mimo}. 

\par \textit{Third}, it is optimal from a Degrees-of-Freedom\footnote{The DoF characterizes the number of interference-free streams that can be transmitted or equivalently the pre-log factor of the rate at high SNR.} (DoF), also known as spatial multiplexing gain,
 perspective in the perfect CSIT setting but not in the presence of imperfect CSIT \cite{RSintro16bruno}. The problem of SDMA design in the presence of imperfect CSIT has been to strive to apply a framework motivated by perfect CSIT to scenarios with imperfect CSIT, not to design a framework motivated by imperfect CSIT from the beginning \cite{RSintro16bruno}. This leads to the well-known severe performance loss of MU--LP in the presence of imperfect CSIT \cite{NJindalMIMO2006}.

\par In view of SC--SIC benefits in a SISO BC, attempts have been made to study multi-antenna NOMA. Two lines of research have emerged that both rely on linearly precoded SC--SIC. 

\par The \textit{first strategy}, which we simply denote as 'SC--SIC', is a direct application of SC--SIC to the MISO BC by degrading the multi-antenna broadcast channel. It consists in ordering users based on their effective scalar channel (after precoding) strengths and enforce receivers to decode messages (and cancel interference) in a successive manner. This is advocated and exemplified for instance in \cite{gc01,ChoiNOMA2015,QSunNOMA2015,QZhangNOMA2016}. This NOMA strategy converts the multi-antenna non-degraded channel into an effective single antenna degraded channel, as at least one receiver ends up decoding all messages. While such a strategy can cope with the deployment of users experiencing aligned channels and different path loss conditions, it comes at the expense of sacrificing and annihilating all spatial multiplexing gains in general settings. By forcing one receiver to decode all streams, the sum DoF is reduced to unity\footnote{This can be easily seen since, for the receiver forced to decode all streams, the model reduces to a Multiple Access Channel (MAC) with a single-antenna receiver, which has a sum-DoF of 1. This was discussed in length in \cite{hamdi2017bruno}.}. This is the same DoF as that achieved by TDMA/single-user beamforming (or OMA). This is significantly smaller than the sum DoF achieved by DPC and MU--LP in a MISO BC with perfect CSIT, which is the minimum of the number of transmit antennas and the number of users\footnote{Recall that this spatial multiplexing gain is the main driver for using multiple antennas in a multi-user setup and the introduction of MU--MIMO in 4G \cite{BrunoLTEMIMO2013}.}. Moreover, this loss in multiplexing gain comes with a significant increase in receiver complexity due to the multi-layer SIC compared to the treat interference as noise strategy of MU--LP. As a remedy to recover the DoF loss, we could envision a dynamic switching between NOMA and SDMA, reminiscent of the dynamic switching between SU--MIMO and MU--MIMO in 4G \cite{BrunoLTEMIMO2013}. One would dynamically choose the best option between NOMA and SDMA as a function of the channel states. A particular instance of this approach is taken in \cite{quasidegrade2016} where a dynamic switching between SC--SIC and Zero-Forcing Beamforming (ZFBF) was investigated.

\par The \textit{second strategy}, which we denote as 'SC--SIC per group', consists in grouping $K$ users into $G$ groups. Users within each group are served using SC--SIC and users across groups are served using SDMA so as to mitigate the inter-group interference. Examples of such a strategy can be found in\cite{NOMA2013YSaito,DingNOMA2016,NOMA2017Choi,wshin2017mimonoma,NOMA2017Nguyen,MZeng2017NOMA}. This strategy can therefore be seen as a combination of SDMA and NOMA where the multi-antenna system is effectively decomposed into $G$ hopefully non-interfering single-antenna NOMA channels. For this 'SC--SIC per group' approach to perform at its best, users within each group need to have their channels aligned and users across groups need to be orthogonal.

\par Similarly to SDMA, multi-antenna NOMA designs also rely on accurate CSIT. In the practical scenario of imperfect CSIT, NOMA design relies on the same above two strategies but optimizes the precoder so as to cope with CSIT imperfection and resulting extra multi-user interference. As an example, the MISO BC channel is again degraded in \cite{QZhangNOMA2016} and precoder optimization with imperfect CSIT is studied.

\par The \textit{benefit of multi-antenna NOMA}, similarly to the single-antenna NOMA, is the potential to cope with an overloaded regime where multiple users experience different channel strengths/path losses and/or are closely aligned with each other.

\par The \textit{limitations of multi-antenna NOMA} are fourfold. 

\par \textit{First}, the use of SC--SIC in NOMA is fundamentally motivated by a degraded BC in which users can be ordered based on their channel strengths. This is the key property of the SISO BC that enables SC--SIC to achieve its capacity region. Unfortunately, motivated by the promising gains of SC--SIC in a SISO BC, the multi-antenna NOMA literature strives to apply SC--SIC to a non-degraded MISO BC. This forces to degrade a non-degraded BC and therefore leads to an inefficient use of the spatial dimensions in general settings, leading to a DoF loss. 

\par \textit{Second}, NOMA is not suited for general user deployments since degrading a MISO BC is efficient when users are sufficiently aligned with each other and exhibit a disparity of channel strengths, not in general settings. 

\par \textit{Third}, multi-antenna NOMA comes with an increase in complexity at both the transmitter and the receivers. Indeed, a multi-layer SIC is needed at the receivers, similarly to the single-antenna NOMA. However, in addition, since there exists no natural order for the users’ channels in multi-antenna NOMA (because we deal with vectors rather than scalars), the precoders, the groups and the decoding orders have to be jointly optimized by the scheduler at the transmitter. Taking as an example the application of NOMA based on 'SC--SIC' to a three-user MISO BC, we need to optimize three precoders, one for each user, along with the six possible decoding orders. Increasing the number of users leads to an exponential increase in the number of possible decoding orders. 'SC--SIC per group' divides users into multiple groups but that approach leads to a joint design of user ordering and user grouping. To decrease the complexity in user ordering and user grouping, multi-antenna NOMA ('SC--SIC' and 'SC--SIC per group') forces users belonging to the same group to share the same precoder (beamforming vector) \cite{NOMA2013YSaito}. Unfortunately, such a restriction can only further hurt the overall performance since it shrinks the overall optimization space.
 
\par \textit{Fourth}, multi-antenna NOMA is subject to the same drawback as SDMA in the presence of imperfect CSIT, namely its design is not motivated by any fundamental limits of a MISO BC with imperfect CSIT. 

\par The key is to recognize that the limitations and drawbacks of SDMA and NOMA originate from the fact that those two multiple accesses fundamentally rely on two extreme interference management strategies, namely fully treat interference as noise and fully decode interference. Indeed, while NOMA relies on some users to fully decode and cancel interference created by other users, SDMA relies on fully treating any residual multi-user interference as noise. In the presence of imperfect CSIT, CSIT inaccuracy results in an additional multi-user interference that is treated as noise by both NOMA (SC--SIC per group) and SDMA.

\subsection{RSMA: Bridging the Extremes}

\par In contrast, with RSMA, we take a different route and depart from the SDMA and NOMA literature and those two extremes of fully decode interference and treat interference as noise. We introduce a more general and powerful multiple access framework based on linearly precoded Rate-Splitting (RS) at the transmitter and SIC at the receivers. This enables to decode part of the interference and treat the remaining part of the interference as noise \cite{RSintro16bruno}. This capability of RSMA to partially decode interference and partially treat interference as noise enables to softly bridge the two extreme strategies of fully treating interference as noise and fully decoding interference. This contrasts sharply with SDMA and NOMA that exclusively rely on the two extremes or a combination thereof.

\par In order to partially decode interference and partially treat interference as noise, RS splits messages into common\footnote{'Common' is sometimes
	referred to as 'public'.}  and private messages and relies on a superimposed transmission of common messages decoded by multiple users, and private messages decoded by their corresponding users (and treated as noise by co-scheduled users). Users rely on SIC to first decode the common messages before accessing the private messages. By adjusting the message split and the power allocation to the common and private messages, RS has the ability to softly bridge the two extreme of fully treat interference as noise and fully decode interference. 

\par The idea of RS dates back to Carleial's work and the Han and Kobayashi (HK) scheme for the two-user single-antenna Interference Channel (IC) \cite{TeHan1981}. However, the use of RS as the building block of RSMA is motivated by recent works that have shown the benefit of RS in multi-antenna BC and the recent progress on characterizing the fundamental limits of a multi-antenna BC (and IC) with imperfect CSIT. Hence, importantly, in contrast with the conventional RS (HK scheme) used for the two-user SISO IC, we here use RS in a different setup, namely 1) in a BC and 2) with multiple antennas. The use and benefits of RS in a multi-antenna BC only appeared in the last few years\footnote{This also contrasts with NOMA, for which the usefulness of SC--SIC in a BC is known for several decades \cite{Tcover1972,tsefundamentalWC2005}.}.

\par The capacity region of the $K$-user MISO BC with imperfect CSIT remains an open problem. As an alternative, recent progress has been made to characterize the DoF region of the underloaded and overloaded MISO BC with imperfect CSIT. In \cite{AG2015}, a novel information theoretic upperbound on the sum DoF of the $K$-user underloaded MISO BC with imperfect CSIT was derived. Interestingly, this sum-DoF coincides with the sum-DoF achieved by a linearly precoded RS strategy at the transmitter with SIC at the receivers \cite{DoF2013SYang,RS2016hamdi}. RS (with SIC) is therefore optimum to achieve the sum DoF of the $K$-user underloaded MISO BC with imperfect CSIT, in contrast with MU--LP that is clearly suboptimum (and so is SC--SIC since it achieves a sum DoF of unity\footnote{Note that in the specific case where we have finite precision CSIT, the sum DoF collapses to 1 \cite{AG2015} and RS, SC--SIC, TDMA all achieve the same optimal DoF.}) \cite{RS2016hamdi}. It turns out that RS with a flexible power allocation is not only optimum for the sum DoF but for the entire DoF region of an underloaded MISO BC with imperfect CSIT \cite{enrico2017bruno}. The DoF benefit of RS in imperfect CSIT settings were also shown in more complicated underloaded networks with multiple transmitters in \cite{chenxi2017brunotopology} and multi-antenna receivers \cite{chenxi2017bruno}. 
Considering user fairness, the optimum symmetric DoF (or max-min DoF), i.e. the DoF that can be achieved by all users simultaneously, of the underloaded MISO BC with imperfect CSIT with MU--LP and RS was studied in \cite{RS2016joudeh}. RS symmetric DoF was shown to outperform that of MU--LP. Finally, moving to the overloaded MISO BC with heterogeneous CSIT qualities, a multi-layer power partitioning strategy that superimposes degraded symbols on top of linearly precoded rate-splitted symbols was shown in \cite{enrico2016bruno} to achieve the optimal DoF
region.

\par The benefits of RS have also appeared in multi-antenna settings with perfect CSIT. In an overloaded multigroup multicast setting with perfect CSIT, considering again fairness, the symmetric DoF achieved by RS, MU--LP and degraded NOMA transmissions (where receivers decode messages and cancel interference in a successive manner as in 'SC--SIC') was studied in \cite{hamdi2017bruno}. It was shown that RS here again outperforms both MU--LP and SC--SIC. 

\par The DoF metric is insightful to identify the multiplexing gains of the MISO BC at high SNR but fails to capture the diversity of channel strengths among users. This limitation is countered by the Generalized DoF (GDoF) framework, which inherits the tractability of the DoF framework while capturing the diversity in channel strengths \cite{Tse2008}. In \cite{AG2016Gdof,AG2017Gdof}, the GDoF of an underloaded MISO BC with imperfect CSIT is studied and here again RS is used as part of the achievability scheme.

\par The DoF (GDoF) superiority of RS over MU--LP and SC--SIC in all those multi-antenna settings (with perfect and imperfect CSIT) comes from the ability of RS to better handle the multi-user interference by evolving in a regime in between the extremes of fully treating it as noise and fully decoding it.

\par Importantly, the rate enhancements of RS over MU--LP, as predicted by the DoF analysis, are reflected in the finite SNR regime as shown in a number of recent works. In \cite{RS2015bruno}, finite SNR rate analysis of RS in MISO BC in the presence of quantized feedback was analyzed and it was shown that RS benefits from a CSI feedback overhead reduction compared to MU--LP. Using optimization methods, the precoder design of RS at finite SNR was investigated in \cite{RS2016hamdi} for the sum-rate and rate region maximization with imperfect CSIT, in \cite{RS2016joudeh} for max-min fair transmission with imperfect CSIT, and in \cite{hamdi2017bruno} for multigroup multicast with perfect CSIT. Moreover, the benefit of RS over MU--LP in the finite SNR regime was shown in Massive MIMO \cite{Mingbo2016}, millimetre-wave systems \cite{minbo2017bruno} and multi-antenna deployments subject to hardware impairments \cite{AP2017bruno}. Finally, the performance benefits of the power-partitioning strategy relying on RS in the overloaded MISO BC with heterogeneous CSIT was confirmed using simulations at finite SNR in the presence of a diversity of channel strengths \cite{enrico2016bruno}. In particular, in contrast to the RS used in \cite{RSintro16bruno,RS2016hamdi,enrico2017bruno,enrico2016bruno,RS2016joudeh,hamdi2017bruno,RS2015bruno,minbo2017bruno,AP2017bruno} that relies on a single common message, \cite{Mingbo2016} (as well as \cite{chenxi2017brunotopology}) showed the benefits in the finite SNR regime of a multi-layer (hierarchical) RS relying on multiple common messages decoded by various groups of users. 

\par In this paper, in view of the limitations of SDMA and NOMA and the above literature on RS in multi-antenna BC, we design a novel multiple access, called Rate-Splitting Multiple Access (RSMA) for downlink communication system\footnote{It is worth noting that Rate-Splitting Multiple Access (RSMA) also exists in the uplink for the SISO Multiple Access Channel \cite{Rimo1996}. Though they share the same name and the splitting of the messages, they have different motivations and structures.}. RSMA is a much more attractive solution (performance and complexity-wise) that retains the benefits of SDMA and NOMA but tackle all the aforementioned limitations of SDMA and NOMA. Considering a MISO BC, we make the following contributions.

\par \textit{First}, we show that RSMA is a more general class/framework of multi-user transmission that encompasses SDMA and NOMA as special cases. RSMA is shown to reduce to SDMA if channels are of similar strengths and  sufficiently orthogonal with each other and to NOMA if channels exhibit sufficiently diverse strengths and are sufficiently aligned with each other. This is the first paper to explicitly recognize that SDMA and NOMA are both subsets of a more general transmission framework based on RS\footnote{As already explained in \cite{RSintro16bruno}, RS can also be seen as a form of non-orthogonal multi-user transmission. Indeed, in its simplest form, the common message in RS can be seen as a non-orthogonal layer added onto the private layers.}.

\par \textit{Second}, we provide a general framework of multi-layer RS design that encompasses existing RS schemes as special cases. In particular, the single-layer RS of \cite{RS2015bruno,RS2016hamdi,RS2016joudeh,AP2017bruno, enrico2017bruno,minbo2017bruno,enrico2016bruno,hamdi2017bruno} and the multi-layer (hierarchical and topological) RS of \cite{Mingbo2016,chenxi2017brunotopology} are special instances of the generalized RS strategy developed here. Moreover the use of RS was primarily motivated by multi-antenna deployments subject to multi-user interference due to imperfect CSIT in those works. The benefit of RS in the presence of perfect CSIT and/or a diversity of channel strengths in a multi-antenna setup, as considered in this paper, is less investigated. RS was shown in \cite{hamdi2017bruno} to boost the performance of overloaded multigroup multicast. However, no attempt has been made so far to identify the benefit of RS in multi-antenna BC with perfect CSIT and/or a diversity of channel strengths.  

\par \textit{Third}, we show that the rate performance (rate region, weighted sum-rate with and without QoS constraints) of RSMA is always equal to or larger than that of SDMA and NOMA. Considering a MISO BC with perfect CSIT and no QoS constraints, RSMA performance comes closer to the optimal DPC region than SDMA and NOMA. In scenarios with QoS constraints or imperfect CSIT, RSMA always outperforms SDMA and NOMA. Since it is motivated by fundamental DoF analysis, RSMA is also optimal from a DoF perspective in both perfect and imperfect CSIT and therefore optimally exploit the spatial dimensions and the availability of CSIT, in contrast with SDMA and NOMA that are suboptimal.

\par \textit{Fourth}, we show that RSMA is much more robust than SDMA and NOMA to user deployments, CSIT inaccuracy and network load. It can operate in a wide range of practical deployments involving scenarios where the user channels are neither orthogonal nor aligned, and exhibit similar strengths or a diversity of strengths; where the CSI is perfectly or imperfectly known to the transmitter; where the network load can vary between the underloaded and the overloaded regimes. In particular, in the overloaded regime, the RSMA framework is shown to be particularly suited to cope with a variety of device capabilities, e.g. high-end devices along with cheap Internet-of-Things (IoT)/Machine-Type Communications (MTC) devices.  Indeed, the RS framework can be used to pack the IoT/MTC traffic in the common message, while still delivering high quality service to high-end devices.

\par \textit{Fifth}, we show that the performance gain can come with a lower computational complexity than NOMA for both the transmit scheduler and the receivers. In contrast to NOMA that requires complicated user grouping and ordering and potential dynamic switching (between SDMA, 'SC--SIC' and 'SC--SIC per group') at the transmit scheduler and multiple layers of SIC at the receivers, a simple one-layer RS that does not require any user ordering, grouping or dynamic switching at the transmit scheduler and a single layer of SIC at the receivers still significantly outperforms NOMA. In contrast to SDMA, RSMA is less sensitive to user pairing and therefore does not require complex user scheduling and pairing\footnote{This benefit of RS was briefly pointed out in \cite{Mingbo2016}.}. However, RSMA comes with a slightly higher encoding complexity than SDMA and NOMA due to the encoding of the common streams on top of the private streams.

\par \textit{Sixth}, though SC--SIC is optimal to achieve the capacity region of SISO BC, we show that a single-layer RS is a low-complexity alternative that only requires a single layer of SIC at each receiver and achieves close to SC--SIC (with multi-layer SIC) performance in a SISO BC deployment.

\par As a takeaway message, we note that the ability of a wireless network architecture to partially decode interference and partially treat interference as noise can lead to enhanced throughput and QoS, increased robustness and lower complexity compared to alternatives that are forced to operate in the extreme regimes of fully treating interference as noise and fully decoding interference.

\par It is also worth making the analogy with other types of channels where the ability to bridge the extremes of treating interference as noise and fully decoding interference has appeared. Considering a two-user SISO IC, interference is fully decoded in the strong interference regime and is treated as noise in the weak interference regime. Between those two extremes, interference is neither strong enough to be fully decoded nor weak enough to be treated as noise. 
The best known strategy for the two-user SISO IC is obtained using RS (so-called HK scheme). RS in this context is well known to be superior to strategies relying on fully treating interference as noise, fully decoding interference or orthogonalization (TDMA, FDMA) \cite{Tse2008,TeHan1981}. Limiting ourselves to those extremes strategies is suboptimal \cite{Tse2008,TeHan1981}.

\par The rest of the paper is organized as follows. The system model is described in Section \ref{system model}. The existing multiple accesses are specified in Section \ref{existing}. In Section \ref{RS}, the proposed RSMA and its low-complexity structures are described and compared with existing multiple accesses. The corresponding Weighted Sum Rate (WSR) problems are formulated and the WMMSE approach to solve the problem is discussed. Numerical results are illustrated in Section \ref{simulation}, followed by conclusions and future works in Section \ref{conclusion}. 

\par \textit{Notations}: The boldface uppercase and lowercase letters are used to represent matrices and vectors.
The superscripts $(\cdot)^T$ and $(\cdot )^H$ respectively denote transpose
and conjugate-transpose operators.
$\mathrm{tr}(\cdot)$ and $\mathrm{diag}(\cdot)$ are the trace and diagonal entries respectively. $\left\vert\cdot\right\vert$ is the absolute value and
$\left\Vert\cdot\right\Vert$ is the Euclidean norm. $\mathbb{E}\{\cdot\}$ refers to the statistical expectation. $\mathbb{C}$ denotes the complex space. $\mathbf{I}$ and $\mathbf{0}$ stand for an identity matrix and an all-zero vector, respectively, with appropriate dimensions.  $\mathcal{CN}(\delta ,\sigma^2)$ represents a complex Gaussian distribution with mean $\delta $ and variance $\sigma^2$. $|\mathcal{A}|$ is the cardinality of the set $\mathcal{A}$.

\section{System Model }
\label{system model}

\par Consider a system where a Base Station (BS) equipped with $N_t$ antennas serves $K$ single-antenna users. The users are indexed by the set $\mathcal{K}=\{1,\ldots,K\}$.
Let $\mathbf{{x}}\in \mathbb{C}^{N_t\times 1}$ denote the signal vector transmitted in a given channel use. It is subject to the power constraint $\mathbb{E}\{\left\Vert  \mathbf{x}\right\Vert^2\}\leq P_{t}$.
The signal received at user-$k$  is
\begin{equation}
\begin{aligned}
y_{k}&=\mathbf{{h}}_{k}^{H}\mathbf{{x}}+n_{k}, \forall k \in \mathcal{K}
\end{aligned}
\end{equation}
where $\mathbf{{h}}_{k}\in\mathbb{C}^{N_{t}\times1}$ is the channel between the BS and user-$k$. $n_{k}\sim\mathcal{CN}(0,\sigma_{n,k}^{2})$ is the Additive White Gaussian Noise (AWGN) at the receiver. Without loss of generality, we assume the noise variances are equal to one for all users. The transmit SNR is equal to the total power consumption $P_t$. We assume CSI of users is perfectly known at the BS in the following model. The imperfect CSIT scenario will be discussed in the proposed algorithm and the numerical results. Channel State Information at the Receivers (CSIR) is assumed to be perfect.

\par In this work, we are interested in beamforming designs for signal $\mathbf{x}$ at the BS. Specifically, the objective of beamforming designs is to maximize the WSR of users subject to a power constraint of the BS and QoS constraints of each user. We firstly state and compare two baseline multi-antenna multiple accesses, namely SDMA and NOMA.  Then RSMA is explained. The WSR problem of each strategy will be formulated and the algorithm adopted to solve the corresponding problem will be stated in the following sections. 
\section{SDMA and NOMA}
\label{existing}
\par In this section, we describe two baseline multiple accesses. The messages $W_1,\ldots,W_K$ intended for users 1 to $K$ respectively are encoded into $K$ independent  data streams $\mathbf{{s}}=[s_1,\ldots,s_K]^{T}$ independently.  Symbols are mapped to the transmit antennas through a precoding matrix denoted by $\mathbf{P}=[\mathbf{p}_1,\ldots,\mathbf{p}_K]$, where $\mathbf{p}_k\in\mathbb{C}^{N_t\times1}$ is the precoder for user-$k$. The superposed signal is 
$
\mathbf{x}=\mathbf{P}\mathbf{{s}}=\sum_{k\in\mathcal{K}}\mathbf{p}_ks_k.
$
Assuming that $\mathbb{E}\{\mathbf{{s}}\mathbf{{s}}^H\}=\mathbf{I}$, the transmit power is constrained by  $\mathrm{tr}(\mathbf{P}\mathbf{P}^{H})\leq P_{t}$.

\subsection{SDMA}
\par SDMA based on MU--LP is a well-established multiple access. Each user only decodes its desired message by treating interference as noise. The Signal-to-Interference-plus-Noise Ratio (SINR) at user-$k$ is given by
\begin{equation}
	\gamma_{k}=\frac{|\mathbf{{h}}_{k}^{H}\mathbf{{p}}_{k}|^{2}}{\sum_{j\neq k,j\in \mathcal{K}}|\mathbf{{h}}_{k}^{H}\mathbf{{p}}_{j}|^{2}+1}.
\end{equation}
For a given weight vector $\mathbf{u}=[u_1,\ldots,u_K]$, the WSR achieved by MU--LP is 
\begin{equation}
\label{opti prob:linear}
\begin{aligned}
R_{\mathrm{MU-LP}}(\mathbf{u})&=\max_{\mathbf{P}}  \sum_{k\in\mathcal{K}}u_kR_k\\
\mbox{s.t.}\quad&\mathrm{tr}(\mathbf{P}\mathbf{P}^{H})\leq P_{t}\\
&R_k\geq R_k^{th}, \forall k \in \mathcal{K}
\end{aligned}
\end{equation}
where $R_k=\log_2(1+\gamma_k)$ is the achievable rate of user-$k$. $u_{k}$ is a nonnegative constant which allows resource allocation to prioritize different users.  $R_k^{th}$ accounts for any potential individual rate constraint for user $k$. It ensures the QoS of each user. The Weighted MMSE (WMMSE) algorithm proposed in \cite{wmmse08} is adopted to solve problem (\ref{opti prob:linear}). The main idea of the WMMSE algorithm is to reformulate the WSR problem into its equivalent WMMSE problem and solve it using the Alternating Optimization (AO) approach. The rate region of the MU--LP strategy is approximated by  $R_{\mathrm{MU-LP}}(\mathbf{u})$ for different rate weight vectors $\mathbf{u}$. The resulting rate region $R_{\mathrm{MU-LP}}$ is the convex hull enclosing the resulting points. In general, solution to  problem (\ref{opti prob:linear}) would provide the optimal MU--LP beamforming strategy for any channel deployment (in between aligned and orthogonal channels and with similar or diverse channel strengths).

%

\subsection{NOMA}
\par NOMA relies on superposition coding at the transmitter and successive interference cancellation at the receiver. As discussed in the introduction, the two main strategies in multi-antenna NOMA are the 'SC--SIC' and 'SC--SIC per group'. SC--SIC can be treated as a special case of SC--SIC per group where there is only one group of users.  
\subsubsection{SC--SIC}
In SC--SIC, the precoders and decoding orders have to be optimized jointly. The decoding order is vital to the rate obtained at each user.
To maximize the WSR, all possible decoding orders of users are required to be considered. 
Denote $\pi$ as one of the decoding orders, 
the message of user-$\pi(k)$ is decoded before the message of user-$\pi(j),\forall k\leq j$. The messages of user-$\pi(k),\forall k\leq i$ are decoded at user-$\pi(i)$ using SIC. The SINR experienced at user-$\pi(i)$ to decode the message of  user-$\pi(k),  k\leq i$  is given by
\begin{equation}
	\gamma_{\pi(i)\rightarrow\pi(k)}=\frac{|\mathbf{{h}}_{\pi(i)}^{H}\mathbf{{p}}_{\pi(k)}|^{2}}{\sum_{j>k,j\in\mathcal{K}}|\mathbf{{h}}_{\pi(i)}^{H}\mathbf{{p}}_{\pi(j)}|^{2}+1}.
\end{equation}

\par For a given weight vector $\mathbf{u}=[u_1,\ldots,u_K]$ and a fixed decoding order $\pi$, the WSR achieved by  SC--SIC is
\begin{equation}
\label{opti prob:noma}
\begin{aligned}
R_{\mathrm{SC-SIC}}(\mathbf{u},\mathbf{\pi})&=\max_{\mathbf{P}}    \sum_{k\in\mathcal{K}}u_{\pi(k)}R_{\pi(k)}\\
\mbox{s.t.}\quad&\mathrm{tr}(\mathbf{P}\mathbf{P}^{H})\leq P_{t}\\
&R_k\geq R_k^{th}, \forall k \in \mathcal{K}
\end{aligned}
\end{equation}
where $R_{\pi(k)}=\min_{i\geq k,i\in \mathcal{K}} \{\log_2(1+\gamma_{\pi(i)\rightarrow\pi(k)})\}$. In \cite{gc01}, the problem (\ref{opti prob:noma}) with equal weights is solved by the approximation technique minorization-maximization algorithm (MMA). To keep a single and unified approach to solve the WSR problem of different beamforming strategies,  we still use the WMMSE algorithm to solve it. By approximating the rate region with a set of rate weights, the rate region $R_{\mathrm{SC-SIC}}(\mathbf{\pi})$ with a certain decoding order $\pi$ is attained. To achieve the rate region of SC--SIC, all decoding orders should be considered. The largest achievable rate region of SC--SIC is defined as the convex hull of the union over all decoding orders as
$
R_{\mathrm{SC-SIC}}=\mathrm{conv}(\cup_{\pi}R_{\mathrm{SC-SIC}}(\mathbf{\pi})).
$

\subsubsection{SC--SIC per group}

\label{sec: sc-sic per group}
\par Assuming the $K$ users are divided into $G$ groups, denoted as $\mathcal{G}=\{1,\ldots,G\}$. In each group, there is a subset of users $\mathcal{K}_g,g\in\mathcal{G}$. The user groups satisfy the following conditions: $\mathcal{K}_g\cap\mathcal{K}_{g'}=\emptyset$, if $g\neq g'$, and $\sum_{g\in\mathcal{G}}|\mathcal{G}_g|=K$. Denote $\pi_g$ as one of the decoding orders of the users in $\mathcal{K}_g$, 
the message of user-$\pi_g(k)$ is decoded before the message of user-$\pi_g(j),\forall k\leq j$. The messages of user-$\pi_g(k),\forall k\leq i$ are decoded at user-$\pi_g(i)$ using SIC. The SINR experienced at user-$\pi_g(i)$ to decode the message of  user-$\pi_g(k),  k\leq i$  is given by
\begin{equation}
\gamma_{\pi_g(i)\rightarrow\pi_g(k)}=\frac{|\mathbf{{h}}_{\pi_g(i)}^{H}\mathbf{{p}}_{\pi_g(k)}|^{2}}{\sum_{j>k,j\in\mathcal{K}_g}|\mathbf{{h}}_{\pi_g(i)}^{H}\mathbf{{p}}_{\pi_g(j)}|^{2}+I_{\pi_g(i)}+1},
\end{equation}
where $I_{\pi_g(i)}=\sum_{g'\in\mathcal{G},g'\neq g}\sum_{j\in\mathcal{K}_{g'}}|\mathbf{{h}}_{\pi_g(i)}^{H}\mathbf{{p}}_{j}|^{2}$ is the inter-group interference suffered at user-$\pi_g(i)$. 
For a given weight vector $\mathbf{u}=[u_1,\ldots,u_K]$, a fixed grouping method $\mathcal{G}$ and a fixed decoding order $\pi=\{\pi_1,\ldots,\pi_G\}$, the WSR achieved by  SC--SIC per group is
\begin{equation}
\label{opti prob:partial noma}
\begin{aligned}
R_{\mathrm{SC-SIC}}^{\mathrm{group}}(\mathbf{u},\mathcal{G}, \mathbf{\pi})&=\max_{\mathbf{P}}    \sum_{g\in\mathcal{G}}\sum_{k\in\mathcal{K}_g}u_{\pi_g(k)}R_{\pi_g(k)}\\
\mbox{s.t.}\quad&\mathrm{tr}(\mathbf{P}\mathbf{P}^{H})\leq P_{t}\\
&R_k\geq R_k^{th}, \forall k \in \mathcal{K}
\end{aligned}
\end{equation}
where $R_{\pi_g(k)}=\min_{i\geq k,i\in \mathcal{K}_g} \{\log_2(1+\gamma_{\pi_g(i)\rightarrow\pi_g(k)})\}$. Similarly to the SC--SIC strategy, the problem can be solved by using the WMMSE algorithm. To maximize the WSR, all possible grouping methods and decoding orders should be considered. 

\par \textit{Remark 1}: As described in the introduction, it is common in the multi-antenna NOMA literature
('SC--SIC' and 'SC--SIC per group') to force users belonging
to the same group to share the same precoder, so as to decrease the complexity in user ordering and user grouping. Note that, in the system model described for both SC--SIC and SC--SIC per group, we consider the most general framework where each message is precoded by its own precoder. Hence, we here do not constrain symbols to be superimposed on the same precoder as this would further reduce the performance of NOMA strategies and therefore leading to even lower performance. Hence the performance obtained with NOMA in this work can be seen as the best possible performance achieved by NOMA.

\section{Rate-Splitting Multiple Access}
\label{RS}
\par In this section, we firstly introduce the idea of RS by introducing a two-user example ($K=2$) and a three-user example ($K=3$).  Then we propose the generalized framework of RS and specify two low-complexity RS strategies. We further compare RSMA with SDMA and NOMA from  the fundamental structure and complexity aspects. 
Finally, we discuss the general optimization framework to solve the WSR problem.

\subsection{Two-user example}
\par We first consider a two-user example. There are two messages $W_1$ and $W_2$ intended for user-1 and user-2, respectively. 
The message of each user is split into two parts, $\{W_1^{12},W_1^1\}$ for user-1 and $\{W_2^{12},W_2^2\}$ for user-2. The messages $W_1^{12}, W_2^{12}$ are encoded together into a common stream $s_{12}$ using a codebook shared by both users. Hence, $s_{12}$ is a common stream required to be decoded by both users.  The messages $W_{1}^1$ and $W_{2}^2$ are encoded into the private stream $s_1$ for user-1 and $s_2$ for user-2, respectively. The overall data streams to be transmitted based on RS is   $\mathbf{{s}}=[
s_{12},s_{1}, s_{2}]^{T}$. The data streams are linearly precoded via precoder $\mathbf{{P}}=[
\mathbf{{p}}_{12}, \mathbf{{p}}_{1}, \mathbf{{p}}_{2}]$ , where $\mathbf{{p}}_{12}\in\mathbb{C}^{N_{t}\times1}$ is the precoder for the common stream $s_{12}$. 
The resulting transmit signal is
$
\begin{aligned}
\mathbf{{x}}=\mathbf{{P}}\mathbf{{s}}=\mathbf{{p}}_{12}s_{12}+\mathbf{{p}}_{1}s_{1}+\mathbf{{p}}_{2}s_{2}.
\end{aligned}
$
We assume that $\mathrm{tr}(\mathbf{{s}}\mathbf{{s}}^H)=\mathbf{I}$ and the total transmit power is constrained by  $\mathrm{tr}(\mathbf{P}\mathbf{P}^{H})\leq P_{t}$. 

\par
\begin{figure}[t!]
	\centering
	\includegraphics[width=3.3in]{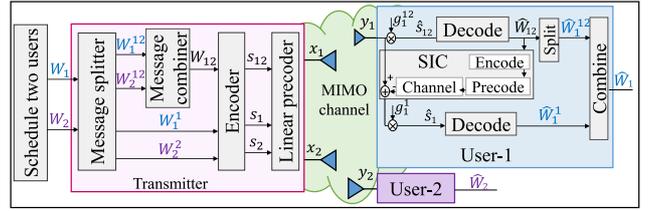}%
	\caption{Two-user transmission model using RS.}
	\label{fig: two-user transmission model}
	\vspace{-2mm}
\end{figure}

 At user sides, both user-1 and user-2 firstly decode the data stream $s_{12}$ by treating the interference from $s_1$ and $s_2$ as noise. Therefore, each user decodes part of the message of the other interfering user encoded in $s_{12}$. The interference is partially decoded at each user. The SINR of the common stream at user-$k$ is 
\begin{equation}
\gamma_{k}^{12}=\frac{\left|\mathbf{{h}}_{k}^{H}\mathbf{{p}}_{12}\right|^{2}}{\left|\mathbf{{h}}_{k}^{H}\mathbf{{p}}_{1}\right|^{2}+\left|\mathbf{{h}}_{k}^{H}\mathbf{{p}}_{2}\right|^{2}+1}.
\end{equation}
Once  $s_{12}$ is successfully decoded, its contribution to the  original received signal  $y_k$ is subtracted. After that, user-$k$ decodes its private stream $s_{k}$ by treating the private stream of user-$j$ ($j\neq k$) as noise. The two-user transmission model using RS is shown in Fig. \ref{fig: two-user transmission model}. The SINR of decoding the private stream $s_{k}$ at user-$k$ is
\begin{equation}
\gamma_{k}=\frac{\left|\mathbf{{h}}_{k}^{H}\mathbf{{p}}_{k}\right|^{2}}{\left|\mathbf{{h}}_{k}^{H}\mathbf{{p}}_{j}\right|^{2}+1}.
\end{equation}

\par The corresponding achievable rates of user-$k$ for the streams $s_{12}$ and $s_{k}$ are
$
\label{eq:user rate common}
R_{k}^{12}=\log_{2}\left(1+\gamma_{k}^{12}\right)
$ and $
\label{eq:user rate private}
R_{k}=\log_{2}\left(1+\gamma_{k}\right)
$.
To ensure that $s_{12}$ is successfully decoded by both users, the achievable common rate shall not exceed $R_{12}=\min\left\{ R_{1}^{12},R_{2}^{12}\right\}$. 
All boundary points for the two-user RS rate region can be obtained by assuming that $R_{12}$ is shared between users such that $C_k^{12}$ is the $k$th user's portion of the common rate
with $C_1^{12}+C_2^{12}=R_{12}$. 
Following the two-user RS structure described above, 
the total achievable rate of user-$k$ is 
$
R_{k,tot}=C_{k}^{12}+R_{k}.
$
For a given pair of weights $\mathbf{u}=[u_1,u_2]$, the WSR achieved by the two-user RS approach is
\begin{subequations}
\label{eq:two users}
	\begin{align}
		R_{\mathrm{RS}_2}(\mathbf{u})&=\max_{\mathbf{{P}}, \mathbf{c}}u_{1}R_{1,tot}+u_{2}R_{2,tot}  \\
		\mbox{s.t.}\quad
		& C_1^{12}+C_2^{12}\leq R_{12}\\
		&	\text{tr}(\mathbf{P}\mathbf{P}^{H})\leq P_{t}\\
		& R_{k,tot}\geq R_k^{th}, k\in\{1,2\}\\
		& \mathbf{c}\geq \mathbf{0}		
	\end{align}
\end{subequations}
where  $\mathbf{c}=[C_1^{12},C_2^{12}]$ is the common rate vector required to be optimized in order to maximize the WSR.
For a fixed pair of weights, problem (\ref{eq:two users}) can be solved using the WMMSE approach in \cite{RS2016hamdi}, except we have perfect CSIT here. 
By calculating $R_{\mathrm{RS}_2}(\mathbf{u})$ for a set of different rate weights $\mathbf{u}$, we obtain the rate region.

\par In contrast to MU--LP and SC--SIC, the RS scheme described above offers a more flexible formulation. In particular, instead of hard switching between MU--LP and SC--SIC, it allows both to operate simultaneously if necessary, and hence smoothly bridges the two. In the extreme of treating multi-user interference as noise, RS boils down to MU--LP\footnote{Note that OMA (single-user beamforming) is a subset of MU--LP and is obtained by allocating power exclusively to $s_1$ or $s_2$.} by simply allocating no power to the common stream $s_{12}$. In the other extreme of fully decoding interference, RS boils down to SC--SIC by forcing one user, say user-1, to fully decode the message of the other user, say user-2. This is achieved by allocating no power to $s_2$, encoding $W_1$ into $s_1$ and encoding $W_2$ into $s_{12}$, such that $\mathbf{x}=\mathbf{p}_{12}s_{12}+\mathbf{p}_1s_1$. User-1 and user-2 decode $s_{12}$ by treating $s_1$ as noise and user-1 decodes $s_1$ after canceling $s_{12}$. A physical-layer multicasting strategy is obtained by encoding both $W_1$ and $W_2$ into $s_{12}$ and allocating no power to $s_{1}$ and $s_{2}$.

\par \textit{Remark 2}: It should be noted that while the RS transmit signal model resembles a broadcasting system with unicast (private) streams and a multicast stream, the role of the common message is fundamentally different. The common message in a unicast-multicast system carries public information intended as a whole to all users in the system, while the common message $s_{12}$ in RS encapsulates parts of private messages, and is not entirely required by all users, although decoded by the two users for interference mitigation purposes \cite{RSintro16bruno}.

\par \textit{Remark 3}: A general framework is adopted where potentially each user can split its message into common and private parts. Note however that depending on the objective function, it is sometimes not needed for all users to split their messages. For instance for sum-rate maximization subject to no individual rate constraint, it is sufficient to have only one user to split its message \cite{RS2016hamdi}. However, when it comes to satisfying some fairness (WSR, QoS constraint, max-min fairness), splitting the message of multiple users appears necessary \cite{RS2016hamdi,RS2016joudeh,hamdi2017bruno}.

\vspace{-3mm}
\subsection{Three-user example}
\label{sec: three-user example}
\par We further consider a three-user example. Different from the two-user case, the message of user-$1$ is split into $\{W_{1}^{123}$, $W_{1}^{12}$, $W_{1}^{13}$, $W_{1}^{1}\}$. Similarly, the message of user-$2$ and user-$3$ are split into $\{W_{2}^{123}, W_{2}^{12},W_{2}^{23},W_{2}^{2}\}$ and $\{W_{3}^{123}, W_{3}^{13},W_{3}^{23},W_{3}^{3}\}$, respectively.
The superscript represents a specific group of users whose messages with the same superscript are going to be encoded together. For example, $W_{1}^{123},W_{2}^{123},W_{3}^{123}$ are encoded into the common stream $s_{123}$ intended for all the three users.  $W_{1}^{12}$ and $W_{1}^{13}$ are correspondingly encoded with the split messages of user-2 $W_{2}^{12}$ and user-3 $W_{3}^{13}$ into data streams $s_{12}$ and $s_{13}$.  $s_{12}$ is the partial common stream intended for user-1 and user-2. Hence, user-1 and user-2 will decode $s_{12}$ while user-3 will decode its intended streams by treating $s_{12}$ as noise. Similarly, we obtain $s_{23}$ partially encoded for user-2 and user-3. $W_1^1, W_2^2, W_3^3$ are respectively encoded into private streams $s_1, s_2$ and $s_3$.

\par
\begin{figure*}[t!]
	\centering
	\includegraphics[width=6.8in]{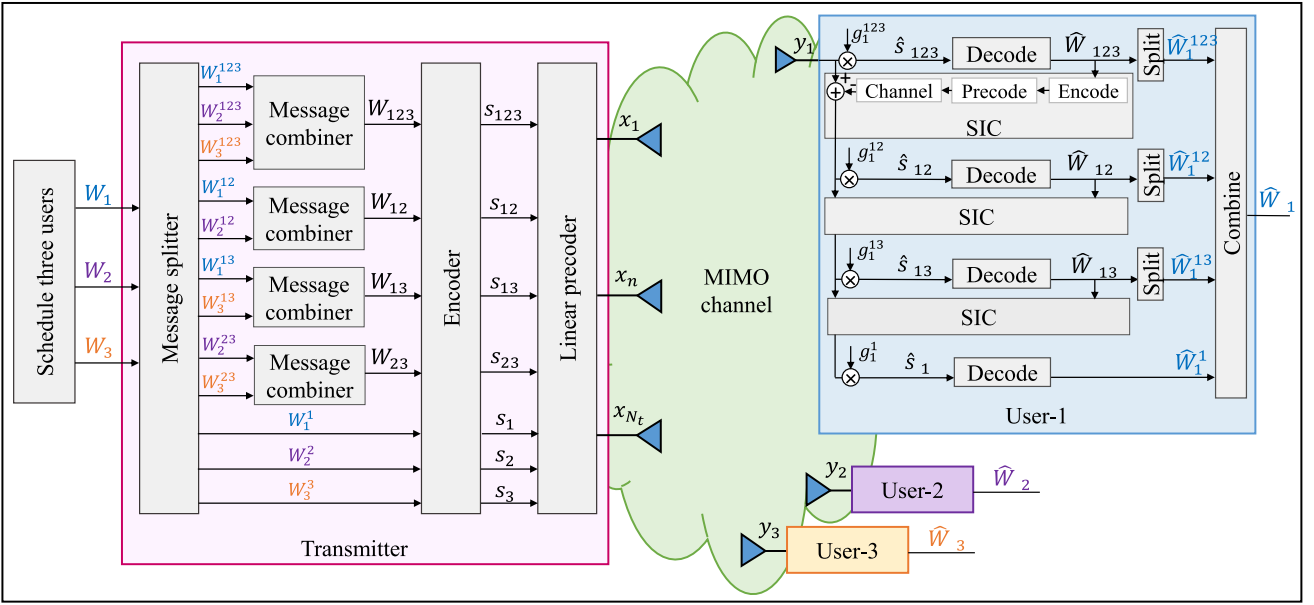}%
	\caption{Three-user transmission model using RS.}
	\label{fig: three-user transmission model}
\end{figure*}

 The vector of data streams to be transmitted is $\mathbf{{s}}=[s_{123},s_{12},s_{13},s_{23},s_{1},s_{2},s_{3}]^T$. After linear precoding using precoder $\mathbf{{P}}=[\mathbf{{p}}_{123},\mathbf{{p}}_{12},\mathbf{{p}}_{13},\mathbf{{p}}_{23},\mathbf{{p}}_{1},\mathbf{{p}}_{2},\mathbf{{p}}_{3}]$, the signals are superposed and broadcast. The decoding procedure when $K=3$ is more complex comparing with that in the two-user example. The main difference lies in decoding partial common streams for two-users. Define the streams to be decoded by $l$ users as \textit{$l$-order streams}. The 2-order streams to be decoded at user-1 are $s_{12}, s_{13}$. The 2-order streams to be decoded at user-2 and user-3 are $s_{12}, s_{23}$ and $s_{13}, s_{23}$, respectively.  As the 1-order and 2-order streams to be decoded at different users are not the same, we take user-1 as an example. The decoding procedure is the same for other users.
User-1 decodes four streams $s_{123},s_{12},s_{13},s_{1}$ based on SIC while treating other streams as noise. The decoding procedure starts from the 3-order stream (common stream) and progresses downwards to the 1-order stream (private stream). Specifically, user-1 first decodes $s_{123}$ and subtracts its contribution from the received signal. 
The SINR of the stream $s_{123}$ at user-$1$ is
\begin{equation}
\small 
\gamma_{1}^{123}=\frac{\left|\mathbf{{h}}_{1}^{H}\mathbf{{p}}_{123}\right|^{2}}{\sum_{i\in\{12,13,23\}}\left|\mathbf{{h}}_{1}^{H}\mathbf{{p}}_{i}\right|^{2}+\sum_{k=1}^3\left|\mathbf{{h}}_{1}^{H}\mathbf{{p}}_{k}\right|^{2}+1}.
\end{equation}
After that, user-1 decodes two streams $s_{12},s_{13}$ and treats interference of $s_{23}$ as noise. Both decoding orders of decoding $s_{12}$ followed by $s_{13}$ and  $s_{13}$ followed by $s_{12}$ should be considered in order to maximize the WSR. Denote $\pi_l$ as one of the decoding order to decode $l$-order streams. There is only one 1-order stream and one 3-order stream to be decoded at each user. Therefore, only one decoding order exists for both $\pi_1$ and $\pi_3$. In contrast, each user is required to decode two 2-order streams. Denote $s_{\pi_{2,k}{(i)}}$ as the $i$th data stream to be decoded at user-$k$ based on the decoding order $\pi_2$. One instance of $\pi_2$ is $12\rightarrow13\rightarrow23$, where $s_{12}$ is decoded before $s_{13}$ and $s_{13}$ is decoded before $s_{23}$ at all users. Since only data streams $s_{12}$ and $s_{13}$ are decoded at user-1, the decoding order at user-1 based on $\pi_2$ is $\pi_{2,1}=12\rightarrow13$. Hence, $s_{\pi_{2,1}{(1)}}=s_{12}$
and $s_{\pi_{2,1}{(2)}}=s_{13}$. The data stream $s_{\pi_{2,1}{(1)}}$ is decoded before  $s_{\pi_{2,1}{(2)}}$.  The SINRs of decoding streams $s_{\pi_{2,1}{(1)}}$ and $s_{\pi_{2,1}{(2)}}$ at user-1 are 
\begin{equation}
\small 
\gamma_{1}^{\pi_{2,1}{(1)}}=\frac{\left|\mathbf{{h}}_{1}^{H}\mathbf{{p}}_{\pi_{2,1}{(1)}}\right|^{2}}{\left|\mathbf{{h}}_{1}^{H}\mathbf{{p}}_{\pi_{2,1}{(2)}}\right|^{2}+\left|\mathbf{{h}}_{1}^{H}\mathbf{{p}}_{23}\right|^{2}+\sum_{k=1}^3\left|\mathbf{{h}}_{1}^{H}\mathbf{{p}}_{k}\right|^{2}+1}.
\end{equation}

\begin{equation}
\small 
\gamma_{1}^{\pi_{2,1}{(2)}}=\frac{\left|\mathbf{{h}}_{1}^{H}\mathbf{{p}}_{\pi_{2,1}{(2)}}\right|^{2}}{\left|\mathbf{{h}}_{1}^{H}\mathbf{{p}}_{23}\right|^{2}+\sum_{k=1}^3\left|\mathbf{{h}}_{1}^{H}\mathbf{{p}}_{k}\right|^{2}+1}.
\end{equation}
User-1 finally decodes $s_1$ by treating other data streams as noise. The three-user RS transmission model with the decoding order $\pi_2= 12\rightarrow13\rightarrow23$ is shown in Fig. \ref{fig: three-user transmission model}. The SINR of decoding $s_1$ at user-1 is
\begin{equation}
\small 
\gamma_{1}=\frac{\left|\mathbf{{h}}_{1}^{H}\mathbf{{p}}_1\right|^{2}}{\left|\mathbf{{h}}_{1}^{H}\mathbf{{p}}_{23}\right|^{2}+\sum_{k=2}^3\left|\mathbf{{h}}_{1}^{H}\mathbf{{p}}_{k}\right|^{2}+1}.
\end{equation}

\par The corresponding rate of each data stream is calculated in the same way as in the two-user example.
To ensure that $s_{123}$ is successfully decoded by all users, the achievable common rate shall not exceed $R_{123}=\min\left\{ R_{1}^{123},R_{2}^{123},R_{3}^{123}\right\}$. To ensure that $s_{12}$ is successfully decoded by user-1 and user-2,  the achievable common rate shall not exceed $R_{12}=\min\left\{ R_{1}^{12},R_{2}^{12}\right\}$. Similarly, we have $R_{13}=\min\left\{ R_{1}^{13},R_{3}^{13}\right\}$ and $R_{23}=\min\left\{ R_{2}^{23},R_{3}^{23}\right\}$.
All boundary points for the three-user RS rate region can be obtained by assuming that $R_{123}$, $R_{12}$, $R_{13}$ and $R_{23}$ are shared by the corresponding group of users. Denote the portion of the common rate allocated to user-$k$ for the message $s_{123}$ as $C_k^{123}$, we have $C_1^{123}+C_2^{123}+C_3^{123}=R_{123}$. Similarly, we have $C_1^{12}+C_2^{12}=R_{12}$, $C_1^{13}+C_3^{13}=R_{13}$, $C_2^{23}+C_3^{23}=R_{23}$.
Following the three-user RS structure described above, 
the total achievable rate of each user is 
$
R_{1,tot}=C_{1}^{123}+C_{1}^{12}+C_{1}^{13}+R_{1}
$
,
$
R_{2,tot}=C_{2}^{123}+C_{2}^{12}+C_{2}^{23}+R_{2}
$
 and 
$
R_{3,tot}=C_{3}^{123}+C_{3}^{13}+C_{3}^{23}+R_{3}.
$
For a given weight vector $\mathbf{u}=[u_1,u_2,u_3]$ and a fixed decoding order $\pi=[\pi_1,\pi_2,\pi_3]$, the WSR achieved by the three-user RS approach is
\begin{subequations}
	\label{eq:three users}
	\begin{align}
	R_{\mathrm{RS}_3}(\mathbf{u},\pi)&=\max_{\mathbf{{P}}, \mathbf{c}} \sum_{k=1}^{3}u_{k}R_{k,tot} \label{o1}\\
	\mbox{s.t.}\quad
	& C_1^{123}+C_2^{123}+C_3^{123}\leq R_{123}\\
	&	C_1^{12}+C_2^{12}\leq R_{12}\\
	&	C_1^{13}+C_3^{13}\leq R_{13}\\
	&	C_2^{23}+C_3^{23}\leq R_{23}\\
	&	\text{tr}(\mathbf{P}\mathbf{P}^{H})\leq P_{t} \\
	&   R_{k,tot}\geq R_k^{th}, k\in\{1,2,3\} \\
	& \mathbf{c}\geq \mathbf{0}
	\label{constraint3}
	\end{align}
\end{subequations}
where   $\mathbf{c}=[C_1^{123},C_2^{123},C_3^{123},C_1^{12},C_2^{12},C_1^{13},C_3^{13},C_2^{23},C_3^{23}]$ is the common rate vector required to be optimized in order to maximize the WSR. 
By calculating $R_{\mathrm{RS}_3}(\mathbf{u},\pi)$ for a set of different rate weights $\mathbf{u}$, we obtain the rate region $R_{\mathrm{RS}_3}(\pi)$ of a certain decoding order $\pi$.  The rate region of the three-user RS is achieved as the convex hull of the union over all decoding orders as 
$
R_{\mathrm{RS}}=\mathrm{conv}\left(\bigcup_{\pi}R_{\mathrm{RS}}(\mathbf{\pi})\right).
$

\par Similar to the two-user case, SC--SIC and MU--LP are again easily identified as special sub-strategies of RS by switching off some of the streams. Problem (\ref{eq:three users}) is non-convex and non-trivial. We propose a WMMSE algorithm to solve it as discussed in Section \ref{sec: algorithm}. 
  
\subsection{Generalized rate-splitting}
\label{sec: generalized RS}
\par We further propose a generalized RS framework for $K$ users. The users are indexed by the set $\mathcal{K}=\{1,\ldots,K\}$.  For any subset $\mathcal{A}$ of the users, $\mathcal{A}\subseteq\mathcal{K}$, the BS transmits a data stream $s_{\mathcal{A}}$ to be decoded by the users in the subset $\mathcal{A}$ while treated as noise by other users. $s_{\mathcal{A}}$ loads messages of all the users in the subset $\mathcal{A}$. The message intended for user-$k$ ($k\in\mathcal{K}$) is split as $\{ W_k^{\mathcal{A}'} | \mathcal{A}' \subseteq \mathcal{K}, k \in \mathcal{A}' \}$. The messages $\{W_{k'}^{\mathcal{A}}|k'\in\mathcal{A}\}$ of users with the same superscript $\mathcal{A}$  are encoded together into the stream  $s_{\mathcal{A}}$. 

\par The stream order defined in Section \ref{sec: three-user example} is applied to the generalized RS.  The stream order of data stream $s_{\mathcal{A}}$ is $|\mathcal{A}|$. For a given $l\in\mathcal{K}$, there are $K\choose l$ distinct $l$-order streams. For example, we have only one $K$-order stream (traditional common stream) while we have $K$ $1$-order streams (private steams).  Define $\mathbf{s}_l\in\mathbb{C}^{{K\choose l}\times 1}$ as the \textit{$l$-order data stream vector} formed by all $l$-order streams in $\{s_{\mathcal{A}'}|\mathcal{A}'\subseteq\mathcal{K},|\mathcal{A}'|=l\}$. Note that when $l=K$, there is a single $K$-order stream. $\mathbf{s}_{K}$ reduces to $s_{\mathcal{K}}$. For example, when $K=3$, the 3-order stream vector is $\mathbf{s}_{3}=s_{123}$. The 1-order and the 2-order stream vectors  are $\mathbf{s}_{1}=[s_{1},s_{2},s_{3}]^T$ and $\mathbf{s}_{2}=[s_{12},s_{13},s_{23}]^T$, respectively.  The data streams are linearly precoded via the precoding matrix $\mathbf{P}_{l}$ formed by $ \{\mathbf{p}_{\mathcal{A}'}|\mathcal{A}'\subseteq\mathcal{K},|\mathcal{A}'|=l\}$. 
The precoded streams are superposed and the resulting transmit signal is
\begin{equation}
\mathbf{{x}}=\sum_{l=1}^{K}\mathbf{{P}}_{{l}}\mathbf{{s}}_{{l}}=\sum_{l=1}^{K}\sum_{\mathcal{A}'\subseteq\mathcal{K},|\mathcal{A}'|=l}\mathbf{{p}}_{\mathcal{A}'}{{s}}_{\mathcal{A}'}.
\end{equation}

\par At user sides, each user is required to decode the intended streams based on SIC. The decoding procedure starts from the $K$-order stream and then goes down to the $1$-order stream.
A given user is involved in multiple $l$-order streams with an exception of the $K$-order and $1$-order streams.  
Denote $\pi_{l}$ as one of the decoding orders to decode the $l$-order data streams $\mathbf{s}_{l}$ for all users.
The $l$-order stream vector to be decoded at user-$k$ based on a certain decoding order  $\pi_{l}$ is $\mathbf{s}_{\pi_{l,k}}=[s_{\pi_{l,k}{(1)}},\cdots,s_{\pi_{l,k}{(|\mathcal{S}_{l,k}|)}}]^H$, where $\mathcal{S}_{l,k}=\{s_{\mathcal{A}'}|\mathcal{A}'\subseteq\mathcal{K},|\mathcal{A}'|=l,k\in\mathcal{A}'\}$ is the set of $l$-order streams to be decoded at user-$k$.  We assume $s_{\pi_{l,k}{(i)}}$ is decoded before $s_{\pi_{l,k}{(j)}}$ if $i<j$.
The SINR of user-$k$ to decode the $l$-order stream ${s}_{\pi_{l,k}{(i)}}$ with a certain decoding order $\pi_{l}$ is 
\begin{equation}
\small 
\gamma_{k}^{\pi_{l,k}{(i)}}=\frac{|\mathbf{{h}}_{k}^{H}\mathbf{{p}}_{\pi_{l,k}{(i)}}|^{2}}{I_{\pi_{l,k}{(i)}}+1},
\end{equation}
where
\[
\begin{aligned}
\small
I_{\pi_{l,k}{(i)}}&=\sum_{j>i}|\mathbf{{h}}_{k}^{H}\mathbf{{p}}_{\pi_{l,k}(j)}|^{2}+\sum_{l'=1}^{l-1}\sum_{j=1}^{|\mathcal{S}_{l',k}|}|\mathbf{{h}}_{k}^{H}\mathbf{{p}}_{\pi_{l',k}(j)}|^{2}\\
&+\sum_{\mathcal{A}'\subseteq\mathcal{K},k\notin\mathcal{A}'}|\mathbf{{h}}_{k}^{H}\mathbf{{p}}_{{\mathcal{A}'}}|^{2}
\end{aligned}\]  
is the interference at user-$k$ to decode  ${s}_{\pi_{l,k}{(i)}}$. $\sum_{j>i}|\mathbf{{h}}_{k}^{H}\mathbf{{p}}_{\pi_{l,k}(j)}|^{2}$ is the interference from the remaining non-decoded $l$-order streams in $\mathbf{s}_{{\pi_{l,k}}}$. $\sum_{l'=1}^{l-1}\sum_{j=1}^{|\mathcal{S}_{l',k}|}|\mathbf{{h}}_{k}^{H}\mathbf{{p}}_{\pi_{l',k}(j)}|^{2}$ is the interference from lower order streams $\mathbf{s}_{{\pi_{l',k}}},\forall l'<l$ to be decoded at user-$k$. $\sum_{\mathcal{A}'\subseteq\mathcal{K},k\notin\mathcal{A}'}|\mathbf{{h}}_{k}^{H}\mathbf{{p}}_{{\mathcal{A}'}}|^{2}$ is the interference from the streams that are not intended for user-$k$. The corresponding achievable rate of user-$k$ for the data stream ${s}_{\pi_{l,k}{(i)}}$ is 
$
R_k^{\pi_{l,k}{(i)}}=\log_{2}(1+\gamma_{k}^{\pi_{l,k}{(i)}}).
$
To ensure that the streams shared by more than two users are successfully decoded by all users, the achievable rate of each user in the subset $\mathcal{A}$ $(\mathcal{A}\in\mathcal{K},2\leq|\mathcal{A}|\leq K)$ to decode the $|\mathcal{A}|$-order stream $s_{\mathcal{A}}$ shall not exceed
\begin{equation}
\label{eq:min rate}
R_{\mathcal{A}}=\min_{k'}\left\{ R_{k'}^{\mathcal{A}}\mid k'\in\mathcal{A}\right\}.
\end{equation} 
For a given $l\in\mathcal{K}$, the $l$-order streams to be decoded at different users are different. $s_{\mathcal{A}}$ is decoded at  user-$k$ $(k\in\mathcal{A})$ based on the decoding order $\pi_{|\mathcal{A}|,k}$.   
$R_{\mathcal{A}}$ becomes the rate of receiving stream $s_{\mathcal{A}}$ at all users in the user group $\mathcal{A}$ with a certain decoding order $\pi_{|\mathcal{A}|}$. 
All boundary points for the $K$-user RS rate region can be obtained by assuming that  $R_{\mathcal{A}}$ is shared by all users in the user group  $\mathcal{A}$. Denote the portion of the common rate allocated to user-$k$ $(k\in \mathcal{A})$ as $C_k^{\mathcal{A}}$, we have $\sum_{k'\in \mathcal{A}}C_{k'}^{\mathcal{A}}=R_{\mathcal{A}}$.
Following the RS structure described above, 
the total achievable rate of user-$k$ is
\begin{equation}
R_{k,tot}=\sum_{\mathcal{A}'\subseteq\mathcal{K},k\in \mathcal{A}'}C_k^{\mathcal{A}'}+R_k,
\end{equation}
where $R_k$ is the rate of the 1-order stream $s_k$. It is intended for user-$k$ only. No common rate sharing is required for $R_k$.
For a given weight vector $\mathbf{u}=[u_1,\cdots,u_{K}]$ and a certain decoding order  $\pi=\{\pi_1,\ldots,\pi_{K}\}$, the WSR achieved by RS  is
\begin{equation}
\label{prob: K-user RS}
\begin{aligned}
R_{\mathrm{RS}}(\mathbf{u},\pi)&=\max_{\mathbf{{P}},\mathbf{c}}\sum_{k\in\mathcal{K}}u_{k}R_{k,tot}\\
\mbox{s.t.}\quad & \sum_{k'\in \mathcal{A}}C_{k'}^{\mathcal{A}}\leq R_{\mathcal{A}}, \forall \mathcal{A}\subseteq\mathcal{K}\\
&\text{tr}(\mathbf{P}\mathbf{P}^{H})\leq P_{t}\\
&   R_{k,tot}\geq R_k^{th}, k\in\mathcal{K}\\
& \mathbf{c}\geq \mathbf{0}
\end{aligned}
\end{equation}
$\mathbf{P}= [\mathbf{P}_{1},\ldots,\mathbf{P}_{K}]$ is the precoding matrix of all order streams. $\mathbf{c}$ is the common rate vector formed by $\{C_k^{\mathcal{A}}| \mathcal{A}\subseteq\mathcal{K}, k\in \mathcal{A}\}$.  For a fixed weight vector, problem (\ref{prob: K-user RS}) can be solved using the WMMSE approach  discussed in Section \ref{sec: algorithm} by establishing Rate-WMMSE relationships for all data streams. By calculating $R_{\mathrm{RS}}(\mathbf{u},\pi)$ for a set of different rate weights $\mathbf{u}$, we obtain the rate region $R_{\mathrm{RS}}(\pi)$ of a certain decoding order $\pi$. To achieve the rate region, all decoding orders should be considered. The capacity region of RS is defined as the convex hull of the union over all decoding orders as
\begin{equation}
R_{\mathrm{RS}}=\mathrm{conv}\left(\bigcup_{\pi}R_{\mathrm{RS}}(\mathbf{\pi})\right).
\end{equation}
\subsection{Structured and low-complexity rate-splitting}
\par The generalized RS described in Section \ref{sec: generalized RS} is able to provide more room for rate and QoS enhancements at the expense of more layers of SIC at receivers. Hence, though the generalized RS framework is very general and can be used to identify the best possible performance, its implementation can be complex due to the large number of SIC layers and common messages involved.  To overcome the problem, we introduce two low-complexity RS strategies for $K$ users, 1-layer RS and 2-layer Hierarchical RS (HRS). Those two RS strategies require the implementation of one and two layers of SIC at each receiver, respectively. 

\subsubsection{1-layer RS}
\label{sec:1-layer RS}
\par
\begin{figure}[t!]
	\centering
	\includegraphics[width=3.3in]{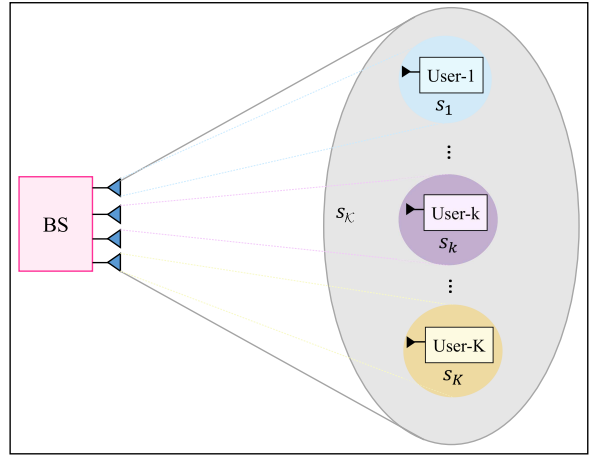}%
	\caption{1-layer RS model of $K$ users. The common stream $s_{\mathcal{K}}$ is shared by all the users.}
	\label{fig: 1-layer RS transmission model}
\end{figure}

 Instead of transmitting all order streams, 1-layer RS transmits the $K$-order common stream and 1-order private streams. Only one SIC is required at each receiver. The message of each user is split into two parts $\{W_{k}^{\mathcal{K}},W_{k}^k\},\forall k\in\mathcal{K}$. The messages $W_{1}^{\mathcal{K}},\ldots,W_{K}^{\mathcal{K}}$ are jointly encoded into the $K$-order stream ${s}_{\mathcal{K}}$ intended to be decoded by all users. $W_{k}^{k}$ is encoded into ${s}_{k}$ to be decoded by user-$k$ only. The overall data streams to be transmitted based on 1-layer RS is   $\mathbf{{s}}=[
s_{\mathcal{K}},s_{1},\ldots, s_{K}]^{T}$. The data streams are linearly precoded via precoder $\mathbf{{P}}=[
\mathbf{{p}}_{\mathcal{K}}, \mathbf{{p}}_{1},\ldots, \mathbf{{p}}_{K}]$.
The resulting transmit signal is
$
\mathbf{{x}}=\mathbf{{P}}\mathbf{{s}}=\mathbf{{p}}_{\mathcal{K}}s_{\mathcal{K}}+\sum_{k\in\mathcal{\mathcal{K}}}\mathbf{{p}}_{k}s_{k}.
$
Fig. \ref{fig: 1-layer RS transmission model} shows a 1-layer RS model. Readers are referred to Fig. 1 in  \cite{RSintro16bruno} for a detailed illustration of the 1-layer RS architecture. 

\par At user sides, all users firstly decode the data stream $s_{\mathcal{K}}$ by treating the interference from $s_1,\ldots,s_K$ as noise. The SINR of the $K$-order stream at user-$k$ is 
\begin{equation}
\gamma_{k}^{\mathcal{K}}=\frac{\left|\mathbf{{h}}_{k}^{H}\mathbf{{p}}_{\mathcal{K}}\right|^{2}}{\sum_{j\in\mathcal{K}}\left|\mathbf{{h}}_{k}^{H}\mathbf{{p}}_{j}\right|^{2}+1}.
\end{equation}
Once  $s_{\mathcal{K}}$ is successfully decoded, its contribution to the  original received signal  $y_k$ is subtracted. After that, user-$k$ decodes its private stream $s_{k}$ by treating the 1-order private streams of other users as noise. The SINR of decoding the private stream $s_{k}$ at user-$k$ is
\begin{equation}
\gamma_{k}=\frac{\left|\mathbf{{h}}_{k}^{H}\mathbf{{p}}_{k}\right|^{2}}{\sum_{j\in\mathcal{K},j\neq k}\left|\mathbf{{h}}_{k}^{H}\mathbf{{p}}_{j}\right|^{2}+1}.
\end{equation}

\par The corresponding achievable rates of user-$k$ for the streams $s_{\mathcal{K}}$ and $s_{k}$ are
$
R_{k}^{\mathcal{K}}=\log_{2}\left(1+\gamma_{k}^{\mathcal{K}}\right)
$ and $
R_{k}=\log_{2}\left(1+\gamma_{k}\right)
$.
To ensure that $s_{\mathcal{K}}$ is successfully decoded by all users, the achievable common rate shall not exceed $R_{\mathcal{K}}=\min\left\{ R_{1}^{\mathcal{K}},\ldots,R_{K}^{\mathcal{K}}\right\}$. 
$R_{\mathcal{K}}$ is shared among users such that $C_k^{\mathcal{K}}$ is the $k$th user's portion of the common rate
with $\sum_{k\in\mathcal{K}}C_k^{\mathcal{K}}=R_{\mathcal{K}}$. 
Following the two-user RS structure described above, 
the total achievable rate of user-$k$ is 
$
R_{k,tot}=C_{k}^{\mathcal{K}}+R_{k}.
$
For a given weight vector $\mathbf{u}=[u_1,\ldots,u_K]$, the WSR achieved by the $K$-user 1-layer RS approach is
\begin{subequations}
	\label{eq:onelayerRS}
	\begin{align}
	R_{\mathrm{1-layer RS}}(\mathbf{u})&=\max_{\mathbf{{P}}, \mathbf{c}}\sum_{k\in\mathcal{K}}u_{k}R_{k,tot}\\
	\mbox{s.t.}\quad
	& \sum_{k\in \mathcal{K}}C_k^{\mathcal{K}}\leq R_{\mathcal{K}}\\
	&	\text{tr}(\mathbf{P}\mathbf{P}^{H})\leq P_{t}\\
    &   R_{k,tot}\geq R_k^{th}, k\in\mathcal{K}\\
    & \mathbf{c}\geq \mathbf{0}
	\end{align}
\end{subequations}
where  $\mathbf{c}=[C_1^{\mathcal{K}},\ldots,C_K^{\mathcal{K}}]$.
For a given weight vector, problem (\ref{eq:onelayerRS}) can be solved using the WMMSE approach in \cite{RS2016hamdi}.

\par In contrast to NOMA, this 1-layer RS does not require any user ordering or grouping at the transmitter side since all users decode the common message (using single layer of SIC) before accessing their respective private messages. We also note that the 1-layer RS is a sub-scheme of the generalized RS and is a super-scheme of MU--LP (since by not allocating any power to the common message, the 1-layer RS boils down to MU--LP). However, for $K>2$, SC--SIC and SC--SIC per group are not sub-schemes of 1-layer RS (even though they were sub-schemes of the generalized RS). This explains why, in \cite{RSintro16bruno}, the authors already contrasted 1-layer RS and NOMA and expressed that the two strategies cannot be treated as extensions or subsets of each other. 
\par This 1-layer RS appeared in many scenarios subject to imperfect CSIT in \cite{RS2016hamdi,enrico2017bruno,enrico2016bruno,RS2016joudeh,hamdi2017bruno,RS2015bruno,minbo2017bruno,AP2017bruno}.

\subsubsection{2-layer HRS} 

\par The $K$ users are divided into $G$ groups $\mathcal{G}=\{1,\ldots,G\}$ with $\mathcal{K}_g,g\in\mathcal{G}$ users in each group. The user groups satisfy the same conditions as in Section \ref{sec: sc-sic per group}.  Besides the $K$-order stream and 1-order streams, 2-layer HRS also allows the transmission of a $|\mathcal{K}_g|$-order stream intended for users in $\mathcal{K}_g$. 
The overall data streams to be transmitted based on 2-layer RS is   $\mathbf{{s}}=[
s_{\mathcal{K}},s_{\mathcal{K}_1},\ldots,s_{\mathcal{K}_G},s_{1},\ldots, s_{K}]^{T}$. The data streams are linearly precoded via precoder $\mathbf{{P}}=[
\mathbf{{p}}_{\mathcal{K}}, \mathbf{{p}}_{\mathcal{K}_1},\ldots,\mathbf{{p}}_{\mathcal{K}_G},\mathbf{{p}}_{1},\ldots, \mathbf{{p}}_{K}]$.
The resulting transmit signal is
$
\mathbf{{x}}=\mathbf{{P}}\mathbf{{s}}=\mathbf{{p}}_{\mathcal{K}}s_{\mathcal{K}}+\sum_{g\in\mathcal{G}}\mathbf{{p}}_{\mathcal{K}_g}s_{\mathcal{K}_g}+\sum_{k\in\mathcal{\mathcal{K}}}\mathbf{{p}}_{k}s_{k}.
$
\begin{figure}[t!]
	\centering
	\includegraphics[width=3.3in]{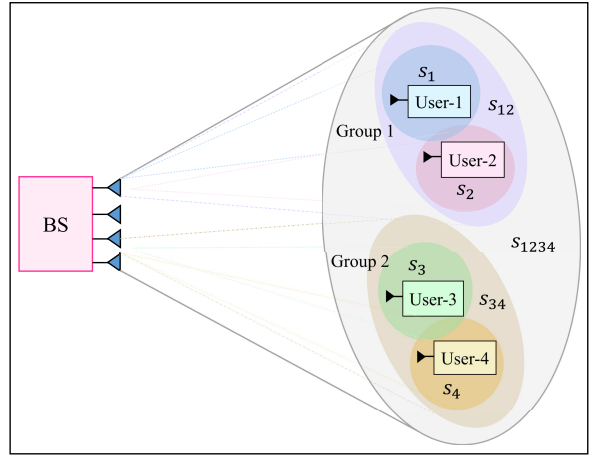}%
	\caption{2-layer HRS example, $K=4$, $G=2$, $\mathcal{K}_1=\{1,2\}$, $\mathcal{K}_2=\{3,4\}$. }
	\label{fig: four-user transmission model}
\end{figure}

\par Fig. \ref{fig: four-user transmission model} shows an example of 2-layer HRS. The users are divided into two groups, $\mathcal{K}_1=\{1,2\}$, $\mathcal{K}_2=\{3,4\}$. $s_{1234}$ is a $4$-order stream intended for all the users while $s_{12}$ and $s_{34}$ are $2$-order streams for users in each group only.

\begin{table*}[t!]
	\centering
	\caption{Comparison of different strategies}
	\label{table 1 fundamental}
	\begin{tabular}{|L{1.5cm}|L{3.5cm}|L{3.5cm}|L{3.5cm}|L{3.5cm}|}
		\hline
		\centering\textbf{Multiple Access}          & \multicolumn{2}{c|}{ \textbf{NOMA}}                                                                                                                                                                                                                     & \centering \textbf{SDMA}                                                         &                               \multicolumn{1}{c|}{ \textbf{RSMA}  }                                                \\ \hline
		\centering\textbf{Strategy}                 & \centering\textbf{SC--SIC}                                                                       & \centering\textbf{SC--SIC per group}                                                                                                                                    &\centering \textbf{MU--LP}                                                       &\multicolumn{1}{c|}{ \textbf{All forms of RS} }                                               \\ \hline
		\textbf{Design Principle}         & Fully decode interference                                                              & Fully decode interference in each group and treat interference between groups as noise                                                                        & Fully treat interference as noise                                     & Partially decode interference and partially treat interference as noise \\ \hline
		\textbf{Decoder architecture}     & SIC at receivers                                                                       & SIC at receivers                                                                                                                                              & Treat interference as noise                                           & SIC at receivers                                                        \\ \hline
		\textbf{User Deployment Scenario} & Users experience aligned channel directions and a large disparity in channel strengths. & Users in each group experience aligned channel directions and a large disparity in channel strengths. Users in different groups experience orthogonal channels. & Users channels are (semi-)orthogonal with similar channel strengths. & Any angle between channels and any disparity in channel strengths       \\ \hline
		\textbf{Network load}             & More suited to overloaded network                                                       & More suited to overloaded network                                                                                                                             & More suited to underloaded network                                    & Suited to any network load                                              \\ \hline
	\end{tabular}
\end{table*}

\par Each user is required to decode three streams $s_{\mathcal{K}}$, $s_{\mathcal{K}_g}$ and $s_{k}$. We assume $k\in\mathcal{K}_g$. The data stream $s_{\mathcal{K}}$ is decoded first by treating the interference from all other streams as noise. The SINR of the $K$-order stream at user-$k$ is
\begin{equation}
\gamma_{k}^{\mathcal{K}}=\frac{\left|\mathbf{{h}}_{k}^{H}\mathbf{{p}}_{\mathcal{K}}\right|^{2}}{\sum_{g\in\mathcal{G}}\left|\mathbf{{h}}_{k}^{H}\mathbf{{p}}_{\mathcal{K}_g}\right|^{2}+\sum_{j\in\mathcal{K}}\left|\mathbf{{h}}_{k}^{H}\mathbf{{p}}_{j}\right|^{2}+1}.
\end{equation}
Once  $s_{\mathcal{K}}$ is successfully decoded, its contribution to the  original received signal  $y_k$ is subtracted. After that, user-$k$ decodes its group common stream $s_{\mathcal{K}_g}$ by treating other group common streams and 1-order private streams as noise. The SINR of decoding the  $|\mathcal{K}_g|$-order stream $s_{\mathcal{K}_g}$ at user-$k$ is
\begin{equation}
\gamma_{k}^{\mathcal{K}_g}=\frac{\left|\mathbf{{h}}_{k}^{H}\mathbf{{p}}_{\mathcal{K}_g}\right|^{2}}{\sum_{g'\in\mathcal{G},g'\neq g}\left|\mathbf{{h}}_{k}^{H}\mathbf{{p}}_{\mathcal{K}_{g'}}\right|^{2}+\sum_{j\in\mathcal{K}}\left|\mathbf{{h}}_{k}^{H}\mathbf{{p}}_{j}\right|^{2}+1}.
\end{equation}
After removing its contribution to the received signal, user-$k$ decodes its private stream $s_{k}$. The SINR of decoding the private stream $s_{k}$ at user-$k$ is
\begin{equation}
\gamma_{k}=\frac{\left|\mathbf{{h}}_{k}^{H}\mathbf{{p}}_{k}\right|^{2}}{\sum_{g'\in\mathcal{G},g'\neq g}\left|\mathbf{{h}}_{k}^{H}\mathbf{{p}}_{\mathcal{K}_{g'}}\right|^{2}+\sum_{j\in\mathcal{K},j\neq k}\left|\mathbf{{h}}_{k}^{H}\mathbf{{p}}_{j}\right|^{2}+1}.
\end{equation}

\par The corresponding achievable rates of user-$k$ for the streams $s_{\mathcal{K}}$, $s_{\mathcal{K}_g}$ and $s_{k}$ are
$
R_{k}^{\mathcal{K}}=\log_{2}\left(1+\gamma_{k}^{\mathcal{K}}\right)
$,
$
R_{k}^{\mathcal{K}_g}=\log_{2}\left(1+\gamma_{k}^{\mathcal{K}_g}\right)
$  and $
R_{k}=\log_{2}\left(1+\gamma_{k}\right)
$.
The achievable common rate of $s_{\mathcal{K}}$ and $s_{\mathcal{K}_g}$ shall not exceed $R_{\mathcal{K}}=\min\left\{ R_{1}^{\mathcal{K}},\ldots,R_{K}^{\mathcal{K}}\right\}$ and $R_{\mathcal{K}_g}=\min_k\left\{ R_{k}^{\mathcal{K}_g}\mid k\in\mathcal{K}_g\right\}$, respectively. 
$R_{\mathcal{K}}$ is shared among users such that $C_k^{\mathcal{K}}$ is the $k$th user's portion of the common rate
with $\sum_{k\in\mathcal{K}}C_k^{\mathcal{K}}=R_{\mathcal{K}}$. $R_{\mathcal{K}_g}$ is shared among users in the group $\mathcal{K}_g$ such that $C_k^{\mathcal{K}_g}$ is the $k$th user's portion of the common rate
with $\sum_{k\in\mathcal{K}_g}C_k^{\mathcal{K}_g}=R_{\mathcal{K}_g}$. 
Following the two-user RS structure described above, 
the total achievable rate of user-$k$ is 
$
R_{k,tot}=C_{k}^{\mathcal{K}}+C_{k}^{\mathcal{K}_g}+R_{k},
$
where $k\in \mathcal{K}_g$.
For a given weight vector $\mathbf{u}=[u_1,\ldots,u_K]$, the WSR achieved by the $K$-user 2-layer HRS approach is
\begin{subequations}
	\label{eq:twolayerRS}
	\begin{align}
	R_{\mathrm{2-layer HRS}}(\mathbf{u})&=\max_{\mathbf{{P}}, \mathbf{c}}\sum_{k\in\mathcal{K}}u_{k}R_{k,tot}\\
	\mbox{s.t.}\quad
	& \sum_{k\in \mathcal{K}}C_k^{\mathcal{K}}\leq R_{\mathcal{K}}\\
	& \sum_{k\in \mathcal{K}_g}C_k^{\mathcal{K}_g}\leq R_{\mathcal{K}_g},\forall g\in\mathcal{G}\\
	&	\text{tr}(\mathbf{P}\mathbf{P}^{H})\leq P_{t}\\
	&   R_{k,tot}\geq R_k^{th}, k\in\mathcal{K}\\
	& \mathbf{c}\geq \mathbf{0}
	\end{align}
\end{subequations}
where $\mathbf{c}$ is the common rate vector formed by $\{C_k^{\mathcal{K}},C_{k'}^{\mathcal{K}_g}|  k\in \mathcal{K}, {k'}\in\mathcal{K}_g, g\in\mathcal{G}\}$.  For a given weight vector, problem (\ref{eq:twolayerRS}) can be solved by simply modifying the WMMSE approach discussed in Section \ref{sec: algorithm}. 

\par Comparing with SC--SIC per group where $|\mathcal{K}_g|-1$ layers of SIC are required at user sides, 2-layer HRS only requires 2 layers of SIC at each user.
Moreover,  the user ordering issue in SC--SIC per group does not exist in 2-layer HRS. The  streams of a higher stream order will always be decoded before the streams of a lower stream order. 1-layer RS is the simplest architecture since only 1 SIC is needed at each user and it is a sub-scheme of the 2-layer HRS. 
We also note that we can obtain a 1-layer RS per group from the 2-layer HRS by not allocating any power to $s_{\mathcal{K}}$. Note that SC--SIC and SC--SIC per group are not necessarily sub-schemes of the 2-layer HRS.
The 2-layer HRS strategy was first introduced in \cite{Mingbo2016}  in the Massive MIMO context.

\begin{table*}[t!]
	\centering
	\caption{Qualitative comparison of the complexity of different strategies}
	\label{table 2 complexity}
	\begin{tabular}{|L{1.4cm}|L{2.55cm}|L{2.8cm}|L{2.75cm}|L{2.7cm}|L{2.75cm}|}
		\hline
		\centering\textbf{Multiple Access}      & \multicolumn{2}{c|}{\textbf{NOMA}}                                                                                                                                                                                                                     & \centering\textbf{SDMA}                                         & \multicolumn{2}{c|}{\textbf{RSMA}}                                                                                                                        \\ \hline
		\textbf{Strategy}             & \textbf{SC--SIC}                                                                          & \textbf{SC--SIC per group}                                                                                                                                 & \textbf{MU--LP}                                       & \textbf{RS}                     & \textbf{1-layer RS}                                                                                                     \\ \hline
		\textbf{Encoder complexity} & encode $K$ streams & encode $K$ streams & encode $K$ streams
		& encode $K$ private streams plus additional common streams                  
		& encode $K+1$ streams \\ \hline
		\textbf{Scheduler complexity} & Very complex as it requires to find aligned users and decide upon suitable user ordering. & Very complex as it requires to divide users into orthogonal groups, with aligned users in each group and decide upon suitable user ordering in each group. & Complex as MU--LP requires to pair together semi-orthogonal users with similar channel gains. 
		& Complex as it requires to decide upon suitable decoding order of the streams with the same stream order                    
		& Simpler user scheduling as RS copes with any user deployment scenario, does not rely on user grouping and user ordering. \\ \hline
		\textbf{Receiver complexity}  & Requires multiple layers of SIC.  Subject to error propagation.                                                        & Requires multiple layers of SIC in each group and a single layer of SIC if groups are made of 2 users.    Subject to error propagation.                                                 & Does not require any SIC.                              & Requires multiple layers of SIC. Subject to error propagation. & Requires a single layer of SIC for all users.      Less subject to error propagation.                                                                     \\ \hline
	\end{tabular}
\end{table*}

\subsection{Encompassing existing NOMA and SDMA}
\par A comparison of NOMA, SDMA and RSMA is shown in Table \ref{table 1 fundamental}. Comparing with NOMA and SDMA, the most important characteristic of RSMA is that it partially decodes interference and partially treats interference as noise through the split into common and privates messages. This capability enables RSMA to maintain a good performance for all user deployment scenarios and all network loads, as it will appear clearer in the numerical results of Section \ref{simulation}.

\par Let us further discuss how the proposed framework of generalized RS in Section \ref{sec: generalized RS} contrasts and encompasses NOMA, SDMA and RS strategies.
We first compare the four-user MIMO--NOMA scheme illustrated in Fig. 5 of \cite{NOMA2013YSaito} with the four-user 2-layer HRS strategy illustrated in Fig. \ref{fig: four-user transmission model}. In Fig. 5 of \cite{NOMA2013YSaito}, user-1 and user-2 are superposed in the same beam. User-3 and user-4 share another beam. The users are decoded based on SC--SIC within each beam. As for the four-user 2-layer HRS strategy  in Fig. \ref{fig: four-user transmission model}, the encoded streams are precoded and transmitted jointly to users. If we set the common message $s_{12}$ to be encoded by the message of user-2 only and decoded by both user-1 and user-2, the common message $s_{34}$ to be encoded by the message of user-4 and decoded by user-3 and user-4, we also set the precoders  $\mathbf{p}_{12}$ and $\mathbf{p}_{1}$ to be equal, the precoders  $\mathbf{p}_{34}$ and $\mathbf{p}_{3}$ to be equal and the precoders of other streams to be $\mathbf{0}$, then the proposed RS scheme reduces to the scheme illustrated in Fig. 5 of \cite{NOMA2013YSaito}.  Similarly, the $K$-user RS model can be reduced to the $K$-user MIMO--NOMA scheme. Therefore, the MIMO--NOMA scheme proposed in \cite{NOMA2013YSaito} is a particular case of our RS framework.

\par In view of the above discussions, it should now be clear that SDMA and the multi-antenna NOMA strategies discussed in the introduction (relying on SC--SIC and SC--SIC per group) are all special instances of the generalized RS framework.

\par In the proposed generalized $K$-user RS model, if we set $\mathbf{P}_{{l}}=\mathbf{0}, \forall l\in\{2,\cdots,K\}$, only 1-order streams (private streams) are transmitted.  Each user only decodes its intended private stream by treating others as noise. Problem (\ref{prob: K-user RS}) is then reduced to the SDMA problem (\ref{opti prob:linear}). If the message of each user is encoded into one stream of distinct stream order, problem (\ref{prob: K-user RS}) is equivalent to the SC--SIC problem (\ref{opti prob:noma}). 
By keeping $1$-order and $K$-order streams, we have the 1-layer RS strategy whose performance benefit in the presence of imperfect CSIT was highlighted in various scenarios in \cite{RS2015bruno,RS2016hamdi,RS2016joudeh,AP2017bruno,enrico2017bruno,minbo2017bruno,enrico2016bruno,hamdi2017bruno}.  There is only one common data stream to be transmitted and decoded by all users before each user decodes its private stream. 
By keeping $1$-order, $K$-order and $l$-order streams, where $l$ is selected from $\{2,\cdots,K-1\}$, the problem becomes  the 2-layer HRS originally proposed in \cite{Mingbo2016} with two-layers of common messages to be transmitted. Another example of such a multi-layer RS has also appeared in the topological RS for MISO networks of \cite{chenxi2017brunotopology}.  Therefore, the formulated $K$-user RS problem is a more general problem. It encompasses SDMA, NOMA and existing RS methods as special cases.

\par Though the current work focuses on MISO BC, the RS framework can be extended to multi-antenna users and the general MIMO BC \cite{chenxi2017bruno} as well as to a general network scenario with multiple transmitters \cite{chenxi2017brunotopology}. Nevertheless the optimization of the precoders in those scenarios remain interesting topics for future research. Applications of this RS framework to relay networks is also worth exploring. Preliminary ideas have appeared in \cite{NOMA2017Zheng}, though joint encoding of the splitted common messages are not taken into account.

\subsection{Complexity of RSMA}

\par We further discuss the complexity of RSMA by comparing it with NOMA and SDMA. A qualitative comparison of NOMA, SDMA and RSMA is shown in Table \ref{table 2 complexity}. In Table \ref{table 2 complexity}, RS refers to the generalized RS of Section \ref{sec: generalized RS}.

\par As mentioned in the introduction,  the complexity of NOMA in the multi-antenna setup is increasing significantly at both the transmitter and the receivers. The optimal decoding order of NOMA is no longer fixed based on the channel gain as in the SISO BC. To maximize the WSR, the decoding order should be optimized together with precoders at the transmitter. Moreover, SC--SIC is suitable for aligned users with large channel gain difference. A proper user scheduling algorithm increases the scheduler complexity. At user sides, $K-1$ layers of SIC are required at each user for a $K$-user SC--SIC system. Increasing the number of users leads to a dramatic increase of the scheduler and receiver complexity, and is subject to more error propagation in the SICs.

\par SC--SIC per group reduces the complexity at user sides. Only $\left\lceil \frac{K}{G}\right\rceil $ layers of SIC are required at each user if we uniformly group the $K$ users into $G$ groups. However, the complexity at the transmitter increases with the number of user groups.  A joint design of user ordering and user grouping for all groups is necessary in order to maximize the WSR. For example, for a 4-user system, if we divide the users into 2 groups with 2 users in each group, we should consider 3 different user grouping methods and 4 different decoding orders for each grouping method. 

\par The complexity of MU--LP is much reduced as it does not require any SIC at user sides. However, as MU--LP is more suitable for users with (semi-)orthogonal channels and similar channel strengths, the transmitter requires accurate CSIT and user scheduling should be carefully designed for interference coordination. The scheduler complexity at the transmitter is still  high. 

\par Comparing with NOMA and SDMA, RSMA is able to balance the performance and complexity better. All forms of RS are suitable for users with any channel gain difference and any channel angle in between, though a multi-layer RS would have more flexibility. Considering the generalized RS, the decoding order of multiple  streams with the same stream order should be optimized together with the precoders  when there are multiple streams of the same stream order intended for each user (e.g.  each user decodes two 2-order streams in the three-user example of Section \ref{sec: three-user example}.). But its special case, 1-layer RS, simplifies both the scheduler and receiver design and it is still able to achieve a good performance in all user deployment scenarios. 1-layer RS requires only 1 SIC at each user. It does not rely on user grouping and user ordering for user scheduling. Therefore, the complexity of the scheduler is much simplified.

\par The cost of RSMA comes with a slightly higher encoding complexity since private and common streams need to be encoded. For the 1-layer RS in a $K$-user MISO BC, $K+1$ streams need to be encoded in contrast to $K$ streams for NOMA and SDMA.  
\subsection{Optimization  of RS}
\label{sec: algorithm}
\par The WMMSE approach proposed in  \cite{wmmse08}  is extended to solve the problem. The WMMSE algorithm to solve the sum rate maximization problem with 1-layer RS (discussed in Section \ref{sec:1-layer RS}) is proposed in \cite{RS2016hamdi}. We further extend it to solve the generalized RS problem (\ref{prob: K-user RS}). To simplify the explanation, we focus on the 3-user problem (\ref{eq:three users}). It can be easily extended to solve the $K$-user generalized RS problem.  

\par As the 1-order and 2-order streams to be decoded at different users are not the same, we take user-1 as an example. The procedure of the WMMSE algorithm is the same for other users. The signal received at user-1 is $y_1=\mathbf{h}_1^H\mathbf{{P}}\mathbf{{s}}+n_1$. It decodes four streams $s_{123}, s_{\pi_{2,1}(1)}, s_{\pi_{2,1}(2)}, s_1$ sequentially using SICs. The 3-order stream $s_{123}$ is decoded first. It is estimated as $\hat{s}_{123}=g_{1}^{123}y_1$, where $g_{1}^{123}$ is the equalizer. After successfully decoding and removing  $s_{123}$ from $y_1$, the estimate of  the 2-order stream $ s_{\pi_{2,1}(1)}$ is $\hat{s}_{\pi_{2,1}(1)}=g_{1}^{\pi_{2,1}(1)}(y_1-\mathbf{h}_1^H\mathbf{{p}}_{123}s_{123})$. Similarly, we calculate the estimates of $\hat{s}_{\pi_{2,1}(2)}$ and $\hat{s}_{1}$ as $\hat{s}_{\pi_{2,1}(2)}=g_{1}^{\pi_{2,1}(2)}(y_1-\mathbf{h}_1^H\mathbf{{p}}_{123}s_{123}-\mathbf{h}_1^H\mathbf{{p}}_{\pi_{2,1}(1)}s_{\pi_{2,1}(1)})$ and $\hat{s}_{1}=g_{1}^{1}(y_1-\mathbf{h}_1^H\mathbf{{p}}_{123}s_{123}-\mathbf{h}_1^H\mathbf{{p}}_{\pi_{2,1}(1)}s_{\pi_{2,1}(1)}-\mathbf{h}_1^H\mathbf{{p}}_{\pi_{2,1}(2)}s_{\pi_{2,1}(2)})$, respectively. $g_{1}^{\pi_{2,1}(1)},g_{1}^{\pi_{2,1}(2)}, g_{1}^{1}$ are the corresponding equalizers at user-1. The Mean Square Error (MSE) of each stream is defined as $\varepsilon_{k}\triangleq\mathbb{E}\{|s_{k}-\hat{{s}_{k}}|^{2}\}$. They are calculated as
\begin{equation}
\label{eq:MSE}
\scalebox{0.95}{\parbox{.5\linewidth}{%
		\begin{align}
		&\varepsilon_{1}^{123}=|g_{1}^{123}|^2T_1^{123}-2\Re\{g_{1}^{123}\mathbf{h}_1^H\mathbf{p}_{123}\}+1,\nonumber\\
		&\varepsilon_{1}^{\pi_{2,1}(1)}=|g_{1}^{\pi_{2,1}(1)}|^2T_1^{\pi_{2,1}(1)}-2\Re\{g_{1}^{\pi_{2,1}(1)}\mathbf{h}_1^H\mathbf{p}_{\pi_{2,1}(1)}\}+1,\nonumber\\
		&\varepsilon_{1}^{\pi_{2,1}(2)}=|g_{1}^{\pi_{2,1}(2)}|^2T_1^{\pi_{2,1}(2)}-2\Re\{g_{1}^{\pi_{2,1}(2)}\mathbf{h}_1^H\mathbf{p}_{\pi_{2,1}(2)}\}+1,\nonumber\\
		&\varepsilon_{1}^{1}=|g_{1}^{1}|^2T_1^{1}-2\Re\{g_{1}^{1}\mathbf{h}_1^H\mathbf{p}_1\}+1 \nonumber
		\end{align}
}}
\end{equation}
where $T_1^{123}\triangleq|\mathbf{h}_1^H\mathbf{p}_{123}|^2+|\mathbf{h}_1^H\mathbf{p}_{12}|^2+|\mathbf{h}_1^H\mathbf{p}_{13}|^2+|\mathbf{h}_1^H\mathbf{p}_{23}|^2+|\mathbf{h}_1^H\mathbf{p}_{1}|^2+|\mathbf{h}_1^H\mathbf{p}_{2}|^2+|\mathbf{h}_1^H\mathbf{p}_{3}|^2+1$ is the receive power at user-1. $T_1^{\pi_{2,1}(1)}\triangleq T_1^{123}-|\mathbf{h}_1^H\mathbf{p}_{123}|^2$, $T_1^{\pi_{2,1}(2)}\triangleq T_1^{\pi_{2,1}(1)}-|\mathbf{h}_1^H\mathbf{p}_{\pi_{2,1}(1)}|^2$, $T_1^{1}\triangleq T_1^{\pi_{2,1}(2)}-|\mathbf{h}_1^H\mathbf{p}_{\pi_{2,1}(2)}|^2$.
The optimum MMSE equalizers are
\begin{equation}
\label{eq:MMSE}
\scalebox{0.95}{\parbox{.5\linewidth}{%
		\begin{align}
		&(g_{1}^{123})^{\mathrm{MMSE}}=(\mathbf{p}_{123})^H\mathbf{h}_1({T}_1^{123})^{-1},\nonumber\\
		&(g_{1}^{\pi_{2,1}(1)})^{\mathrm{MMSE}}=(\mathbf{p}_{\pi_{2,1}(1)})^H\mathbf{h}_1({T}_1^{\pi_{2,1}(1)})^{-1},\nonumber\\
		&(g_{1}^{\pi_{2,1}(2)})^{\mathrm{MMSE}}=(\mathbf{p}_{\pi_{2,1}(2)})^H\mathbf{h}_1({T}_1^{\pi_{2,1}(2)})^{-1},\nonumber\\
		&(g_{1}^{1})^{\mathrm{MMSE}}=(\mathbf{p}_1)^H\mathbf{h}_1({T}_1^{1})^{-1}.\nonumber
		\end{align}
}}
\end{equation}
They are calculated by solving \scalebox{0.85}{$\frac{\partial\varepsilon_{1}^{123}}{\partial g_{1}^{123}}=0$}, \scalebox{0.85}{$\frac{\partial\varepsilon_{1}^{\pi_{2,1}(1)}}{\partial g_{1}^{\pi_{2,1}(1)}}=0$}, \scalebox{0.85}{$\frac{\partial\varepsilon_{1}^{\pi_{2,1}(2)}}{\partial g_{1}^{\pi_{2,1}(2)}}=0,\frac{\partial\varepsilon_{1}^{1}}{\partial g_{1}^{1}}=0$}. Substituting (\ref{eq:MMSE}) into (\ref{eq:MSE}), the MMSEs become
\begin{equation}
\label{eq:opt MMSE}
\scalebox{0.95}{\parbox{.5\linewidth}{%
		\begin{align}
		&(\varepsilon_{1}^{123})^{\textrm{MMSE}}\triangleq\min_{g_{1}^{123}} \varepsilon_{1}^{123} =({T}_1^{123})^{-1}{I}_1^{123},\nonumber\\
		&(\varepsilon_{1}^{\pi_{2,1}(1)})^{\textrm{MMSE}}\triangleq\min_{g_{1}^{\pi_{2,1}(1)}} \varepsilon_{1}^{\pi_{2,1}(1)} =({T}_1^{\pi_{2,1}(1)})^{-1}{I}_1^{\pi_{2,1}(1)},\nonumber\\
		&(\varepsilon_{1}^{\pi_{2,1}(2)})^{\textrm{MMSE}}\triangleq\min_{g_{1}^{\pi_{2,1}(2)}} \varepsilon_{1}^{\pi_{2,1}(2)} =({T}_1^{\pi_{2,1}(2)})^{-1}{I}_1^{\pi_{2,1}(2)},\nonumber\\
		&(\varepsilon_{1}^{1})^{\textrm{MMSE}}\triangleq\min_{g_{1}^{1}} \varepsilon_{1}^{1} =({T}_1^{1})^{-1}{I}_1^{1}, \nonumber
		\end{align}
}}
\end{equation}
where ${I}_1^{123}=T_1^{\pi_{2,1}(1)}$, ${I}_1^{\pi_{2,1}(1)}=T_1^{\pi_{2,1}(2)}$, 
${I}_1^{\pi_{2,1}(2)}=T_1^{1}$,
${I}_1^{1}=T_1^{1}-|\mathbf{h}_1^H\mathbf{p}_{1}|^2$. Based on (\ref{eq:opt MMSE}), the SINRs of decoding the intended streams at user-1 can be expressed as $\gamma_1^{123}={1}/{(\varepsilon_{1}^{123})^{\textrm{MMSE}}}-1$,
$\gamma_1^{\pi_{2,1}(1)}={1}/{(\varepsilon_{1}^{\pi_{2,1}(1)})^{\textrm{MMSE}}}-1$,
$\gamma_1^{\pi_{2,1}(2)}={1}/{(\varepsilon_{1}^{\pi_{2,1}(2)})^{\textrm{MMSE}}}-1$,
$\gamma_1^{1}={1}/{(\varepsilon_{1}^{1})^{\textrm{MMSE}}}-1$.
The corresponding rates are rewritten as  $R_1^{123}=-\log_{2}((\varepsilon_{1}^{123})^{\textrm{MMSE}})$,  $R_1^{\pi_{2,1}(1)}=-\log_{2}((\varepsilon_{1}^{\pi_{2,1}(1)})^{\textrm{MMSE}})$, $R_1^{\pi_{2,1}(2)}=-\log_{2}((\varepsilon_{1}^{\pi_{2,1}(2)})^{\textrm{MMSE}})$, $R_1^{1}=-\log_{2}((\varepsilon_{1}^{1})^{\textrm{MMSE}})$. 
The augmented WMSEs are 
\begin{equation}
\label{eq: augmented WMEs}
\scalebox{0.95}{\parbox{.5\linewidth}{%
		\begin{align}
		&\xi_{1}^{123}=u_1^{123}\varepsilon_{1}^{123}-\log_{2}(u_1^{123}),\nonumber\\
		&\xi_{1}^{\pi_{2,1}(1)}=u_1^{\pi_{2,1}(1)}\varepsilon_{1}^{\pi_{2,1}(1)}-\log_{2}(u_1^{\pi_{2,1}(1)}),\nonumber\\
		&\xi_{1}^{\pi_{2,1}(2)}=u_1^{\pi_{2,1}(2)}\varepsilon_{1}^{\pi_{2,1}(2)}-\log_{2}(u_1^{\pi_{2,1}(2)}),\nonumber\\
		&\xi_{1}^{1}=u_1^{1}\varepsilon_{1}^{1}-\log_{2}(u_1^{1}),\nonumber
		\end{align}
}}
\end{equation}
where $u_1^{123},u_1^{\pi_{2,1}(1)},u_1^{\pi_{2,1}(2)},u_1^{1}$ are weights associated with each stream at user-1. By solving \scalebox{0.9}{$\frac{\partial\xi_{1}^{123}}{\partial g_{1}^{123}}=0,\frac{\partial\xi_{1}^{\pi_{2,1}(1)}}{\partial g_{1}^{\pi_{2,1}(1)}}=0,\frac{\partial\xi_{1}^{\pi_{2,1}(2)}}{\partial g_{1}^{\pi_{2,1}(2)}}=0,\frac{\partial\xi_{1}^{1}}{\partial g_{1}^{1}}=0$}, we derive the optimum equalizers as $(g_{1}^{123})^*=(g_{1}^{123})^{\textrm{MMSE}}$, $(g_{1}^{\pi_{2,1}(1)})^*=(g_{1}^{\pi_{2,1}(1)})^{\textrm{MMSE}}$, 
$(g_{1}^{\pi_{2,1}(2)})^*=(g_{1}^{\pi_{2,1}(2)})^{\textrm{MMSE}}$, 
$(g_{1}^{1})^*=(g_{1}^{1})^{\textrm{MMSE}}$. Substituting the optimum equalizers into (\ref{eq: augmented WMEs}), we obtain
\begin{equation}
\label{eq: augmented WMSs opt g}
\scalebox{0.88}{\parbox{.5\linewidth}{%
		\begin{align}
		&\xi_{1}^{123}\left((g_{1}^{123})^{\textrm{MMSE}}\right)=u_1^{123}(\varepsilon_{1}^{123})^{\textrm{MMSE}}-\log_{2}(u_1^{123}),\nonumber\\
		&\xi_{1}^{\pi_{2,1}(1)}\left((g_{1}^{\pi_{2,1}(1)})^{\textrm{MMSE}}\right)=u_1^{\pi_{2,1}(1)}(\varepsilon_{1}^{\pi_{2,1}(1)})^{\textrm{MMSE}}-\log_{2}(u_1^{\pi_{2,1}(1)}),\nonumber\\
		&\xi_{1}^{\pi_{2,1}(2)}\left((g_{1}^{\pi_{2,1}(2)})^{\textrm{MMSE}}\right)=u_1^{\pi_{2,1}(2)}(\varepsilon_{1}^{\pi_{2,1}(2)})^{\textrm{MMSE}}-\log_{2}(u_1^{\pi_{2,1}(2)}),\nonumber\\
		&\xi_{1}^{1}\left((g_{1}^{1})^{\textrm{MMSE}}\right)=u_1^{1}(\varepsilon_{1}^{1})^{\textrm{MMSE}}-\log_{2}(u_1^{1}).\nonumber
		\end{align}
}}
\end{equation}
By further solving \scalebox{0.8}{$\frac{\partial\xi_{1}^{123}\left((g_{1}^{123})^{\textrm{MMSE}}\right)}{\partial u_{1}^{123}}=0$}, \scalebox{0.8}{$\frac{\partial\xi_{1}^{\pi_{2,1}(1)}\left((g_{1}^{\pi_{2,1}(1)})^{\textrm{MMSE}}\right)}{\partial u_{1}^{\pi_{2,1}(1)}}=0$}, \scalebox{0.8}{$\frac{\partial\xi_{1}^{\pi_{2,1}(2)}\left((g_{1}^{\pi_{2,1}(2)})^{\textrm{MMSE}}\right)}{\partial u_{1}^{\pi_{2,1}(2)}}=0$}, \scalebox{0.8}{$\frac{\partial\xi_{1}^{1}\left((g_{1}^{1})^{\textrm{MMSE}}\right)}{\partial u_{1}^{1}}=0$}, we obtain the optimum MMSE weights as
\begin{equation}
\label{eq: opt u}
\scalebox{0.95}{\parbox{.5\linewidth}{%
		\begin{align}
		&(u_1^{123})^*=(u_1^{123})^{\textrm{MMSE}}\triangleq((\varepsilon_{1}^{123})^{\textrm{MMSE}})^{-1},\nonumber\\
		&(u_1^{\pi_{2,1}(1)})^*=(u_1^{\pi_{2,1}(1)})^{\textrm{MMSE}}\triangleq((\varepsilon_{1}^{\pi_{2,1}(1)})^{\textrm{MMSE}})^{-1},\nonumber\\
		&(u_1^{\pi_{2,1}(2)})^*=(u_1^{\pi_{2,1}(2)})^{\textrm{MMSE}}\triangleq((\varepsilon_{1}^{\pi_{2,1}(2)})^{\textrm{MMSE}})^{-1},\nonumber\\
		&(u_1^{1})^*=(u_1^{1})^{\textrm{MMSE}}\triangleq((\varepsilon_{1}^{1})^{\textrm{MMSE}})^{-1}.\nonumber
		\end{align}
}}
\end{equation}
Substituting (\ref{eq: opt u}) into (\ref{eq: augmented WMSs opt g}), we establish the Rate-WMMSE relationship as
\begin{equation}
\label{eq: rate-wmmse}
\scalebox{0.95}{\parbox{.5\linewidth}{%
		\begin{align}
		&(\xi_{1}^{123})^{\textrm{MMSE}}\triangleq\min_{u_{1}^{123},g_{1}^{123}}\xi_{1}^{123}=1-R_1^{123},\nonumber\\
		&(\xi_{1}^{\pi_{2,1}(1)})^{\textrm{MMSE}}\triangleq\min_{u_{1}^{\pi_{2,1}(1)},g_{1}^{\pi_{2,1}(1)}}\xi_{1}^{\pi_{2,1}(1)}=1-R_1^{\pi_{2,1}(1)},\nonumber\\
		&(\xi_{1}^{\pi_{2,1}(2)})^{\textrm{MMSE}}\triangleq\min_{u_{1}^{\pi_{2,1}(2)},g_{1}^{\pi_{2,1}(2)}}\xi_{1}^{\pi_{2,1}(2)}=1-R_1^{\pi_{2,1}(2)},\nonumber\\
		&(\xi_{1}^{1})^{\textrm{MMSE}}\triangleq\min_{u_{1}^{1},g_{1}^{1}}\xi_{1}^{1}=1-R_1^{1}.\nonumber
		\end{align}
}}
\end{equation}
Similarly, we can establish the Rate-WMMSE relationships for user-2 and user-3. Motivated by the Rate-WMMSE relationship in (\ref{eq: rate-wmmse}), we reformulate the optimization problem (\ref{eq:three users}) as
\begin{subequations}
	\label{eq:three users wmmse}
	\begin{align}
	&\min_{\mathbf{{P}}, \mathbf{x},\mathbf{u},\mathbf{g}} \sum_{k=1}^{3}u_{k}\xi_{k,tot} \label{object wmmse}\\
	\mbox{s.t.}\quad
	& X_{1}^{123}+X_{2}^{123}+X_{3}^{123}+1\geq \xi_{123}\\
	&X_{1}^{12}+X_{2}^{12}+1\geq \xi_{12}\\
	&X_{1}^{13}+X_{3}^{13}+1\geq \xi_{13}\\
	&X_{2}^{23}+X_{3}^{23}+1\geq \xi_{23}\\
	&	\text{tr}(\mathbf{P}\mathbf{P}^{H})\leq P_{t} \\
	&   \xi_{k,tot}\leq 1-R_k^{th}, k\in\{1,2,3\} \\
	& \mathbf{x}\leq \mathbf{0}
	\end{align}
\end{subequations}
where \scalebox{0.9}{ $\mathbf{x}=[X_{1}^{123},X_{2}^{123},X_{3}^{123},X_{1}^{12},X_{2}^{12},X_{1}^{13},X_{3}^{13},X_{2}^{23},X_{3}^{23}]$}.  \scalebox{0.97}{$\mathbf{u}=[u_1^{123},u_2^{123},u_3^{123},u_1^{12},u_2^{12},u_1^{13},u_3^{13},u_2^{23},u_3^{23},u_1^{1},u_2^{2},u_3^{2}]$}. \scalebox{0.99}{$\mathbf{g}=[g_1^{123},g_2^{123},g_3^{123},g_1^{12},g_2^{12},g_1^{13},g_3^{13},g_2^{23},g_3^{23},g_1^{1},g_2^{2},g_3^{2}]$}. \scalebox{0.96}{$\xi_{1,tot}=X_{1}^{123}+X_{1}^{12}+X_{1}^{13}+\xi_{1}^{1}$}, $\xi_{tot}=X_{2}^{123}+X_{2}^{12}+X_{2}^{23}+\xi_{2}^{2}$ and
\scalebox{0.96}{$\xi_{3,tot}=X_{3}^{123}+X_{3}^{13}+X_{3}^{23}+\xi_{3}^{3}$} are individual WMSEs. 
\scalebox{0.94}{$\xi_{123}=\max\left\{ \xi_{1}^{123},\xi_{2}^{123},\xi_{3}^{123}\right\}$}, 
\scalebox{0.93}{$\xi_{12}=\max\left\{ \xi_{1}^{12},\xi_{2}^{12}\right\}$}, \scalebox{0.93}{$\xi_{13}=\max\left\{ \xi_{1}^{13},\xi_{3}^{13}\right\}$}, \scalebox{0.93}{$\xi_{23}=\max\left\{ \xi_{2}^{23},\xi_{3}^{23}\right\}$} are the achievable WMSEs of the corresponding streams. 

\par It can be easily shown that by minimizing (\ref{object wmmse}) with respect to $\mathbf{u}$ and $\mathbf{g}$, respectively, we obtain the MMSE solutions \scalebox{0.94}{$(\mathbf{u}^{\mathrm{MMSE}}, \mathbf{g}^{\mathrm{MMSE}})$} formed by the corresponding MMSE equalizers and weights. They satisfy the KKT optimality conditions of (\ref{eq:three users wmmse}) for $\mathbf{P}$. Therefore, according to the Rate-WMMSE relationship (\ref{eq: rate-wmmse}) and the common rate transformation $\mathbf{c}=-\mathbf{x}$, problem (\ref{eq:three users wmmse})  can be  transformed to problem (\ref{eq:three users}). For any point ($\mathbf{x}^*,\mathbf{P}^*,\mathbf{u}^*,\mathbf{g}^*$) satisfying the KKT optimality conditions of (\ref{eq:three users wmmse}), the solution given by ($\mathbf{c}^*=-\mathbf{x}^*,\mathbf{P}^*$) satisfies the KKT optimality conditions of (\ref{eq:three users}). 
The WSR problem (\ref{eq:three users}) is then transformed into the WMMSE problem (\ref{eq:three users wmmse}). The problem (\ref{eq:three users wmmse}) is still non-convex for the joint optimization of ($\mathbf{x},\mathbf{P},\mathbf{u},\mathbf{g}$). We have derived that when ($\mathbf{x},\mathbf{P},\mathbf{u}$) are fixed, the optimal equalizer is the MMSE equalizer $\mathbf{g}^{\mathrm{MMSE}}$. When ($\mathbf{x},\mathbf{P},\mathbf{g}$) are fixed, the optimal weight is the MMSE weight $\mathbf{u}^{\mathrm{MMSE}}$. When ($\mathbf{u},\mathbf{g}$) are fixed, $(\mathbf{x},\mathbf{P})$ are coupled in the optimization problem (\ref{eq:three users wmmse}), closed form solution can not be derived. But it is a convex Quadratically Constrained Quadratic Program (QCQP) which can be solved using interior-point methods. These properties motivates us to use AO to solve the problem. In $n$th iteration of the AO algorithm, the equalizers and weights are firstly  updated using the precoders obtained in the $n-1$th iteration $(\mathbf{u},\mathbf{g})=\left(\mathbf{u}^{\mathrm{MMSE}}(\mathbf{P}^{[n-1]}), \mathbf{g}^{\mathrm{MMSE}}(\mathbf{P}^{[n-1]})\right)$. With the updated $(\mathbf{u},\mathbf{g})$, $(\mathbf{x},\mathbf{P})$ can then be updated by solving the problem (\ref{eq:three users wmmse}). $(\mathbf{u},\mathbf{g})$ and $(\mathbf{x},\mathbf{P})$ are iteratively updated until the WSR converges. The details of the AO algorithm is shown in Algorithm \ref{WMMSE algorithm}, where $\textrm{WSR}^{[n]}$ is the WSR calculated based on the updated $(\mathbf{x},\mathbf{P})$ in $n$th iteration. $\epsilon$ is the tolerance of the algorithm. The AO algorithm is guaranteed to converge as the WSR is increasing in each iteration and it is bounded above for a given power constraint.

\begin{algorithm}[h!]
	
	\textbf{Initialize}: $n\leftarrow0$, $\mathbf{P}^{[n]}$, $\mathrm{WSR}^{[n]}$\;
	\Repeat{$|\mathrm{WSR}^{[n]}-\mathrm{WSR}^{[n-1]}|\leq \epsilon$}{
		$n\leftarrow n+1$\;
		$\mathbf{P}^{[n-1]}\leftarrow \mathbf{P}$\;
		$\mathbf{u}\leftarrow\mathbf{u}^{\mathrm{MMSE}}(\mathbf{P}^{n-1})$\; $\mathbf{g}\leftarrow\mathbf{g}^{\mathrm{MMSE}}(\mathbf{P}^{n-1})$\;
		update $(\mathbf{x},\mathbf{P})$ by solving (\ref{eq:three users wmmse}) using the updated $\mathbf{u}$ and $\mathbf{g}$;

	}
	
	\caption{Alternating Optimization Algorithm}
	\label{WMMSE algorithm}
\end{algorithm}

\par When considering imperfect CSIT, we follow the robust approach proposed in \cite{RS2016hamdi} for 1-layer RS with imperfect CSIT. The precoders are optimized based on the available channel estimate to maximize a conditional Averaged Weighted Sum Rate (AWSR) metric, computed using partial CSIT knowledge. The stochastic AWSR problem was transformed into a deterministic counter part using the Sample Average Approximated (SAA) method. Then the Rate-WMMSE relationship is applied to transform the AWSR problem into a convex form and solved using an AO algorithm.  
The robust approach for 1-layer RS in \cite{RS2016hamdi} can be easily extended to solve the $K$-user generalized RS problem based on our proposed Algorithm 1, which will not be explained here.

\section{Numerical Results}
\label{simulation}
\par In this section, we evaluate the performance of  SDMA, NOMA  and RSMA  in a wide range of network loads (underloaded and overloaded regimes) and user deployments (with a diversity of channel directions, channel strengths and qualities of Channel State Information at the Transmitter). 
We first illustrate the rate region of different strategies in the two-user case followed by the WSR comparisons of the three-user, four-user and ten-user cases.
\subsection{Underloaded two-user deployment with perfect CSIT}

\par  When $K=2$, the rate region of all strategies can be explicitly compared in a two-dimensional figure. As mentioned earlier, the rate region is the set of all achievable points. Its boundary is calculated by varying the weights assigned to users.  In this work, the weight of user-1 is fixed to $u_1=1$.  The weight of user-$2$ is varied as $u_2 = 10^{[-3, -1,-0.95,\cdots ,0.95,1, 3]}$, which is the same as in \cite{wmmse08}.  To investigate the largest achievable rate region, the individual rate constraints are set to 0 in all strategies $R_k^{th}=0,\forall k\in \{1,2\}$.

 \par In the perfect CSIT scenario, the capacity region is achieved by DPC. Therefore, we compare the rate regions of different beamforming strategies with the DPC region.
The DPC region is generated using the algorithm in \cite{DPCregion}.
  Since the WSR problems for all beamforming strategies described earlier are non-convex, the initialization of $\mathbf{P}$ is vital to the final result. 
  It has been observed in \cite{RS2016hamdi} that Maximum Ratio Transmission (MRT) combined with Singular Value Decomposition (SVD) provides good overall performance over various channel realizations. It is used in this work for precoder initialization of RS. The precoders for the private message $\mathbf{p}_k$ is initialized as $\mathbf{p}_k=p_k\frac{\mathbf{h}_k}{\left\Vert  \mathbf{h}_k\right\Vert}$, where $p_k=\frac{\alpha P_t}{2}$ and $0 \leq \alpha\leq 1$. The precoder for the common message is initialized as $\mathbf{p}_{12}=p_{12}\mathbf{u}_{12}$, where $p_{12}=(1-\alpha)P_t$ and $\mathbf{u}_{12}$ is the largest left singular vector of the channel matrix $\mathbf{H}=[\mathbf{h}_1, \mathbf{h}_2]$. It is calculated as $\mathbf{u}_{12}=\mathbf{U}(:,1)$. $\mathbf{U}$ is derived based on the SVD of $\mathbf{H}$, i.e., $\mathbf{H}=\mathbf{U}\mathbf{S}\mathbf{V}^{H}$. To ensure a fair comparison, the precoders of MU--LP are initialized based on MRT. For SC--SIC,  the precoder of the user decoded first is initialized based on SVD and that of the user decoded last is initialized based on MRT.

\subsubsection{Random channel realizations}

\begin{figure}[t!]
	\centering
	\includegraphics[width=3.4in]{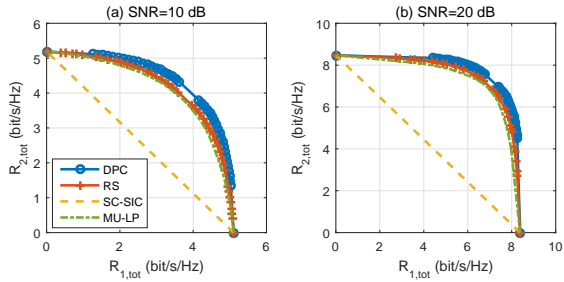}%
	\caption{Achievable rate region comparison of different strategies in underloaded two-user deployment with perfect CSIT, averaged over 100 random channel realizations, $\sigma_1^2=1, \sigma_2^2=1$, $N_t=4$.}
	\label{fig: random bias11}
\end{figure}

\begin{figure}[t!]
	\centering
	\includegraphics[width=3.4in]{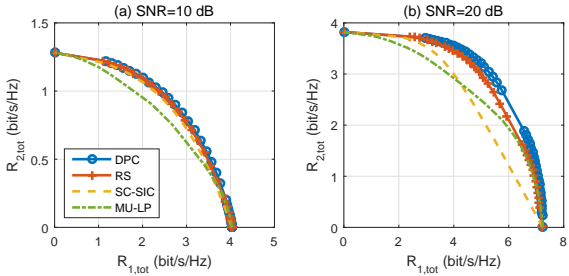}%
	\caption{Achievable rate region comparison of different strategies in underloaded two-user deployment with perfect CSIT, averaged over 100 random channel realizations, $\sigma_1^2=1, \sigma_2^2=0.09$, $N_t=2$.}
	\label{fig: random bias103}
\end{figure}
\par We firstly consider the scenarios when the channel of each user $\mathbf{h}_k$ has independent and identically distributed (i.i.d.)  complex Gaussian entries with a certain variance, i.e., $\mathcal{CN}(0,\sigma_k^2)$. The BS is equipped with two or four antennas ($N_t=2,4$) and serves two single-antenna users. Fig. \ref{fig: random bias11} shows the average rate regions of different strategies over 100 random channel realizations when $\sigma_1^2=1, \sigma_2^2=1$, $N_t=4$. SNRs are 10 dB and 20 dB, respectively. When the number of transmit antenna is larger than the number of users, MU--LP achieves a good performance. The generated precoders of the users tend to be more orthogonal as the number of transmit antennas increases. In contrast, the average rate region achieved by SC--SIC is small. When $\sigma_1^2=1, \sigma_2^2=1$, there is no disparity of average channel strengths. SC--SIC is not able to achieve a good performance in such scenario.  As the SC--SIC strategy is motivated by leveraging the channel strength difference among users, it achieves a good performance when the channels are degraded.  Specifically, when the channels of users are close to alignment, SC--SIC works better than MU--LP if the users have asymmetric channel strengths. However, for the general non-degraded MISO-BC, SC--SIC often yields a performance loss \cite{quasidegrade2016}. 
The simulation results when $\sigma_1^2=1, \sigma_2^2=0.09$, $N_t=2$ is illustrated in Fig. \ref{fig: random bias103}.  The average channel gain difference between the users increases to 5 dB and the number of the transmit antenna reduces to two. In such scenario, the rate region gap between RS and MU--LP increases while the rate region gap between RS and SC--SIC decreases. It shows that SC--SIC is more suited to the scenarios where the users experience a large disparity in channel strengths. In both Fig.  \ref{fig: random bias11} and Fig. \ref{fig: random bias103}, the rate region gaps among different strategies increase with SNR. RS achieves a larger rate region than SC--SIC and MU--LP and it is closer to the capacity region achieved by DPC.

 \begin{figure}[t!]
 	\centering
 	\includegraphics[width=3.4in]{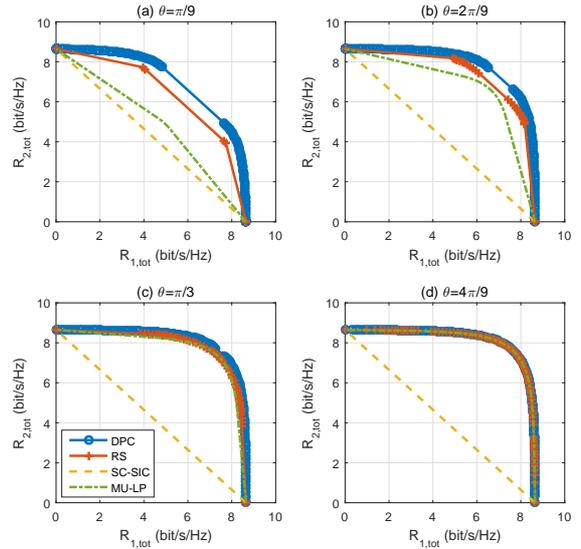}%
 	\caption{Achievable rate region comparison of different strategies in underloaded two-user deployment with perfect CSIT, $\gamma=1$, $N_t=4$, SNR=20 dB.}
 	\label{fig: snr20 bias1 nt4}
 \end{figure}
 
 \begin{figure}[t!]
 	\centering
 	\includegraphics[width=3.4in]{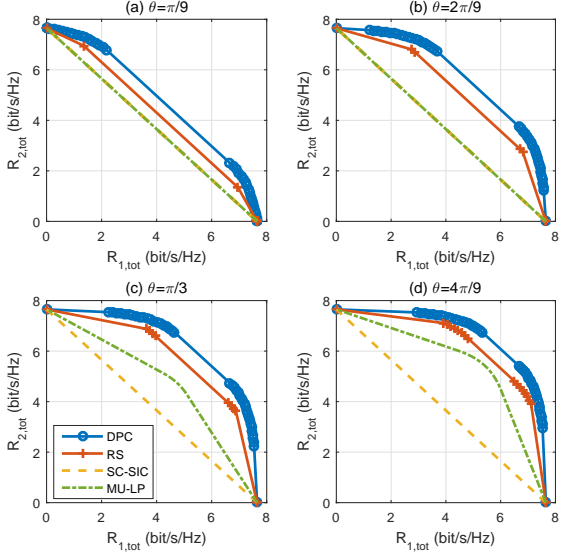}%
 	\caption{Achievable rate region comparison of different strategies in underloaded two-user deployment with perfect CSIT, $\gamma=1$, $N_t=2$, SNR=20 dB.}
 	\label{fig: snr20 bias1 nt2}
 \end{figure}
 \subsubsection{Specific channel realizations}
 \par In order to have a better insight into the benefits of RS over MU--LP and SC--SIC, we investigate the influence of user angle and channel strength on the performance.  When $N_t=4$, the channels of users are realized as 
 \begin{equation}
 	\begin{aligned}
 		& \mathbf{h}_1=\left[1, 1, 1, 1\right]^H,\\
 		& \mathbf{h}_2=\gamma\times\left[
 		1,e^{j\theta},e^{j2\theta},e^{j3\theta}\right]^H.
 	\end{aligned}
 \end{equation}
In above channel realizations,
$\gamma$ and  $\theta$ are control variables. $\gamma$ controls the channel strength of user-$2$. If $\gamma=1$, the channel strength of user-$1$ is equal to that of user-$2$.  If $\gamma=0.3$, user-$2$ suffers from an additional $5$ dB path loss compared to user-$1$.  $\theta$ controls the angle between the channels of user-$1$ and user-$2$. It varies from $0$ to $\frac{\pi }{2}$. If $\theta=0$, the channel of user-$1$ is aligned with that of user-$2$. If $\theta=\frac{\pi }{2}$, the channels of user-$1$ and user-$2$ are orthogonal to each other. 
In the following results,  $\gamma=1, 0.3$, which corresponds to 0 dB, 5 dB channel strength difference, respectively. For each  $\gamma$,  $\theta$ adopts value from $\theta=\left[\frac{\pi }{9},\frac{2\pi }{9},\frac{\pi }{3},\frac{4\pi }{9}\right]$.
Intuitively,  when $\theta$ is less than $\frac{\pi }{9}$, the channels of users are sufficiently aligned and SC--SIC performs well. When $\theta$ is larger than $\frac{4\pi }{9}$, the channels of users are sufficiently orthogonal to each other and MU--LP is more suitable. Therefore, we consider angles within the range of  $\left[\frac{\pi }{9}, \frac{4\pi }{9}\right]$.  SNR is fixed to 20 dB. When $N_t=2$, the channels of user-1 and user-2 are realized as $ \mathbf{h}_1=\left[1, 1\right]^H$ and $\mathbf{h}_2=\gamma\times\left[
1,e^{j\theta}\right]^H$, respectively. The same values of $\gamma$ and $\theta$ are adopted in $N_t=2$ as used in $N_t=4$.\footnote{Note that for a given $\theta$,  the users' Direction of Arrival (DoA) are the same for $N_t=2$ and $N_t=4$ scenarios while the channel angle  is more orthogonal when $N_t=4$ comparing with that when $N_t=2$. }

\par Fig. \ref{fig: snr20 bias1 nt4} shows the results when $\gamma=1$, $N_t=4$.  In all subfigures, the rate region achieved by RS is equal to or larger than that of SC--SIC and MU--LP.
When $\gamma=1$ and $ \theta=\frac{\pi }{9}$, the channels of user-$1$ and user-$2$ almost coincide. 
 RS exhibits a clear rate region improvement over SC--SIC and MU--LP. SC--SIC cannot achieve a good performance due to the equal channel gain while the performance of MU--LP is poor when the user channels are closely aligned to each other.
 As $\theta$ increases, the gap between the rate regions of RS and MU--LP  reduces as the performance of MU--LP  is better when the channels of users are more orthogonal to each other while the gap between the rate regions of  MU--LP and SC--SIC increases. The rate regions of RS and MU--LP tend to the capacity region achieved by DPC as $\theta$ increases. As shown in Fig. \ref{fig: snr20 bias1 nt4}(d), when the channels of users are sufficiently orthogonal to each other, the rate regions of DPC, RS and MU--LP  are almost identical. In such an orthogonal scenario, RS reduces to MU--LP.

\par  Fig. \ref{fig: snr20 bias1 nt2} shows the results when $\gamma=1$, $N_t=2$. In all subfigures, RS outperforms MU--LP and SC--SIC.  Comparing with the results of  $N_t=4$, the rate region gap between RS and MU--LP is enlarged when $N_t=2$. When the number of transmit antenna decreases, it becomes more difficult for MU--LP to design orthogonal precoders for users. MU--LP is more suited to underloaded scenarios ($N_t>K$).  In both Fig. \ref{fig: snr20 bias1 nt4} and Fig. \ref{fig: snr20 bias1 nt2}, the rate region of SC--SIC is the worst due to the equal channel gain. In contrast, RS performs well for any angle between user channels.

\begin{figure}[t!]
	\centering
	\includegraphics[width=3.4in]{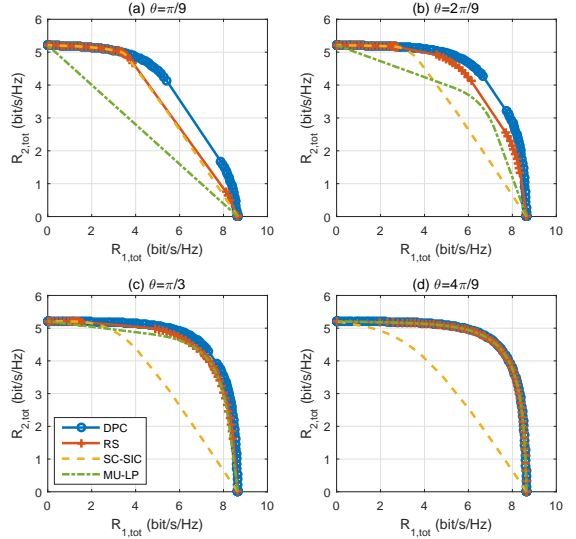}%
	\caption{Achievable rate region comparison of different strategies in perfect CSIT, $\gamma=0.3$, $N_t=4$, SNR=20 dB.}
	\label{fig: snr20 bias03 nt4}
\end{figure}

\begin{figure}[t!]
	\centering
	\includegraphics[width=3.4in]{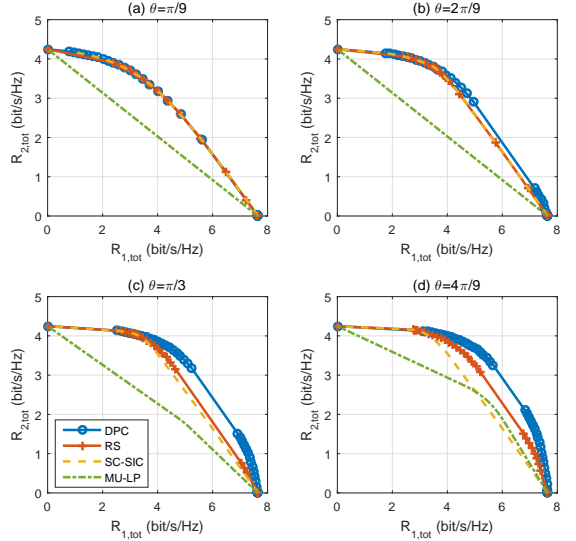}%
	\caption{Achievable rate region comparison of different strategies in perfect CSIT, $\gamma=0.3$, $N_t=2$, SNR=20 dB.}
	\label{fig: snr20 bias03 nt2}
\end{figure}

\par Fig. \ref{fig: snr20 bias03 nt4}  shows the rate region comparison of DPC, RS, SC--SIC and MU--LP transmission schemes with $5$ dB channel strength difference between the two users, i.e., $\gamma=0.3$ and $N_t=4$. RS and SC--SIC are much closer to the DPC region in the setting of Fig. \ref{fig: snr20 bias03 nt4} compared to Fig. \ref{fig: snr20 bias1 nt4} because of the $5$ dB channel strength difference. Fig. \ref{fig: snr20 bias03 nt4}(b) and \ref{fig: snr20 bias03 nt4}(c) are interesting as SC--SIC and MU--LP outperform each other at one part of the rate region. There is a crosspoint between the two schemes in each figure mentioned. The rate region of RS is equal to or larger than the convex hull of the rate regions of SC--SIC and MU--LP.


\par Fig. \ref{fig: snr20 bias03 nt2}  shows the rate region comparison when $\gamma=0.3$, $N_t=2$. Comparing Fig. \ref{fig: snr20 bias03 nt2} with Fig. \ref{fig: snr20 bias03 nt4}, SC--SIC achieves a relatively  better performance when the number of transmit antenna reduces. The WSRs of RS and SC--SIC are overlapped and they almost achieve the capacity region when $\theta=\frac{\pi}{9}$. However, as $\theta$ increases, the rate region gap between RS and SC--SIC increases despite the 5 dB channel gain difference. 
 Both  SC--SIC and RS rely on one SIC when there are two users in  the system. Though the receiver complexity of SC--SIC and RS are the same, RS achieves explicit performance gain over SC--SIC in most investigated scenarios.  Comparing with MU--LP and SC--SIC, RS is suited to any channel angles and channel gain difference.

\par More results of  underloaded two-user deployments with perfect CSIT are given in Appendix A. We further illustrate the rate regions of different strategies when SNR is 10 dB. Comparing the corresponding figures of 10 dB and 20 dB, we conclude that as SNR increases, the gaps among the rate regions of different schemes increase, with RS exhibiting further performance benefits. In all investigated scenarios, RS always outperforms MU--LP and SC--SIC.

\subsection{Underloaded two-user deployment with imperfect CSIT}
\par Next we investigate the rate region of different transmission schemes in the presence of imperfect CSIT. We assume the users are able to estimate the channel perfectly while the instantaneous channel estimated at the BS is imperfect. We assume the estimated channel of user-1 and user-2 are   $
\nonumber
\widehat{\mathbf{h}}_{1}=\left[
	1,1,1,1\right]^H
$
and 
$
\nonumber
\widehat{\mathbf{h}}_2=\gamma\times\left[
1, e^{j\theta}, e^{j2\theta}, e^{j3\theta}\right]^H
$ when $N_t=4$. 
For the given channel estimate at the BS, the channel realization  is $\mathbf{h}_k=\widehat{\mathbf{h}}_{k}+\widetilde{\mathbf{h}}_{k},\forall k\in\{1,2\}$, where $\widetilde{\mathbf{h}}_{k}$ is the estimation error of user-$k$. $\widetilde{\mathbf{h}}_{k}$ has i.i.d. complex Gaussian entries drawn from $\mathcal{CN}(0,\sigma_{e,k}^2)$. The error covariance of user-1 and user-2 are $\sigma_{e,1}^2=P_t^{-0.6}$ and $\sigma_{e,2}^2=\gamma P_t^{-0.6}$, respectively. The precoders are initialized and designed using the estimated channels $\widehat{\mathbf{h}}_{1},\widehat{\mathbf{h}}_{2}$ and the same methods as stated in perfect CSIT scenarios. 1000 different channel error samples are generated for each user. Each point in the rate region is the average rate\footnote{The readers are referred to \cite{RS2016hamdi} for a rigorous discussion about the notion of average rate.} over the generated 1000 channels. SNR is fixed to 20 dB.

\par
\begin{figure}[t!]
	\centering
	\includegraphics[width=3.3in]{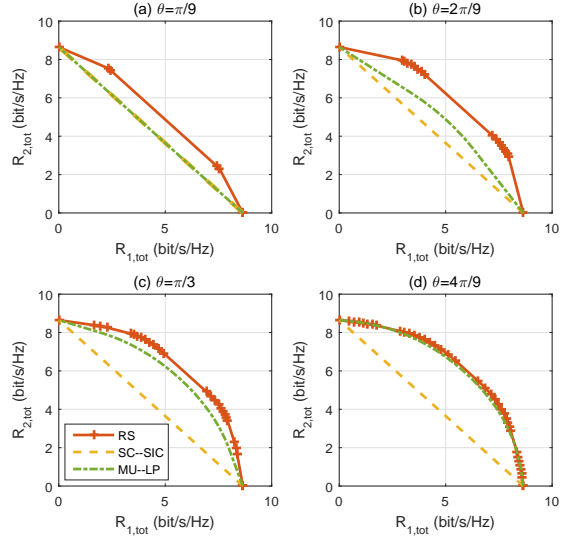}%
	\caption{ Average rate region comparison of different strategies in imperfect CSIT, $\gamma=1$, $N_t=4$, SNR=20 dB.}
	\label{fig: snr20 bias1 nt4 imperfect}
\end{figure}

\begin{figure}[t!]
	\centering
	\includegraphics[width=3.3in]{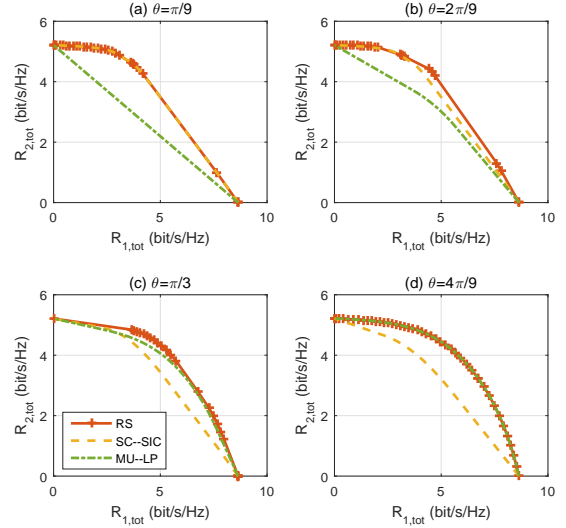}%
	\caption{ Average rate region comparison of different strategies in imperfect CSIT, $\gamma=0.3$, $N_t=4$, SNR=20 dB.}
	\label{fig: snr20 bias03 nt4 imperfect}
\end{figure}

  Fig. \ref{fig: snr20 bias1 nt4 imperfect} and Fig. \ref{fig: snr20 bias03 nt4 imperfect} show the results when $\gamma=1$ and $\gamma=0.3$, respectively.  Similarly to the results in perfect CSIT,  the gaps between the rate regions of RS and MU--LP reduce as $\theta$ increases in both figures. When $\theta=\frac{4\pi}{9}$, the channels of the two users are sufficiently orthogonal. The rate regions of RS and MU--LP are almost identical. SC--SIC achieves a good performance when the channels of users are sufficiently aligned with enough channel gain difference, as shown in Fig. \ref{fig: snr20 bias03 nt4 imperfect}(a).

\par Comparing Fig. \ref{fig: snr20 bias1 nt4 imperfect} and Fig. \ref{fig: snr20 bias1 nt4}, the rate region gap between RS and MU--LP increases in imperfect CSIT due to the residual interference introduced. The interference-nulling in MU--LP is distorted and yields residual interference at the receiver, which jeopardizes the achievable rate. 
In contrast, the rate region gap between RS and SC--SIC slightly reduces in imperfect CSIT, as observed by comparing Fig. \ref{fig: snr20 bias03 nt4 imperfect} with Fig. \ref{fig: snr20 bias03 nt4}. SC--SIC is less sensitive to CSIT inaccuracy comparing with MU--LP. However, the rate region gap between RS and SC--SIC is still obvious.
In comparison, RS is more flexible and robust to multi-user interference originating from the imperfect CSIT, as evidenced by the recent literature on RS with imperfect CSIT \cite{DoF2013SYang,RS2015bruno,Mingbo2016,RS2016hamdi,RS2016joudeh,chenxi2017brunotopology,AP2017bruno,chenxi2017bruno,enrico2017bruno,minbo2017bruno,enrico2016bruno}. With RS, the amount of interference decoded by both users (through the presence of common stream) is adjusted dynamically to the channel conditions (channel directions and strengths) and CSIT inaccuracy.

\par More results of  underloaded two-user deployments with imperfect CSIT are given in Appendix B. The rate regions of different strategies for varied SNR, $N_t$ and $\gamma$ are illustrated. We further show that the performance of RS is stable in a wide range of parameters, namely number of transmit antennas, user deployments and CSIT inaccuracy. RS achieves equal or better performance than MU--LP and SC--SIC in all simulated channels.

\subsection{Underloaded three-user deployment with perfect CSIT}

\par When $K=3$, the rate region of each strategy is a three-dimensional surface. The gaps among rate regions of different strategies are difficult to display. As each point of the rate region is derived by solving the WSR problem with a fixed weight vector $\mathbf{u}$, the WSRs instead of the rate regions of different transmission strategies are compared in the three-user case.

\par Two RS schemes are investigated in three-user deployments. RS refers to the generalized RS strategy of Section \ref{sec: three-user example} and 1-layer RS refers to the low-complexity RS strategy of Section \ref{sec:1-layer RS}. We compare the WSR of RS, 1-layer RS, DPC, SC--SIC and MU--LP.  The beamforming initialization of different strategies is extended based on the methods adopted in the two-user case. There are three streams of distinct stream orders in RS (1/2/3-order streams). The precoders of the streams are initialized differently. The transmit power $P_t$ is divided into three parts $\alpha_1P_t$, $\alpha_2P_t$, $\alpha_3P_t$  for   streams of three distinct stream orders, where $\alpha_1,\alpha_2,\alpha_3\in[0,1]$ and $\alpha_1+\alpha_2+\alpha_3=1$. The precoder $\mathbf{p}_k, \forall  k\in\{1,2,3\}$ of the 1-order stream (private stream) $s_k$ is initialized as $\mathbf{p}_k=p_k\frac{\mathbf{h}_k}{\left\Vert  \mathbf{h}_k\right\Vert}$, where $p_k=\frac{\alpha_1 P_t}{3}$ is the allocated power. The precoders  $\mathbf{p}_{12},\mathbf{p}_{13},\mathbf{p}_{23}$ of the 2-order streams are respectively initialized  as $\mathbf{p}_{12}=p_{12}\mathbf{u}_{12},\mathbf{p}_{13}=p_{13}\mathbf{u}_{13},\mathbf{p}_{23}=p_{23}\mathbf{u}_{23}$, where $p_{12}=p_{13}=p_{23}=\frac{\alpha_2P_t}{3}$ and $\mathbf{u}_{12}$ is the largest left singular vector of the channel matrix $\mathbf{H}_{12}=[\mathbf{h}_1, \mathbf{h}_2]$. Similarly, $\mathbf{u}_{13}$ and $\mathbf{u}_{23}$ are the largest left singular vectors of the channel matrices $\mathbf{H}_{13}=[\mathbf{h}_1, \mathbf{h}_3]$ and $\mathbf{H}_{23}=[\mathbf{h}_2, \mathbf{h}_3]$, respectively. The precoder  $\mathbf{p}_{123}$ of the 3-order stream (conventional common stream) $s_{123}$ is initialized  as $\mathbf{p}_{123}=p_{123}\mathbf{u}_{123}$, where $p_{123}=\alpha_3P_t$ and $\mathbf{u}_{123}$ is the largest left singular vector of the channel matrix $\mathbf{H}_{123}=[\mathbf{h}_1, \mathbf{h}_2, \mathbf{h}_3]$. 
The beamforming initialization of 1-layer RS is similar as RS except we have $\mathbf{p}_{123}$ and $\mathbf{p}_k,\forall k\in\{1,2,3\}$ only. By setting $\alpha_2=0$, the initialization of RS is applied to 1-layer RS. To ensure a fair comparison, the precoders of MU--LP are initialized based on MRT. For SC--SIC,  the precoder of the user decoded first $\mathbf{p}_{\pi(1)}$ is initialized as $\mathbf{p}_{\pi(1)}=p_{\pi(1)}\mathbf{u}_{\pi(1)}$, where $p_{\pi(1)}=\alpha_3P_t$ and $\mathbf{u}_{\pi(1)}$ is the largest left singular vector of the channel matrix $\mathbf{H}_{123}=[\mathbf{h}_1, \mathbf{h}_2, \mathbf{h}_3]$. The precoder of the user decoded secondly $\mathbf{p}_{\pi(2)}$ is initialized as $\mathbf{p}_{\pi(2)}=p_{\pi(2)}\mathbf{u}_{\pi(2)}$, where $p_{\pi(2)}=\alpha_2P_t$ and $\mathbf{u}_{\pi(2)}$ is the largest left singular vector of the channel matrix $\mathbf{H}_{\pi({23})}=[\mathbf{h}_{\pi({2})}, \mathbf{h}_{\pi({3})}]$. The user decoded last is initialized based on MRT. 

\begin{figure}[t!]
	\centering
	\includegraphics[width=3.4in]{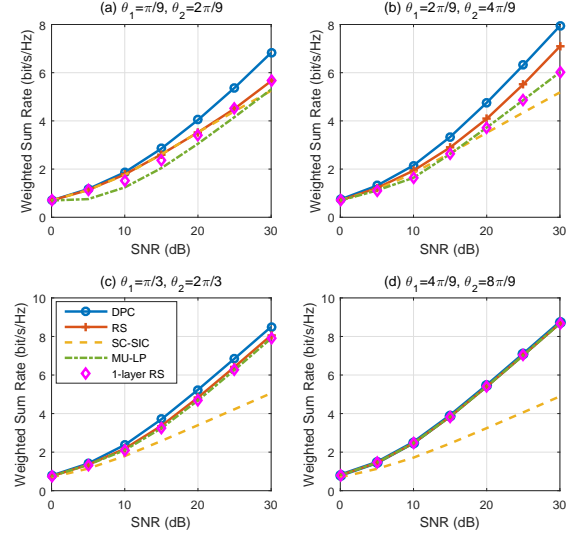}%
	\caption{Weighted sum rate versus SNR comparison of different strategies for underloaded three-user deployment with perfect CSIT, $\gamma_1=1, \gamma_2=0.3$, $u_1=0.2, u_2=0.3, u_3=0.5$, $N_t=4$, $R_k^{th}=0, k\in\{1,2,3\}$.}
	\label{fig: three user bias1103 020305 perfect}
\end{figure}

\par We firstly consider an underloaded scenario. The BS is equipped with four transmit antennas ($N_t=4$) and serves three single-antenna users in all simulations. The individual rate constraint is set to 0, $R_k^{th}=0,\forall k\in\{1,2,3\}$. The channel of users are realized as
 \begin{equation}
 \label{eq: channel three users}
\begin{aligned}
& \mathbf{h}_1=\left[1, 1, 1, 1\right]^H,\\
&\mathbf{h}_2=\gamma_1\times\left[
1, e^{j\theta_1},e^{j2\theta_1}, e^{j3\theta_1}\right]^H,\\
&\mathbf{h}_3=\gamma_2\times\left[
1, e^{j\theta_2}, e^{j2\theta_2}, e^{j3\theta_2}\right]^H.
\end{aligned}
\end{equation}
$\gamma_1, \gamma_2$ and  $\theta_1,\theta_2$ are control variables as discussed in the two-user case. 
For a given set of $\gamma_1,\gamma_2$,  $\theta_1$ adopts value from $\theta_1=\left[\frac{\pi }{9},\frac{2\pi }{9},\frac{\pi }{3},\frac{4\pi }{9}\right]$ and $\theta_2=2\theta_1$.
When $\theta_1=\frac{\pi }{9}, \theta_2=\frac{2\pi }{9}$, the channels of user-$1$ and user-$2$, user-$2$ and user-$3$ are sufficiently aligned. When $\theta_1=\frac{4\pi }{9}, \theta_2=\frac{8\pi }{9}$, the channels of user-$1$ and user-$2$, user-$2$ and user-$3$ are sufficiently orthogonal. We consider SNRs within the range 0 dB to 30 dB.  We assume the sum of the weights allocated to users is equal to one, i.e., $u_1+u_2+u_3=1$.

\begin{figure}[t!]
	\centering
	\includegraphics[width=3.6in]{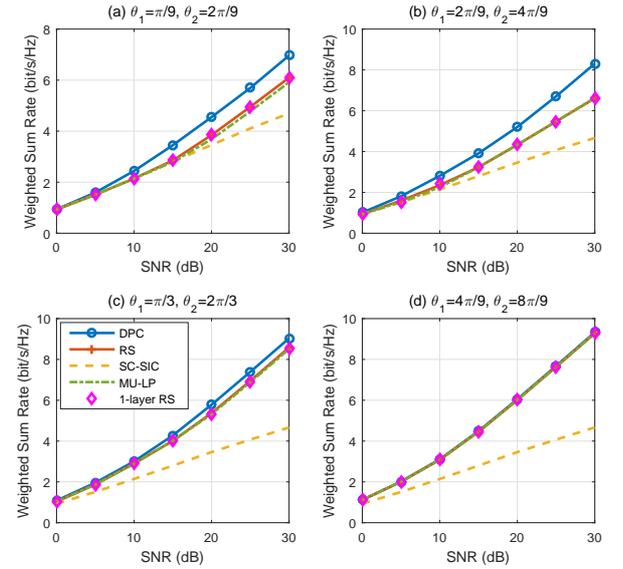}%
	\caption{Weighted sum rate versus SNR comparison of different strategies for underloaded three-user deployment with perfect CSIT, $\gamma_1=1, \gamma_2=0.3$, $u_1=0.4, u_2=0.3, u_3=0.3$, $N_t=4$, $R_k^{th}=0, k\in\{1,2,3\}$.}
	\label{fig: three user bias1103 040303 perfect}
\end{figure}

\par Fig.  \ref{fig: three user bias1103 020305 perfect}  and Fig.  \ref{fig: three user bias1103 040303 perfect} show the results when  the weight vectors are $\mathbf{u}=[0.2, 0.3, 0.5]$ and $\mathbf{u}=[0.4, 0.3, 0.3]$, respectively. In both figures, $\gamma_1=1, \gamma_2=0.3$. There is a 5 dB channel gain difference between user-$1$ and user-$3$ as well as between user-$2$ and user-$3$. In all scenarios and SNRs, RS always outperforms MU--LP and SC--SIC. Comparing with Fig.  \ref{fig: three user bias1103 040303 perfect}, the WSR improvement of RS is more explicit in Fig.  \ref{fig: three user bias1103 020305 perfect}. It implies that RS provides better enhancement of system throughput and user fairness. The performance of SC--SIC is the worst in most subfigures. This is due to the underloaded user deployments where $N_t>K$. One of the three users are required to decode all the messages and all the spatial multiplexing gains are sacrificed. Therefore, the sum DoF of SC--SIC is reduced to 1,  resulting in the deteriorated performance of SC--SIC in underloaded scenarios. In comparison, the performance of MU--LP is better than SC--SIC except in Fig.  \ref{fig: three user bias1103 040303 perfect}(a). MU--LP is more likely to serve the users with higher weights and channel gains by turning off the users with poor weights and  channel gains  when there is no individual rate constraints. It cannot deal efficiently with user fairness when a higher weight is allocated to the user with weaker channel strength. In contrast, SC--SIC works better when user fairness is considered.
 The WSR achieved by low-complexity 1-layer RS is equal to or larger than that of MU--LP and SC--SIC in most subfigures.	
 Comparing with SC--SIC and MU--LP, 1-layer RS is more robust to different user deployments and only a single SIC is required at each user. Moreover, the WSR of 1-layer RS is approaching that of RS in all user deployments. Considering the trade-off between performance and complexity, 1-layer RS is a good alternative to RS.

\par
\begin{figure}[t!]
	\centering
	\includegraphics[width=3.4in]{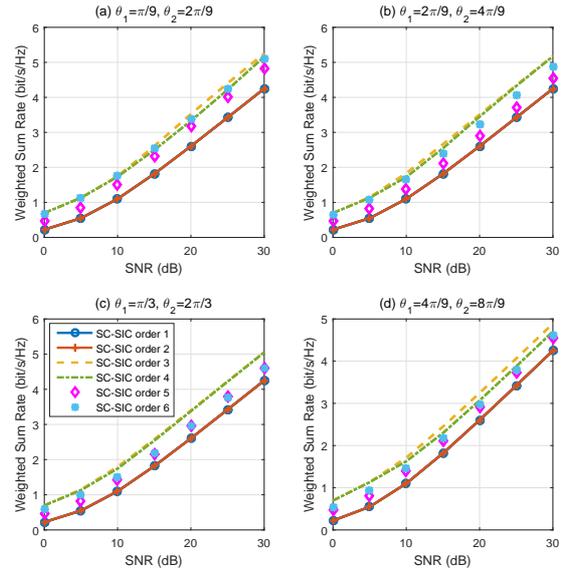}%
	\caption{Weighted sum rate versus SNR comparison of different decoding order of SC--SIC for underloaded three-user deployment with perfect CSIT, $\gamma_1=1, \gamma_2=0.3$, $u_1=0.2, u_2=0.3, u_3=0.5$, $N_t=4$, $R_k^{th}=0, k\in\{1,2,3\}$.}
	\label{fig: three user bias103 020305 noma ordering}
\end{figure}

 In all three-user deployments of SC--SIC, the decoding order is required to be optimized together with the precoder. To investigate the influence of different decoding orders, we compare the WSRs of SC--SIC using different decoding orders when $u_1=0.2, u_2=0.3, u_3=0.5$. There are in total 6 different decoding orders:
 \begin{equation*}
\begin{aligned}
& \textrm{SC-SIC order 1: } s_1\rightarrow s_2\rightarrow s_3\\
& \textrm{SC-SIC order 2: } s_2\rightarrow s_1\rightarrow s_3\\
& \textrm{SC-SIC order 3: } s_1\rightarrow s_3\rightarrow s_2\\
& \textrm{SC-SIC order 4: } s_3\rightarrow s_1\rightarrow s_2\\
& \textrm{SC-SIC order 5: } s_2\rightarrow s_3\rightarrow s_1\\
& \textrm{SC-SIC order 6: } s_3\rightarrow s_2\rightarrow s_1
\end{aligned}
\end{equation*}
In  Fig. \ref{fig: three user bias103 020305 noma ordering}, the WSR of 6 different decoding orders are illustrated in the circumstance where  there is a 5 dB channel gain difference between user-1/2 and user-3.    When $\gamma_1=1, \gamma_2=0.3$, it is typical to decode the message of user-3 first as the channel gain of user-3 is the worst. However, we notice that the optimal decoding order in Fig. \ref{fig: three user bias103 020305 noma ordering} is order 3, user-1 is decoded first. This is due to the smallest weight allocated to user-1, $u_1=0.2$. It implies that the weights assigned to users will affect the optimal decoding order. The scheduler complexity of SC--SIC becomes extremely high in order to find the optimal decoding order.
In contrast, 1-layer RS has a much lower scheduling complexity and does not rely on any user ordering at the transmitter. Moreover, it only requires a single SIC at each receiver.

\par More results of  underloaded three-user deployments with perfect CSIT and imperfect CSIT are given in Appendix C and Appendix E, respectively. The WSRs of different strategies for varied SNR, $N_t$, $\gamma_1, \gamma_2$ and $\mathbf{u}$ are illustrated. In all figures, RS outperforms SC--SIC and MU--LP. Though the scheduler and receiver complexity of 1-layer RS is low, it achieves equal or better performance than SC--SIC and MU--LP in most figures of perfect CSIT and  all figures of imperfect CSIT. All forms of RS are robust to a wide range of CSIT inaccuracy,  channel gain difference and channel angles among users.

\subsection{Overloaded three-user deployment with perfect CSIT}
\subsubsection{Two transmit antenna deployment}
\par We first consider an overloaded scenario where the BS is equipped with two antennas ($N_t=2$), and serves three single-antenna users. The channel realizations and beamforming initialization follows the methods used in the underloaded three-user deployment. The channel of users are realized as $
\nonumber
\mathbf{h}_1=\left[
1, 1\right]^H
$,
$
\nonumber
\mathbf{h}_2=\gamma_1\times\left[
1, e^{j\theta_1}\right]^H
$ and 
$
\nonumber
\mathbf{h}_3=\gamma_2\times\left[
1, e^{j\theta_2}\right]^H
$.
In overloaded scenarios, to guarantee some QoS, we add individual rate constraints to users as the system has otherwise a tendency to turn off some users. In all simulations of two transmit antenna deployment, we assume the rate threshold of each user is equal $R_1^{th}=R_2^{th}=R_3^{th}$.  Since the BS is able to serve users with higher QoS requirements as SNR increases, the rate threshold is assumed to  increase with SNR.  The rate threshold increases as  $\mathbf{r}_{th}=[0.02,0.08,0.19,0.3,0.4,0.4,0.4]$ bit/s/Hz for $\mathrm{SNR}=[0,5,10,15,20,25,30]$ dBs. 

\begin{figure}[t!]
	\centering
	\includegraphics[width=3.4in]{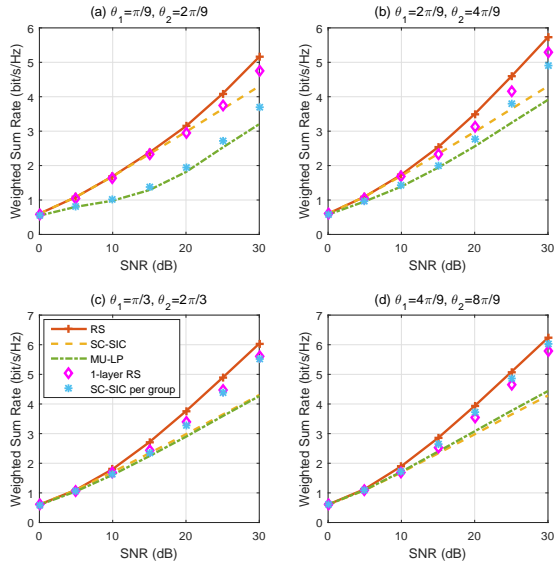}%
	\caption{Weighted sum rate versus SNR comparison of different strategies for overloaded three-user deployment with perfect CSIT, $\gamma_1=1, \gamma_2=0.3$, $u_1=0.4, u_2=0.3, u_3=0.3$, $N_t=2$, $\mathbf{r}_{th}=[0.02,0.08,0.19,0.3,0.4,0.4,0.4]$ bit/s/Hz.}
	\label{fig: three user bias1103 040303 perfect overloaded}
\end{figure}

\par We compare  the performance of RS, 1-layer RS, SC--SIC, MU--LP and SC--SIC per group in the overloaded three-user deployment. In SC--SIC per group, we consider a fixed grouping method. We assume user-1 is in group 1 while user-2 and user-3 are in group 2. The decoding order will be optimized together with the precoder. The beamforming initialization of SC--SIC per group is different from SC--SIC. In group 1, the precoder of user-1 is initialized based on MRT. In group 2, the precoder of the user decoded first $\mathbf{p}_{\pi(1)}$ is initialized as $\mathbf{p}_{\pi(1)}=p_{\pi(1)}\mathbf{u}_{\pi(1)}$ and $\mathbf{u}_{\pi(1)}$ is the largest left singular vector of the channel matrix $\mathbf{H}_{23}=[\mathbf{h}_2, \mathbf{h}_3]$. The precoder of the user decoded secondly is initialized based on MRT.

\par RS exhibits a clear WSR gain over SC--SIC, SC--SIC per group and MU--LP in Fig. \ref{fig: three user bias1103 040303 perfect overloaded}, where  $\gamma_1=1,\gamma_2=0.3$ and  $\mathbf{u}=[0.4, 0.3, 0.3]$. The WSR of MU--LP deteriorates in such overloaded scenario. When the individual rate constraints are not zero and $N_t<K$, MU--LP cannot coordinate the multi-user interference coming from all the users served simultaneously.  
When the angles of channels are large enough (subfigure (c) and subfigure (d) of Fig. \ref{fig: three user bias1103 040303 perfect overloaded}), the WSR of SC--SIC per group is better than SC--SIC. This is due to its ability to combine treating interference as noise (to tackle inter-group interference) with decoding interference (to tackle intra-group interference). 
However, as the angles of channels decreases, the performance of SC--SIC becomes better while that of SC--SIC per group is worse. Whether SC--SIC outperforms SC--SIC per group depends on SNR and user deployments. To ensure the WSR of the NOMA system is maximized,  a joint optimization of NOMA strategies based on switching between SC--SIC and SC--SIC per group on top of deciding the user grouping and user ordering is required. Such switching method has high scheduler and receiver complexity while its achieved performance is still lower than the simple 1-layer RS in most user deployments.

\subsubsection{Single transmit antenna deployment}
\par In a SISO BC, there is no need to split the messages into common and private parts since the capacity region is achieved by SC--SIC. Nevertheless, in view of the benefit of 1-layer RS in the MISO BC, we may wonder whether RS can be of any help in a SISO BC, especially when it comes to reducing the complexity of the receivers and the number of SIC needed.

\par
\begin{figure}[t!]
	\centering
	\includegraphics[width=3.0in]{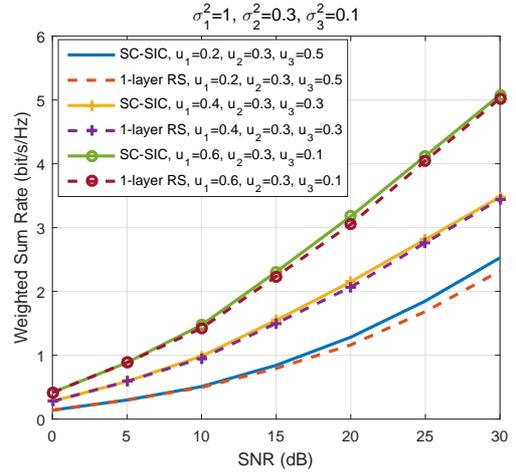}%
	\caption{Weighted sum rate versus SNR comparison of different strategies for overloaded three-user deployment with perfect CSIT, $\sigma_1^2=1, \sigma_2^2=0.3,\sigma_3^2=0.1$, $N_t=1$, $\mathbf{r}_{th}=[0,0,0.01,0.03,0.1,0.2,0.3]$ bit/s/Hz.}
	\label{fig: three user bias0301 siso overloaded}
\end{figure}

 We therefore compare the performance of  1-layer RS with SC--SIC in a 3-user SISO BC. We note that SC--SIC requires 2 layers of SIC while 1-layer RS requires a single SIC for all users. The channel of each user ${h}_k$ has an i.i.d. complex Gaussian entry with a certain variance, i.e., $\mathcal{CN}(0,\sigma_k^2)$.  Fig. \ref{fig: three user bias0301 siso overloaded} shows the average WSRs of different strategies over 10 random channel realizations when $\sigma_1^2=1, \sigma_2^2=0.3, \sigma_3^2=0.1$. 1-layer RS is able to achieve very close performance to SC--SIC.  Comparing with SC--SIC, the complexity of 1-layer RS is much reduced. There is no ordering issue at the BS and only one SIC is required at each user. Jointly considering the performance and complexity of the system, 1-layer RS is an attractive alternative to SC--SIC. 

\par More results of  overloaded three-user deployments with perfect CSIT and imperfect CSIT are given in Appendix D and Appendix F, respectively.  The WSRs of different strategies for varied SNR, $N_t$, $\gamma_1, \gamma_2$ and $\mathbf{u}$ are illustrated. We further show that RS exhibits a clear WSR gain over SC--SIC, SC--SIC per group and MU--LP in all simulated channels and weights.  1-layer RS outperforms SC--SIC, SC--SIC per group and MU--LP in most simulated scenarios. It is more robust and achieves a nearly equivalent WSR to that of RS in all user deployments. We also show that 1-layer RS achieves near optimal performance in various channel conditions of SISO BC.

\subsection{Overloaded four-user deployment with perfect CSIT}
\par We further investigate the four-user system model shown in 
Fig. \ref{fig: four-user transmission model}, where user-1 and user-2 are in group 1 while user-3 and user-4 are in group 2. We compare the 2-layer HRS, 1-layer RS per group, 1-layer RS, SC--SIC per group and MU--LP. In 2-layer HRS, the intra-group interference is mitigated using the intra-group common streams $s_{12}$, $s_{34}$ and the inter-group interference is mitigated using the inter-group common stream  $s_{1234}$. 1-layer RS and 1-layer RS per group are two special strategies of 2-layer HRS. All users in 1-layer RS are treated as single group. Only the 4-order common stream $s_{1234}$ and 1-order private streams are active. No power is allocated to $s_{12}$ and $s_{34}$. In contrast, 1-layer RS per group only allocate power to the intra-group common stream $s_{12}$, $s_{34}$ and 1-order private streams. No power is allocated to the inter-group common stream $s_{1234}$. Users within each group are served using RS and users across groups are served using SDMA so as to mitigate the inter-group interference.

\par We consider an overloaded scenario. The BS is equipped with two antennas and serves four single-antenna users. The channel of users are realized as
\begin{equation}
\begin{aligned}
& \mathbf{h}_1=\left[1, 1\right]^H,\\
&\mathbf{h}_2=\gamma_1\times\left[
1, e^{j\theta_1}\right]^H,\\
&\mathbf{h}_3=\gamma_2\times\left[
1, e^{j\theta_2}\right]^H,\\
&\mathbf{h}_4=\gamma_3\times\left[
1, e^{j\theta_3}\right]^H.
\end{aligned}
\end{equation}
$\gamma_1, \gamma_2, \gamma_3$ and  $\theta_1,\theta_2,\theta_3$ are control variables. $\theta_1$ is the channel angle between user-1 and user-2. It is denoted as intra-group angle of group 1. $\theta_2$ is the channel angle between user-1 and user-2. $\theta_2-\theta_1$ is the channel angle between user-2 and user-3, denoted as inter-group angle. $\theta_3$ is the channel angle between user-1 and user-3. $\theta_3-\theta_2$ is the channel angle between user-3 and user-4. It is the intra-group angle of group 2.  
In the following, we assume the intra-group angle of group 1 is the same as that of group 2.  We have $\theta_3=\theta_1+\theta_2$.  In each figure, the intra-group angle is varied as $\theta_1=\left[0,\frac{\pi }{18},\frac{\pi }{9},\frac{\pi }{6}\right]$.
The individual rate constraint is set to $\mathbf{r}_{th}=[0.03,  0.1,  0.2,  0.3,  0.4,  0.4,  0.4]$ bit/s/Hz for $\mathrm{SNR}=[0,5,10,15,20,25,30]$ dBs. The weights of users are assumed to be equal, i.e., $u_1=u_2=u_3=u_4=0.25$. We also assume the channel gain difference within each group is equal. The channel gain of user-3 is equal to that of user-1 ($\gamma_2=1$) and the channel gain of user-4 is equal to that of user-2 ($\gamma_3=\gamma_1$).

\begin{figure}[t!]
	\centering
	\includegraphics[width=3.3in]{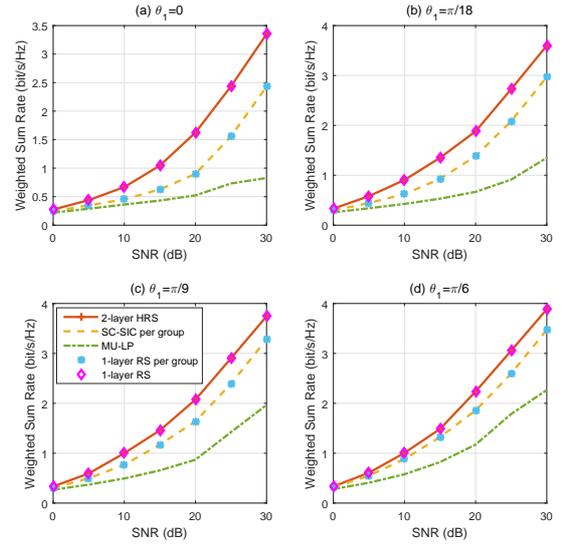}%
	\caption{Weighted sum rate versus SNR comparison of different strategies for overloaded four-user deployment with perfect CSIT, $\gamma_1=0.3$,  $\theta_2=\theta_1+\frac{\pi}{9}$, $\mathbf{r}_{th}=[0.03,  0.1,  0.2,  0.3,  0.4,  0.4,  0.4]$ bit/s/Hz.}
	\label{fig: bias103 weights025025 angle20}
\end{figure}
\begin{figure}[t!]
	\centering
	\includegraphics[width=3.3in]{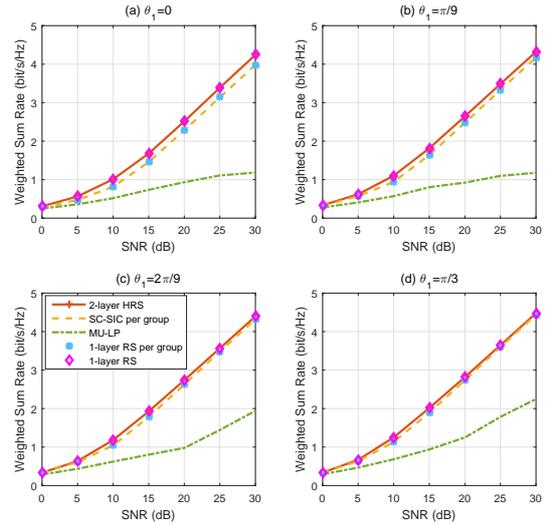}%
	\caption{Weighted sum rate versus SNR comparison of different strategies for overloaded four-user deployment with perfect CSIT, $\gamma_1=0.3$,  $\theta_2=\theta_1+\frac{\pi}{3}$, $\mathbf{r}_{th}=[0.03,  0.1,  0.2,  0.3,  0.4,  0.4,  0.4]$ bit/s/Hz.}
	\label{fig: bias103 weights025025 angle60}
\end{figure}	
\par Fig. \ref{fig: bias103 weights025025 angle20} and Fig. \ref{fig: bias103 weights025025 angle60} show the results when $\gamma_1=0.3$. The inter-group angles are $\frac{\pi}{9}$ and $\frac{\pi}{3}$, respectively. The WSR achieved by 2-layer HRS is equal to 1-layer RS in both figures, which means that 2-layer HRS reduces to 1-layer RS in these user deployments. 2-layer HRS and 1-layer RS outperform all other schemes. The inter-group and intra-group interference can be jointly mitigated by one layer common message.  As the inter-group angle increases, the WSR gaps between 2-layer HRS and 1-layer RS per group reduces. The inter-group interference can be coordinated by SDMA when the inter-group angle is sufficiently large. 1-layer RS  per group has the same WSR as SC--SIC per group in both figures. It reduces to SC--SIC per group because SC--SIC is more suitable when the intra-group angle is sufficiently small and the channel gain difference between users within each group is sufficiently large.

\par More results of  overloaded four-user deployments with perfect CSIT  are given in Appendix G.  The WSRs of different strategies when there is no channel gain difference ($\gamma_{1}=1$) are illustrated. We further show that 2-layer HRS, 1-layer RS and 1-layer RS per group achieve equal or better performance than SC--SIC per group and MU--LP in all simulated channel conditions.

\subsection{Overloaded ten-user deployment with perfect CSIT}
\par We further consider an extremely overloaded scenario subject to QoS constraints. The BS is equipped with two antennas ($N_t=2$) and serves 10 users. The channel of each user $\mathbf{h}_k$ has  i.i.d. complex Gaussian entries with a certain variance, i.e., $\mathcal{CN}(0,\sigma_k^2)$. The rate of each user is averaged over the 10 randomly generated channels. We compare 1-layer RS,  MU--LP, multicast and SC--SIC with a certain decoding order.  There are $10!$ different decoding orders of SC--SIC in the ten-user case. The optimal decoding order of SC--SIC is intractable.   In the following simulations, only the  decoding order based on the ascending channel gain is considered for WSR calculation in SC--SIC. It is the optimal decoding order in SISO BC. Multicast can be regarded as a special scheme of 1-layer RS with only the $10$-order stream to be transmitted to all users. The weight of each user is assumed to be equal to 1.

\begin{figure}[t!]
	\centering
	\includegraphics[width=3.4in]{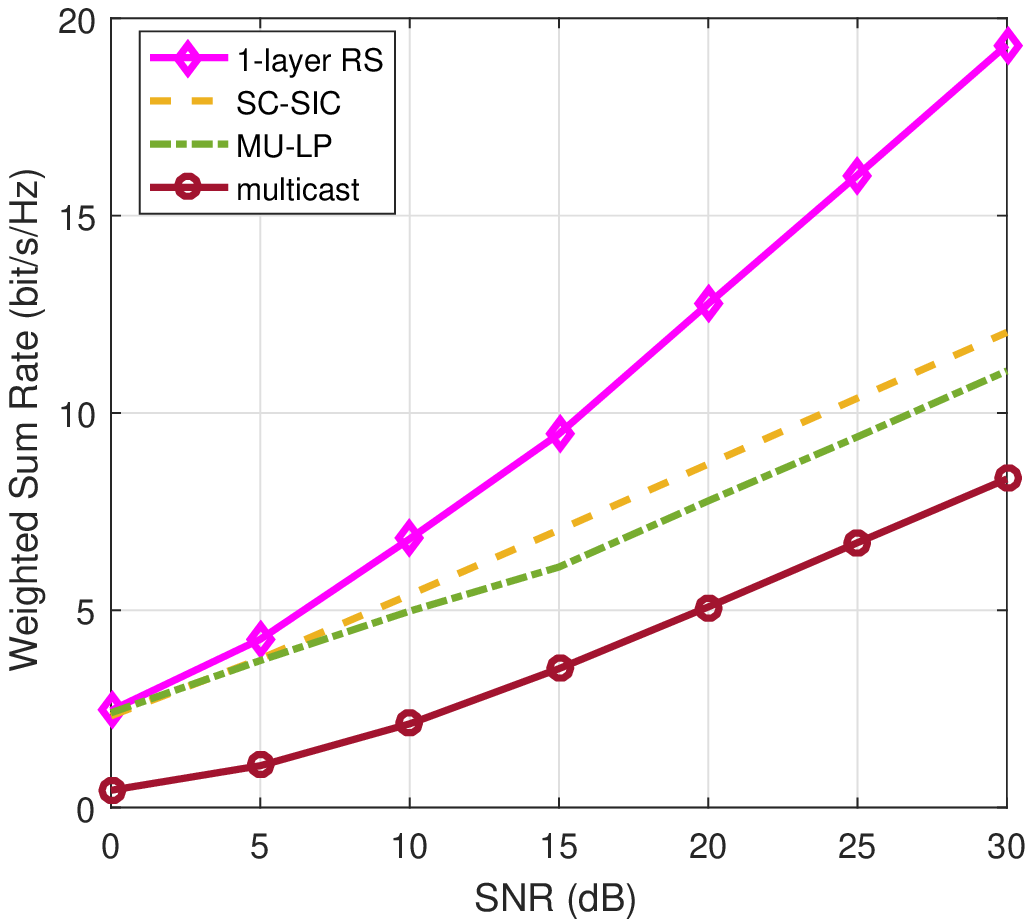}%
	\caption{Weighted sum rate versus SNR comparison of different strategies for overloaded ten-user deployment with perfect CSIT, $\sigma_1^2=\sigma_2^2=\ldots=\sigma_{10}^2=1$, $N_t=2$, SNR=30 dB, $\mathbf{r}_{th}=[0.01,  0.03,  0.05,  0.1,  0.1,  0.1,  0.1]$ bit/s/Hz.}
	\label{fig: ten user equal covariance high threshold}
\end{figure}

\par
\begin{figure}[t!]
	\centering
	\includegraphics[width=3.5in]{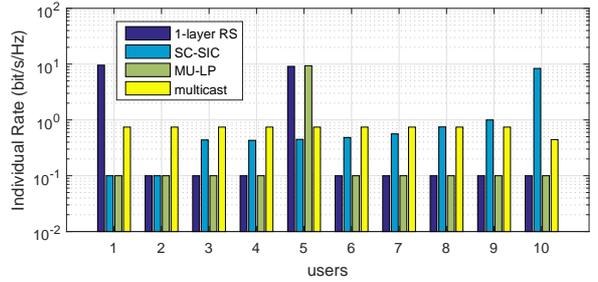}%
	\caption{Individual rate comparison of different strategies for overloaded ten-user deployment with perfect CSIT for 1 randomly generated channel estimate, SNR=30 dB, $N_t=2$, $\mathbf{r}_{th}=[0.01,  0.03,  0.05,  0.1,  0.1,  0.1,  0.1]$ bit/s/Hz.}
	\label{fig: ten user equal covariance individual rate}
\end{figure}

 Fig. \ref{fig: ten user equal covariance high threshold} shows the WSRs of different strategies when $\sigma_1^2=\sigma_2^2=\ldots=\sigma_{10}^2=1$, $\mathbf{r}_{th}=[0.01,  0.03,  0.05,  0.1,  0.1,  0.1,  0.1]$ bit/s/Hz.  The WSR achieved by the multicast scheme is the worst. In such an overloaded user deployment, the spectral efficiency of multicast is low as it is difficult for a single beamformer to satisfy all users.  Under the rate constraint $\mathbf{r}_{th}$, the WSR of SC--SIC is better than that of MU--LP while the slopes of the WSRs are the same for large SNRs. It implies that SC--SIC and MU--LP achieve the same DoF of 1. In contrast, 1-layer RS shows an obvious WSR improvement over all other strategies and exhibits a DoF of 2. This highlights the RS exploits the maximum DoF of the considered deployments (that is limited by 2 given the 2 transmit antennas). To further investigate the reason behind the results, we focus on one random channel realization. The WSRs achieved by all strategies when SNR=30 dB are compared as shown in Fig. \ref{fig: ten user equal covariance individual rate}. The optimized common rate vector of one-layer RS is $\mathbf{c}=[0, 0.1, 0.1, 0.1, 0, 0.1, 0.1, 0.1, 0.1, 0.1]$ bit/s/Hz. 
 No common rate is allocated to user-1 and user-5. But in Fig. \ref{fig: ten user equal covariance individual rate}, we can observe that the rate allocated to user-1 and user-5 are the highest. It implies that RS uses the common message to pack messages from eight users and uses two transmit antennas to deliver private messages to user-1 and user-5. RS achieves a sum DoF of 2 in the overloaded regime. In contrast, MU--LP and SC--SIC allocate most of power to single user. The rate achieved by user-5 when using MU--LP and the rate achieved by user-10 when using SC--SIC is much higher than other users in Fig. \ref{fig: ten user equal covariance individual rate}. The DoFs achieved by MU--LP and SC--SIC are limited to 1 in such circumstance.
 
\par  Note that results here show the usefulness of the RS framework for massive IoT or MTC services. Those devices are typically cheap. In the example above, user-1 and user-5 could be high-end devices, for which RS would be implemented. Those devices would therefore perform SIC. All other devices could be IoT or MTC devices, who would not need to implement RS, nor SIC, but simply decode the common message. Hence the RS framework can be used to pack the IoT/MTC traffic in the common message.

\par More results of  overloaded ten-user deployments with perfect CSIT  are given in Appendix H.  We further illustrate WSRs of different strategies when the rate threshold $\mathbf{r}_{th}$ and channel gain difference are changed. We show that the when the rate threshold of each user is 0, MU--LP is able to achieve a DoF of 2. However, as the rate threshold increases, MU--LP cannot coordinate the inter-user interference and its achieved DoF drops to 1. In the extremely overloaded scenario, the WSR gap between RS and SC--SIC is still large. SC--SIC makes an inefficient use of the transmit antennas and achieves a DoF of 1.

\section{Conclusions and Future Works}
\label{conclusion}
\par To conclude, we propose a new multiple access called Rate-Splitting Multiple Access (RSMA). We compare the proposed RSMA with SDMA and NOMA by solving the problem of maximizing WSR in MISO-BC systems with QoS constraints.
Both perfect and imperfect CSIT are investigated.  WMMSE and its modified algorithms are adopted to solve the respective optimization problems. We show that SDMA and NOMA are subject to many limitations, including high system complexity  and  a lack of robustness to user deployments, network load and CSIT inaccuracy. We propose a general multiple access framework based on rate-splitting (RS), where the common symbols decoded by different groups of users are transmitted on top of private symbols decoded by the corresponding users only. Thanks to its ability of partially decoding interference and partially treating interference as noise, RSMA softly bridges and outperforms SDMA and NOMA in any user deployments, CSIT inaccuracy and network load. The simplified RS forms, such as 1-layer RS and 2-layer HRS, show great potential to reduce the scheduler and receiver complexity but maintain good and robust performance in any user deployments, CSIT inaccuracy and network load. Particularly, we show that 1-layer RS is an attractive alternative to SC--SIC in a SISO BC deployment due to its near optimal performance and very low complexity.  Therefore, RSMA is a more general and powerful multiple access for downlink multi-antenna systems that encompasses SDMA and NOMA as special cases.

\par RSMA has the potential to change the design of the physical layer and MAC layer of next generation communication systems by unifying existing approaches and relying on a superposed transmission of common and private messages. Many interesting problems are left for future research, including among others the role played by RSMA to achieve the fundamental limits of broadcast, interference and relay channels in the presence of imperfect CSIT and disparity of channel strengths, optimization (robust design, sum-rate maximization, max-min fairness, QoS constraints) of RSMA, performance analysis of RSMA, RSMA design for multi-user/Massive/Millimeter-wave/multi-cell/network MIMO, modulation and coding for RSMA, RSMA with multi-carrier transmissions, RSMA with/vs. non-linear precoding, resource allocation and cross-layer design of RSMA, security provisioning in RSMA, RSMA design for cellular and satellite communication networks, prototyping and experimentation of RSMA, standardization issues (link/system-level evaluations, receiver implementation, transmission schemes/modes, CSI feedback mechanisms, downlink and uplink signaling) of RSMA.

\section*{Appendix}
\subsection{Underloaded two-user deployment with perfect CSIT}
\label{appendix: two-user perfect CSIT}
\par
\begin{figure}[h!]
	\centering
	\includegraphics[width=3.4in]{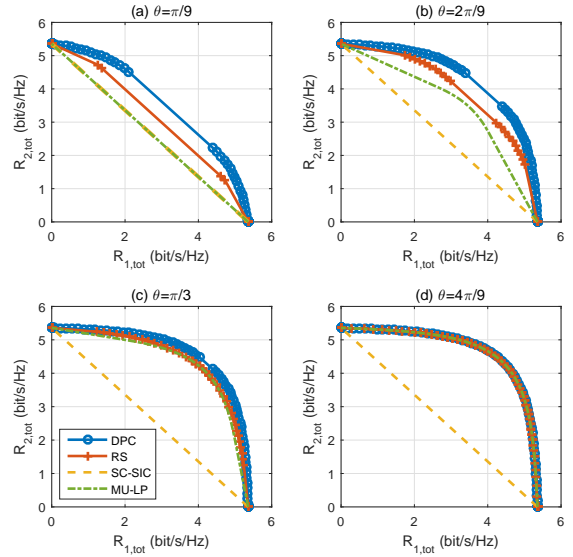}%
	\caption{Achievable rate region comparison of different strategies in underloaded two-user deployment with perfect CSIT, $\gamma=1$, $N_t=4$, SNR=10 dB.}
	\label{fig: snr10 bias1 nt4}
\end{figure}

\begin{figure}[h!]
	\centering
	\includegraphics[width=3.4in]{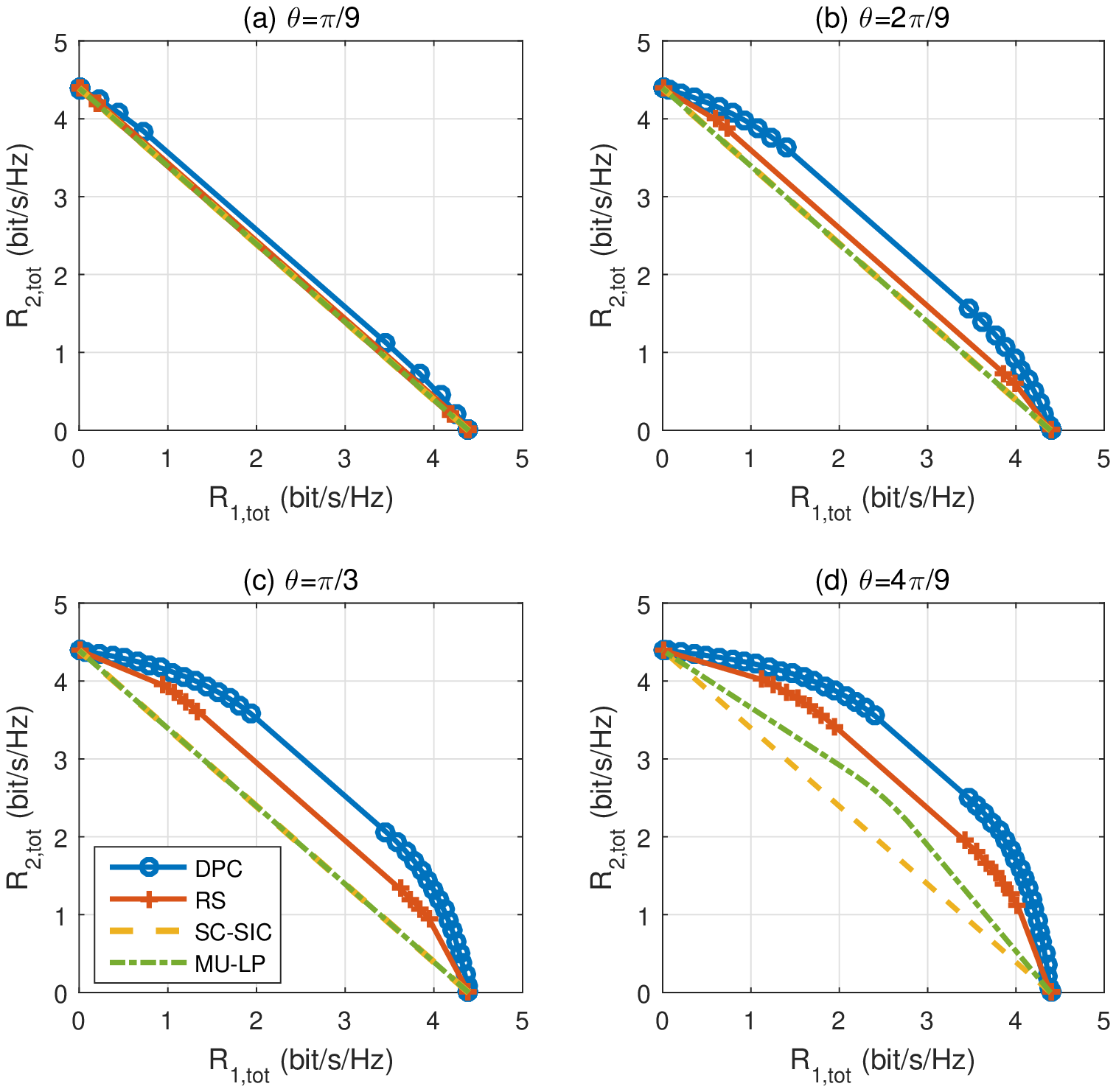}%
	\caption{Achievable rate region comparison of different strategies in underloaded two-user deployment with perfect CSIT, $\gamma=1$, $N_t=2$, SNR=10 dB.}
	\label{fig: snr10 bias1 nt2}
\end{figure}

\begin{figure}[h!]
	\centering
	\includegraphics[width=3.4in]{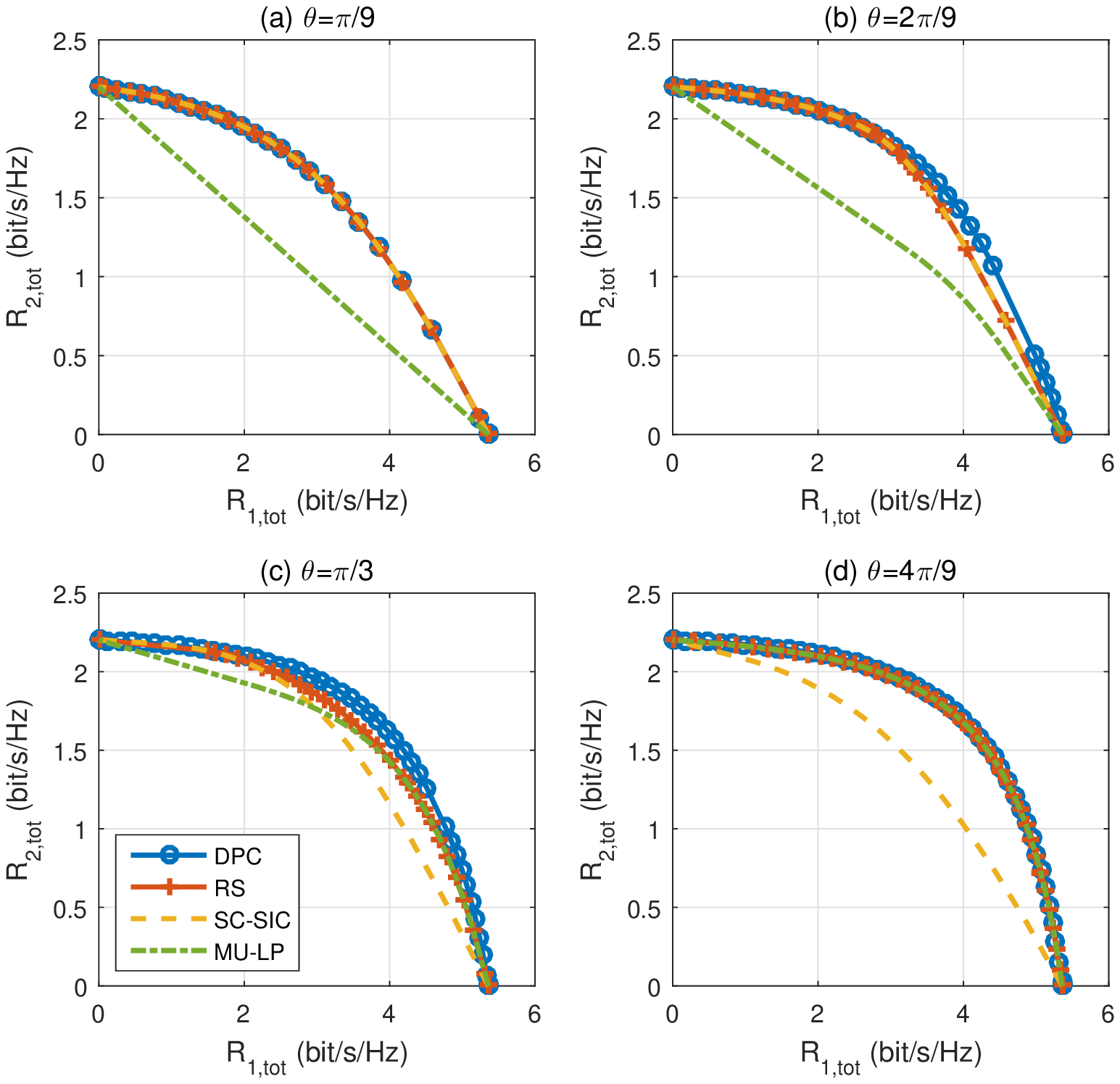}%
	\caption{Achievable rate region comparison of different strategies in perfect CSIT, $\gamma=0.3$, $N_t=4$, SNR=10 dB.}
	\label{fig: snr10 bias03 nt4}
\end{figure}

\begin{figure}[t!]
	\centering
	\includegraphics[width=3.4in]{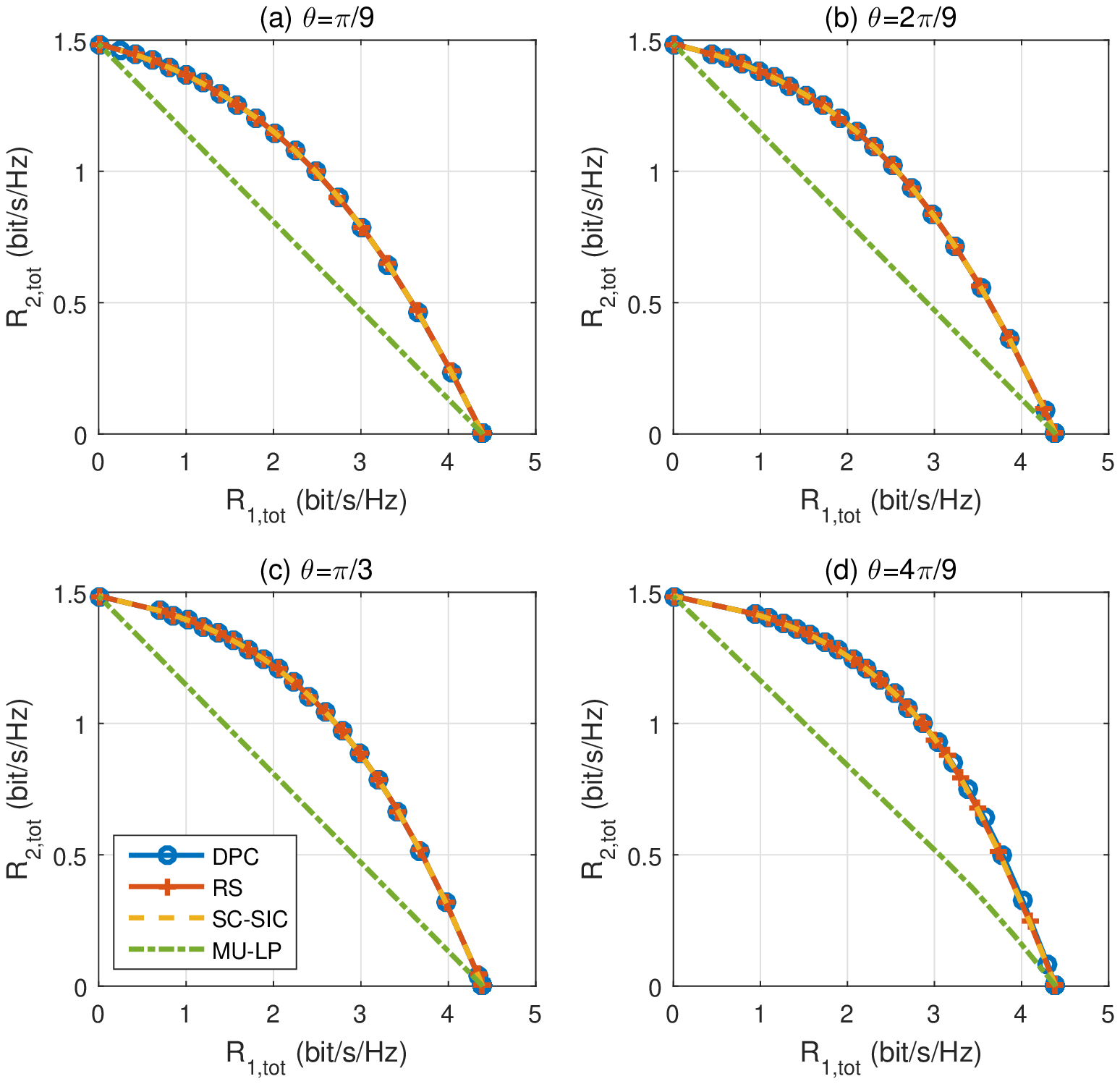}%
	\caption{Achievable rate region comparison of different strategies in perfect CSIT, $\gamma=0.3$, $N_t=2$, SNR=10 dB.}
	\label{fig: snr10 bias03 nt2}
\end{figure}

 To further investigate the influence of SNR, we illustrate the rate region of different strategies when SNR is 10 dB  in Fig. \ref{fig: snr10 bias1 nt4}--\ref{fig: snr10 bias03 nt2} and compare with the results when SNR is 20 dB in Fig. \ref{fig: snr20 bias1 nt4}--\ref{fig: snr20 bias03 nt2}.  Comparing the corresponding figures of 10 dB and 20 dB, we observe that the rate region gaps among different schemes grow with SNR. As SNR increases, the performance improvement of RS becomes more obvious. Specifically, SC--SIC and MU--LP outperform each other at one part of the rate region in Fig. \ref{fig: snr20 bias03 nt4}(b) and Fig. \ref{fig: snr20 bias03 nt2}(d) and the rate region of RS encompasses the convex hull of the rate regions of SC--SIC and MU--LP. However, as SNR decreases to 10 dB, the crosspoints  disappear in Fig. \ref{fig: snr10 bias03 nt4}(b) and Fig. \ref{fig: snr10 bias03 nt2}(d). The rate regions of SC--SIC overlap with that of RS. RS reduces to SC--SIC and they outperform MU--LP in the whole rate region.

\subsection{Underloaded two-user deployment with imperfect CSIT} 

\begin{figure}[t!]
	\centering
	\includegraphics[width=3.4in]{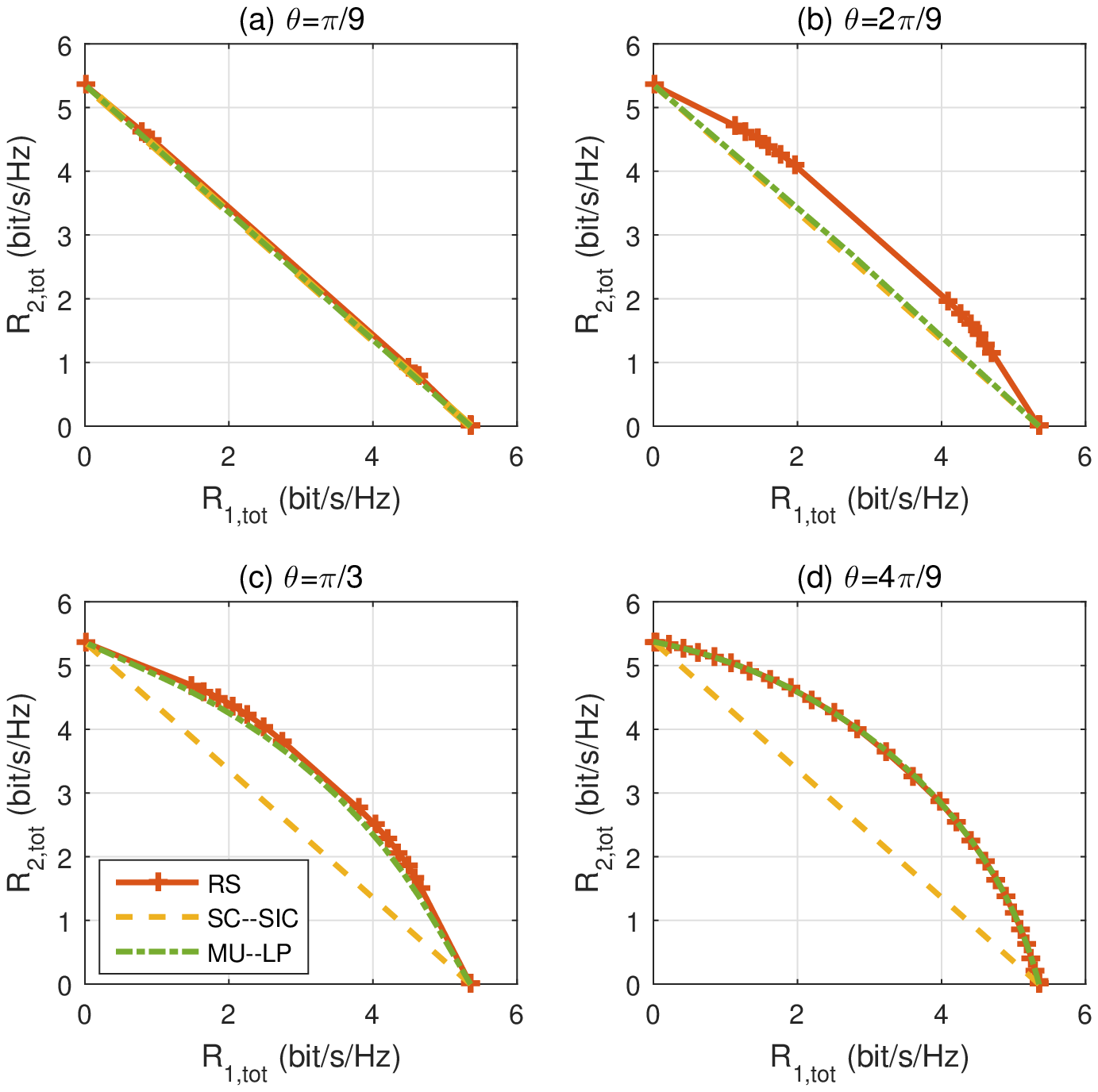}%
	\caption{ Average rate region comparison of different strategies in imperfect CSIT, $\gamma=1$, $N_t=4$, SNR=10 dB.}
	\label{fig: snr10 bias1 nt4 imperfect}
\end{figure}
\begin{figure}[t!]
	\centering
	\includegraphics[width=3.4in]{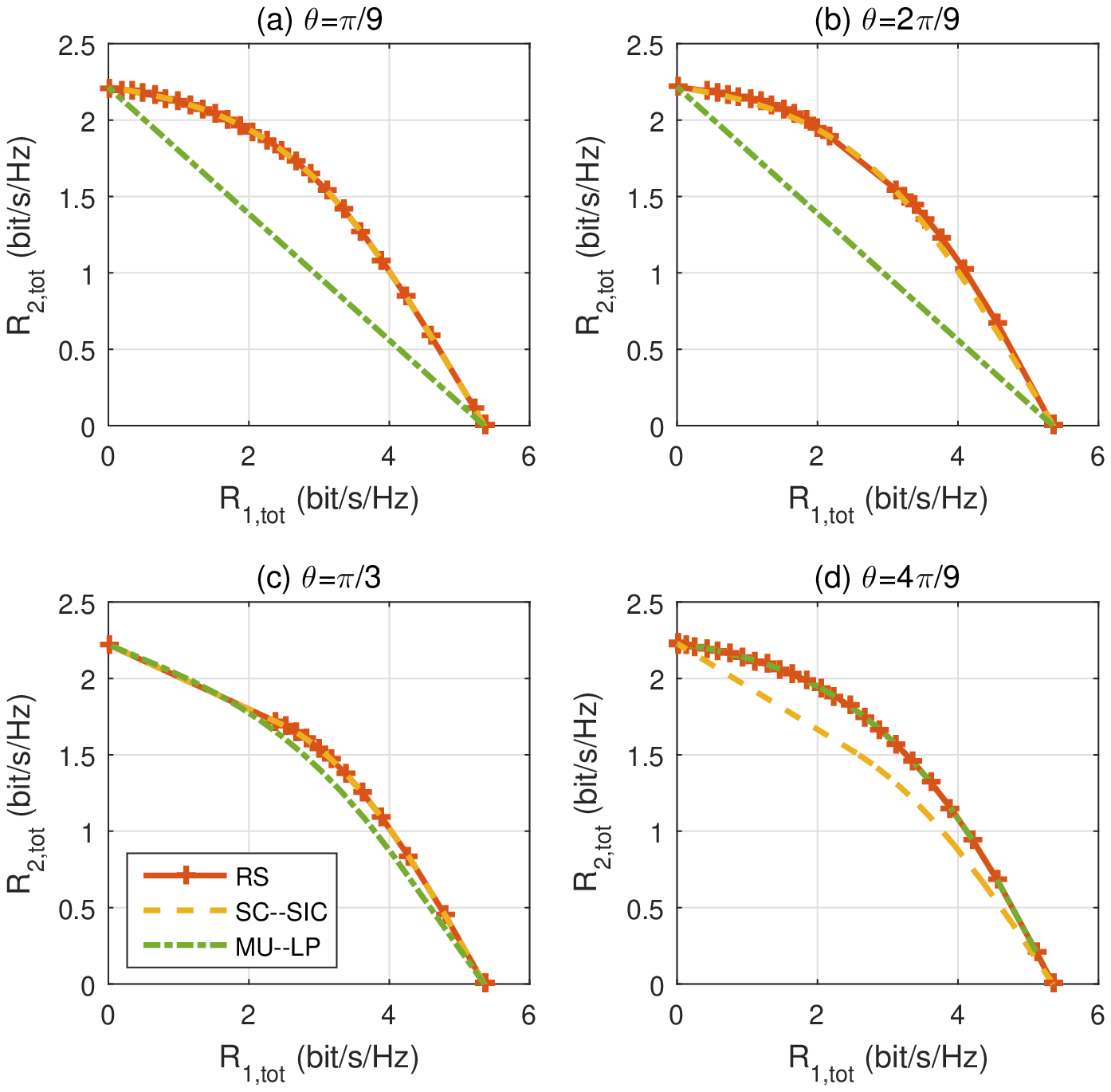}%
	\caption{ Average rate region comparison of different strategies in imperfect CSIT, $\gamma=0.3$, $N_t=4$, SNR=10 dB.}
	\label{fig: snr10 bias03 nt4 imperfect}
\end{figure}
\begin{figure}[t!]
	\centering
	\includegraphics[width=3.4in]{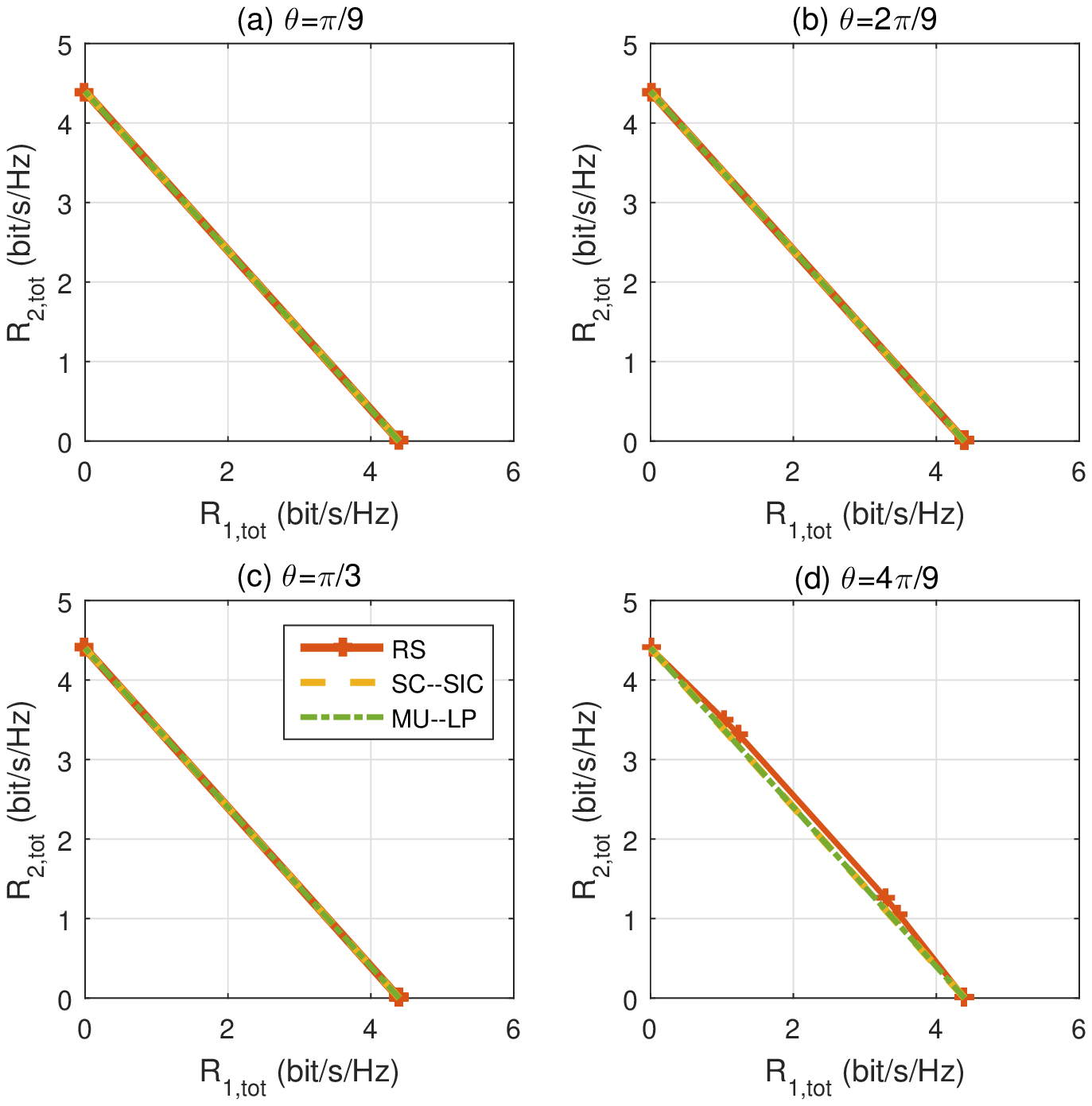}%
	\caption{ Average rate region comparison of different strategies in imperfect CSIT, $\gamma=1$, $N_t=2$, SNR=10 dB.}
	\label{fig: snr10 bias1 nt2 imperfect}
\end{figure}
\begin{figure}[t!]
	\centering
	\includegraphics[width=3.4in]{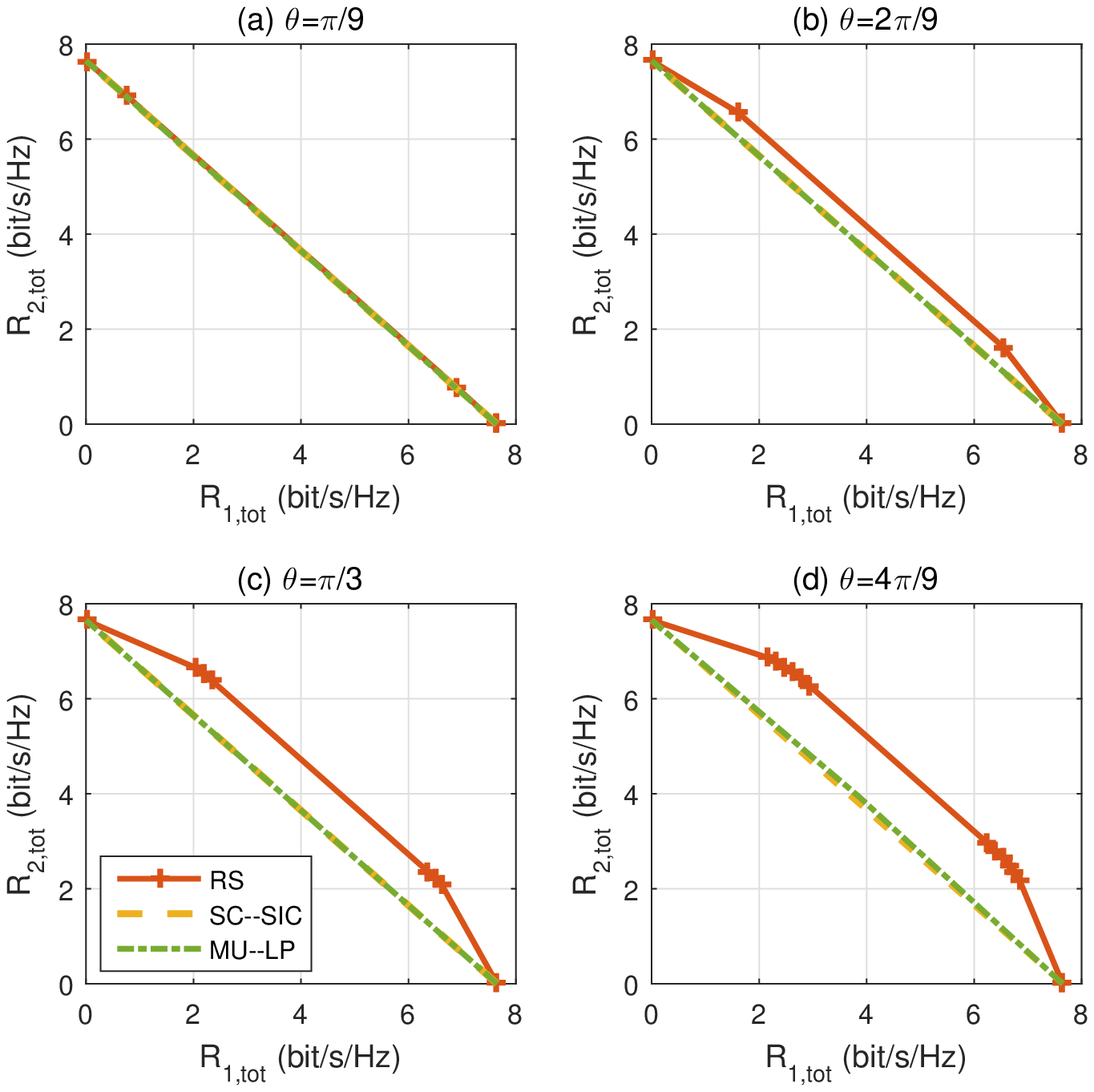}%
	\caption{Average rate region comparison of different strategies in imperfect CSIT, $\gamma=1$, $N_t=2$, SNR=20 dB.}
	\label{fig: snr20 bias1 nt2 imperfect}
\end{figure}

\begin{figure}[t!]
	\centering
	\includegraphics[width=3.4in]{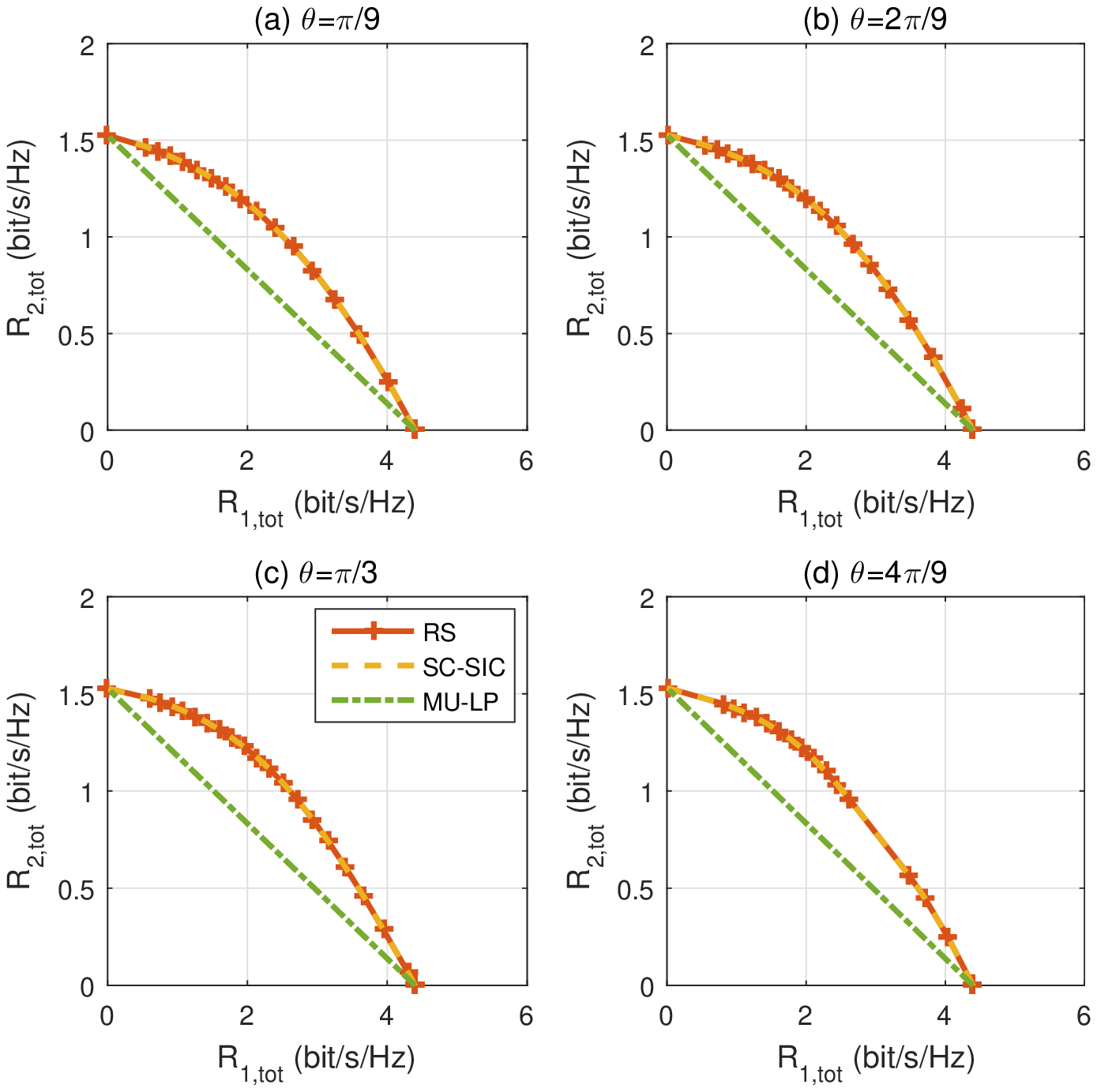}%
	\caption{ Average rate region comparison of different strategies in imperfect CSIT, $\gamma=0.3$, $N_t=2$, SNR=10 dB.}
	\label{fig: snr10 bias03 nt2 imperfect}
\end{figure}
\begin{figure}[t!]
	\centering
	\includegraphics[width=3.4in]{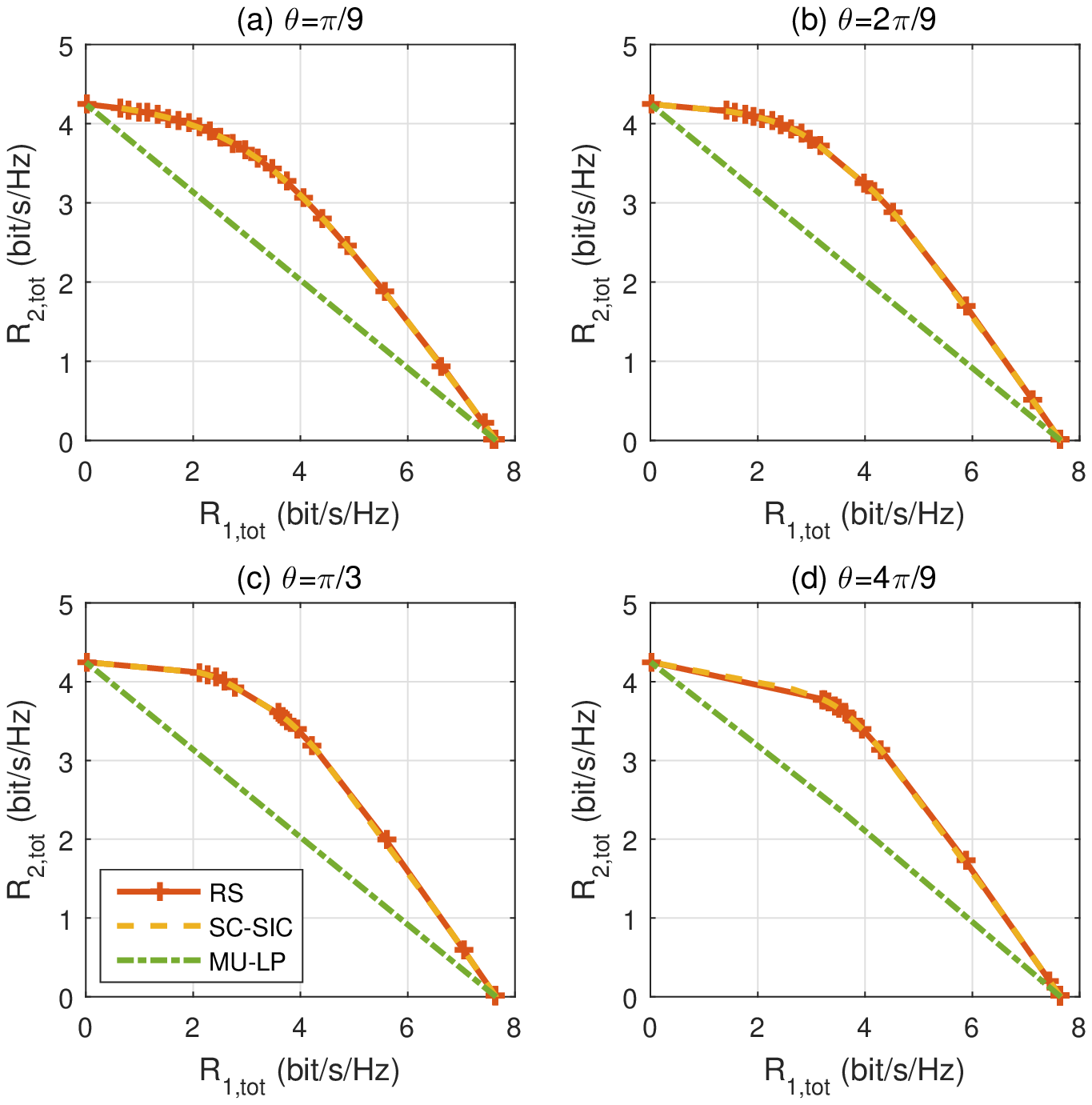}%
	\caption{ Average rate region comparison of different strategies in imperfect CSIT, $\gamma=0.3$, $N_t=2$, SNR=20 dB.}
	\label{fig: snr20 bias03 nt2 imperfect}
\end{figure}

\par To further study the influence of CSIT inaccuracy, SNR, number of transmit antennas and user deployments, we illustrate the rate region of different strategies when SNR, $N_t$ and $\gamma$ are varied in Fig. \ref{fig: snr10 bias1 nt4 imperfect}--\ref{fig: snr20 bias03 nt2 imperfect}.
 
\par Fig. \ref{fig: snr10 bias1 nt4 imperfect} and Fig. \ref{fig: snr10 bias03 nt4 imperfect} show the corresponding results of Fig. \ref{fig: snr20 bias1 nt4 imperfect} and Fig. \ref{fig: snr20 bias03 nt4 imperfect} when SNR decreases to 10 dB. The rate region gaps among users decreases when SNR decreases.

\par Fig. \ref{fig: snr10 bias1 nt2 imperfect} and Fig. \ref{fig: snr20 bias1 nt2 imperfect} show the results when $\gamma=1$, $N_t=2$. When SNR is 10 dB, the rate regions of the three schemes are very close to each other. When SNR is 20 dB, the rate region of RS shows explicit improvement over the rate regions of MU--LP and SC--SIC. Comparing Fig. \ref{fig: snr20 bias1 nt2 imperfect} with Fig. \ref{fig: snr20 bias1 nt2}, the performance of MU--LP is worse when CSIT is imperfect. It shows that MU--LP requires accurate CSIT to design precoders.   There is no crosspoint between SC--SIC and MU--LP in Fig. \ref{fig: snr10 bias03 nt4 imperfect}(c) and Fig. \ref{fig: snr20 bias03 nt4 imperfect}(b) compared respectively with Fig. \ref{fig: snr10 bias03 nt4}(c) and Fig. \ref{fig: snr20 bias03 nt4}(b). 

\par Fig. \ref{fig: snr10 bias03 nt2 imperfect} and Fig. \ref{fig: snr20 bias03 nt2 imperfect} show the results when $\gamma=0.3$. SNR is 10 dB and 20 dB, respectively. The rate region gap between RS and SC--SIC reduces in imperfect CSIT, as observed by comparing Fig. \ref{fig: snr20 bias03 nt2 imperfect} with Fig. \ref{fig: snr20 bias03 nt2}. Comparing with MU--LP, SC--SIC is less sensitive to CSIT inaccuracy.

\subsection{Underloaded three-user deployment with perfect CSIT}

\begin{figure}[t!]
	\centering
	\includegraphics[width=3.4in]{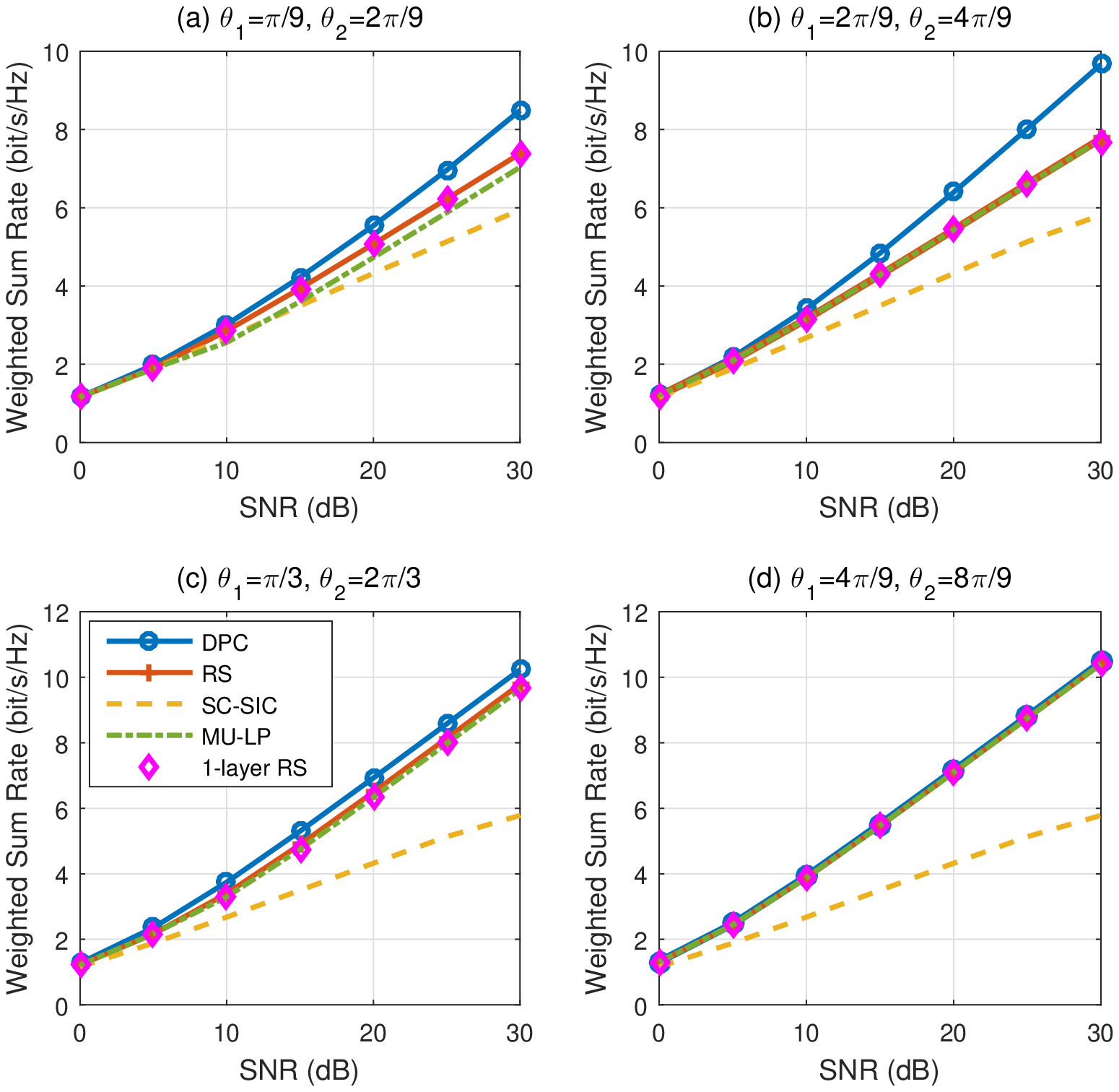}%
	\caption{Weighted sum rate versus SNR comparison of different strategies for underloaded three-user deployment with perfect CSIT, $\gamma_1=\gamma_2=1$, $u_1=0.2, u_2=0.3, u_3=0.5$, $N_t=4$, $R_k^{th}=0, k\in\{1,2,3\}$.}
	\label{fig: three user bias1 020305 perfect}
\end{figure}
\begin{figure}[t!]
	\centering
	\includegraphics[width=3.4in]{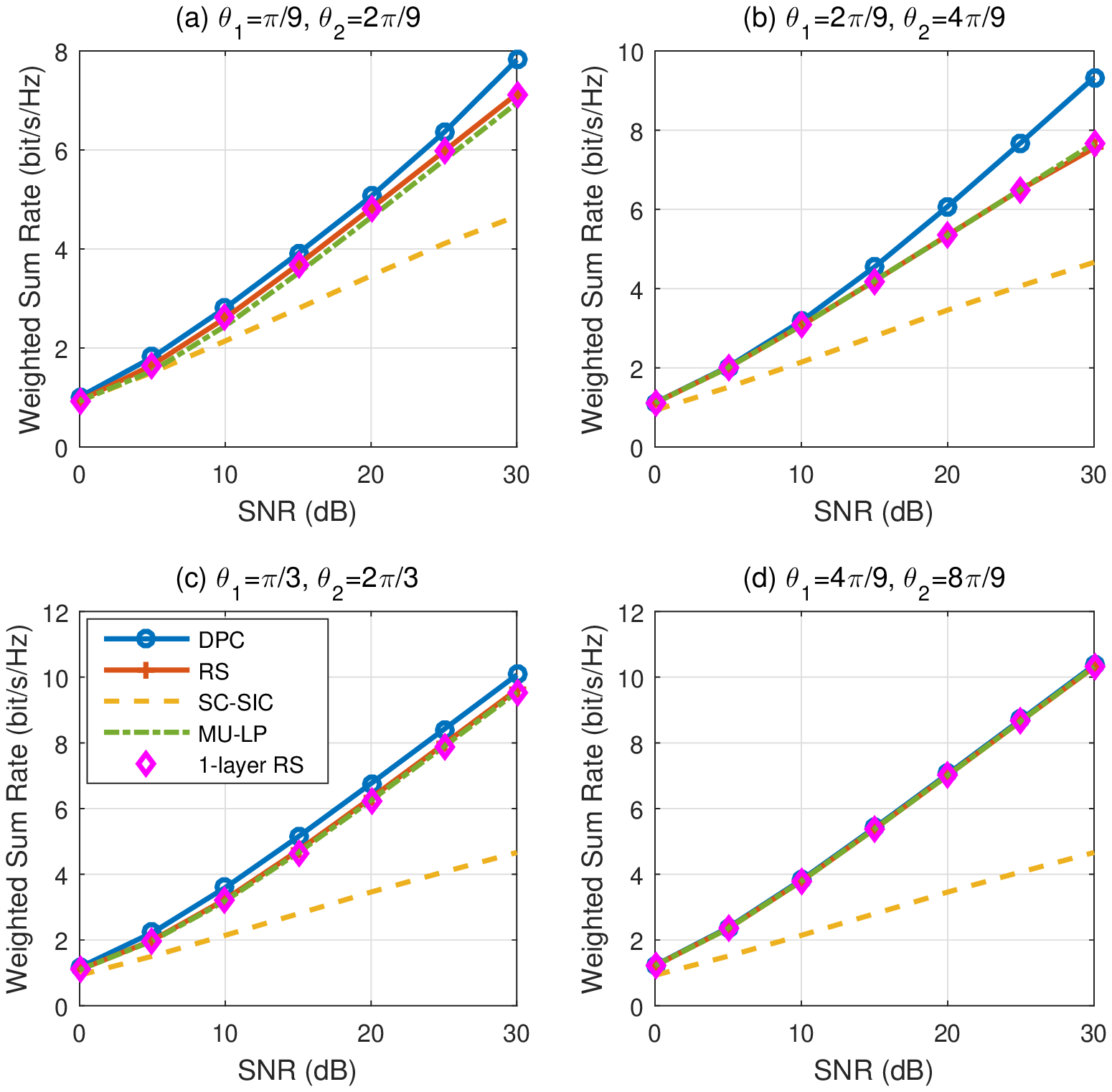}%
	\caption{Weighted sum rate versus SNR comparison of different strategies for underloaded three-user deployment with perfect CSIT, $\gamma_1=\gamma_2=1$, $u_1=0.4, u_2=0.3, u_3=0.3$, $N_t=4$, $R_k^{th}=0, k\in\{1,2,3\}$.}
	\label{fig: three user bias1 040303 perfect}
\end{figure}
\par
We consider three different sets of $\gamma_1, \gamma_2$. When $\gamma_1=\gamma_2=1$, the three users have no channel strength difference. When $\gamma_1=1, \gamma_2=0.3$, there is a 5 dB channel strength difference between user-$1$ and user-$3$ as well as between user-$2$ and user-$3$.  When $\gamma_1=0.3, \gamma_2=0.1$, there is a 5 dB channel strength difference between user-$1$ and user-$2$ as well as user-$2$ and user-$3$. The channel strength difference between user-$1$ and user-$3$ is 10 dB.  We consider three different weight vectors for each set of $\gamma_1,\gamma_2$, i.e., $\mathbf{u}=[0.2, 0.3, 0.5]$, $\mathbf{u}=[0.4, 0.3, 0.3]$ and $\mathbf{u}=[0.6, 0.3, 0.1]$. 

\par
\begin{figure}[t!]
	\centering
	\includegraphics[width=3.4in]{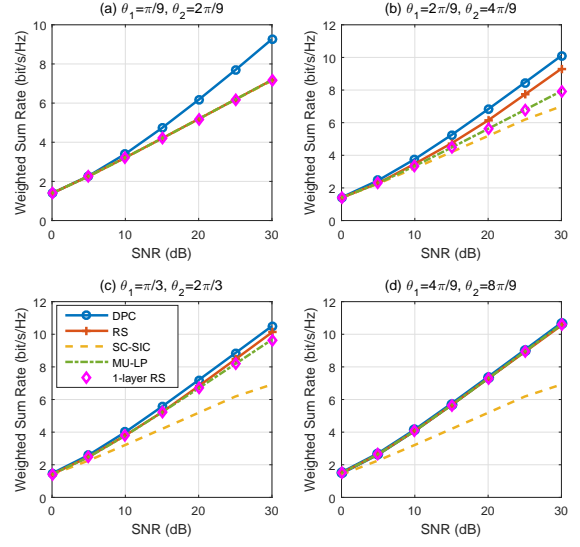}%
	\caption{Weighted sum rate versus SNR comparison of different strategies for underloaded three-user deployment with perfect CSIT, $\gamma_1=\gamma_2=1$, $u_1=0.6, u_2=0.3, u_3=0.1$, $N_t=4$, $R_k^{th}=0, k\in\{1,2,3\}$.}
	\label{fig: three user bias1 060301 perfect}
\end{figure}

\begin{figure}[t!]
	\centering
	\includegraphics[width=3.4in]{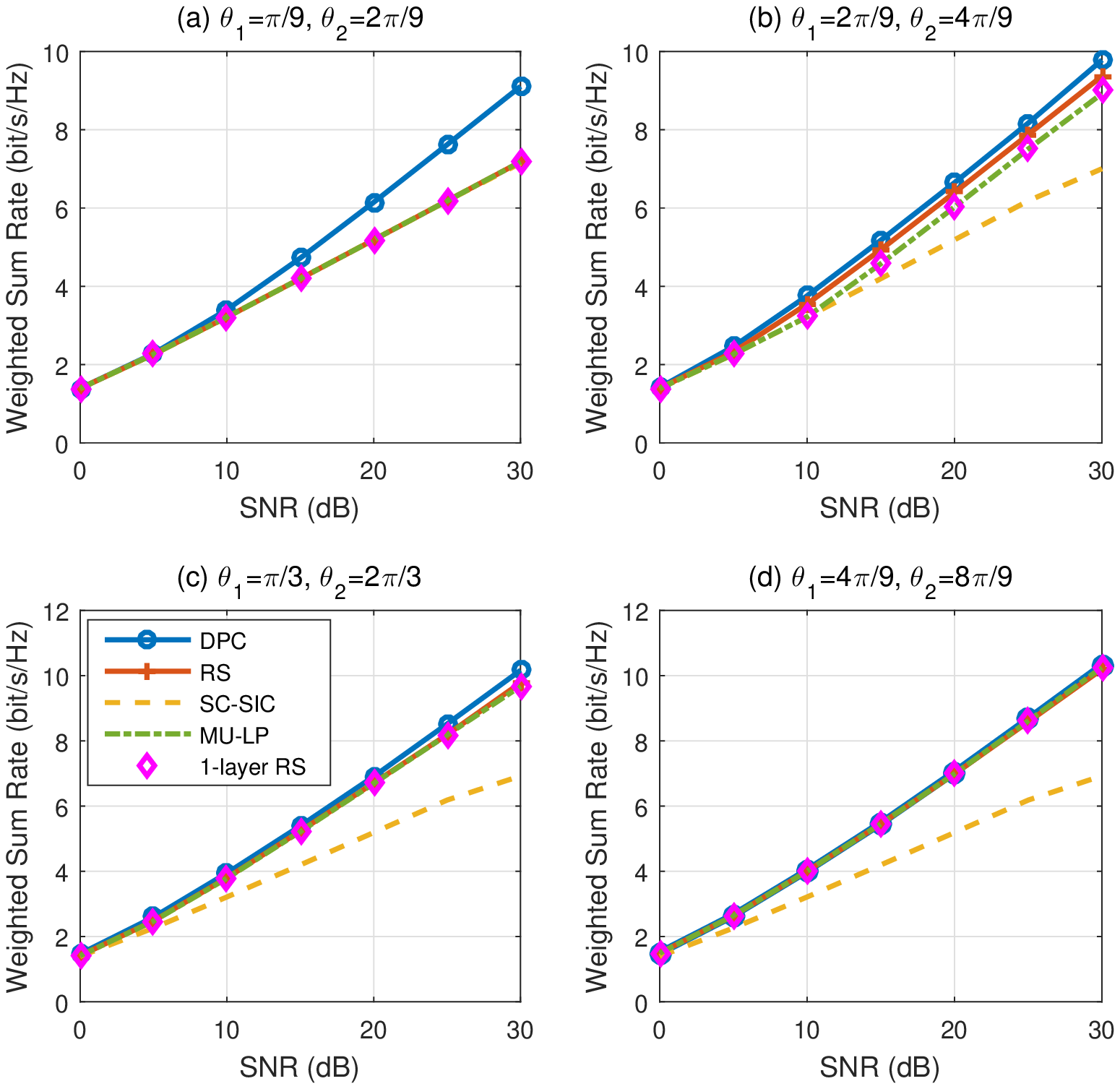}%
	\caption{Weighted sum rate versus SNR comparison of different strategies for underloaded three-user deployment with perfect CSIT, $\gamma_1=1, \gamma_2=0.3$, $u_1=0.6, u_2=0.3, u_3=0.1$, $N_t=4$, $R_k^{th}=0, k\in\{1,2,3\}$.}
	\label{fig: three user bias1103 060301 perfect}
\end{figure}

\begin{figure}[t!]
	\centering
	\includegraphics[width=3.4in]{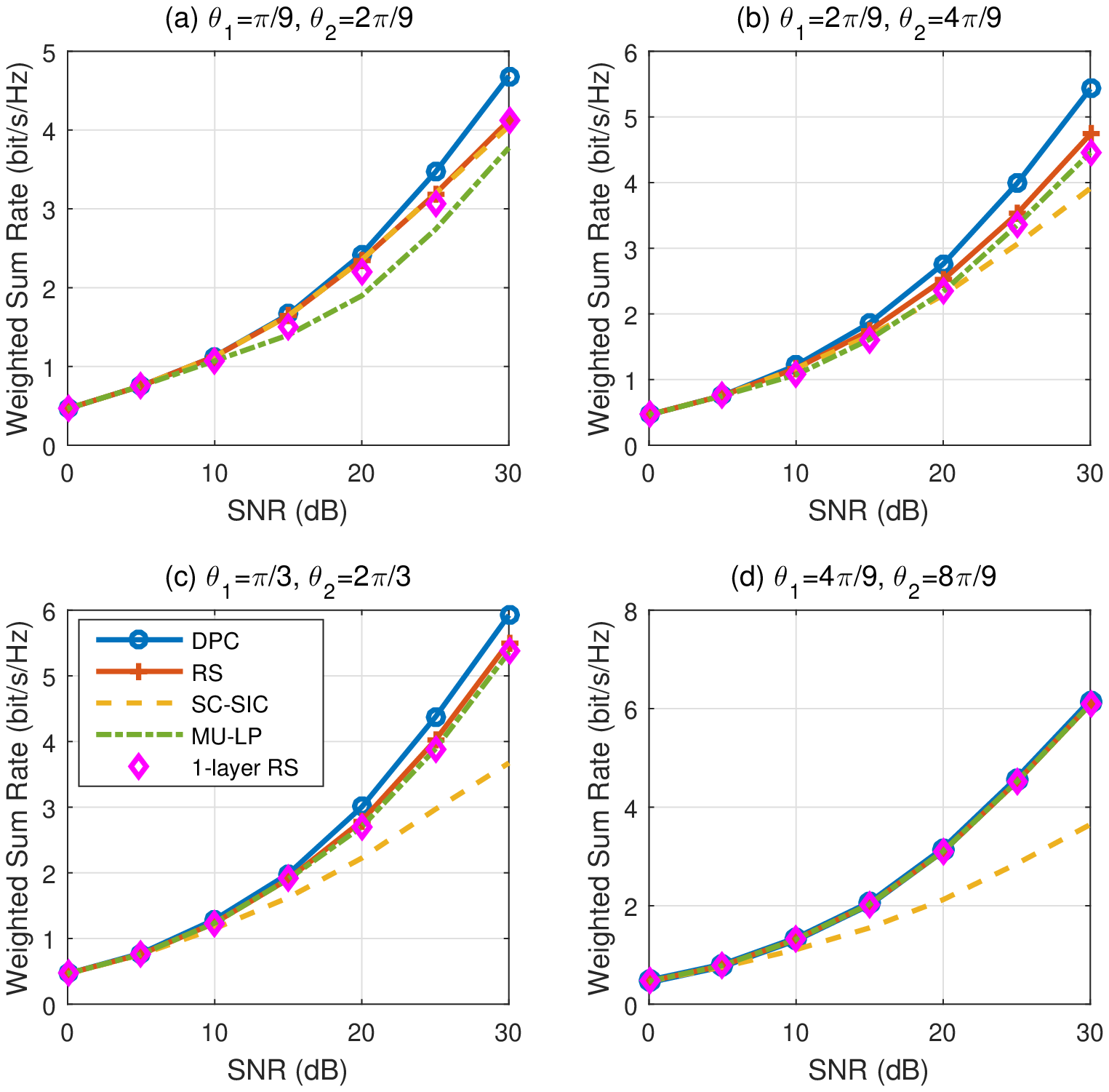}%
	\caption{Weighted sum rate versus SNR comparison of different strategies for underloaded three-user deployment with perfect CSIT, $\gamma_1=0.3, \gamma_2=0.1$, $u_1=0.2, u_2=0.3, u_3=0.5$, $N_t=4$, $R_k^{th}=0, k\in\{1,2,3\}$.}
	\label{fig: three user bias10301 020305 perfect}
\end{figure}

\begin{figure}[t!]
	\centering
	\includegraphics[width=3.4in]{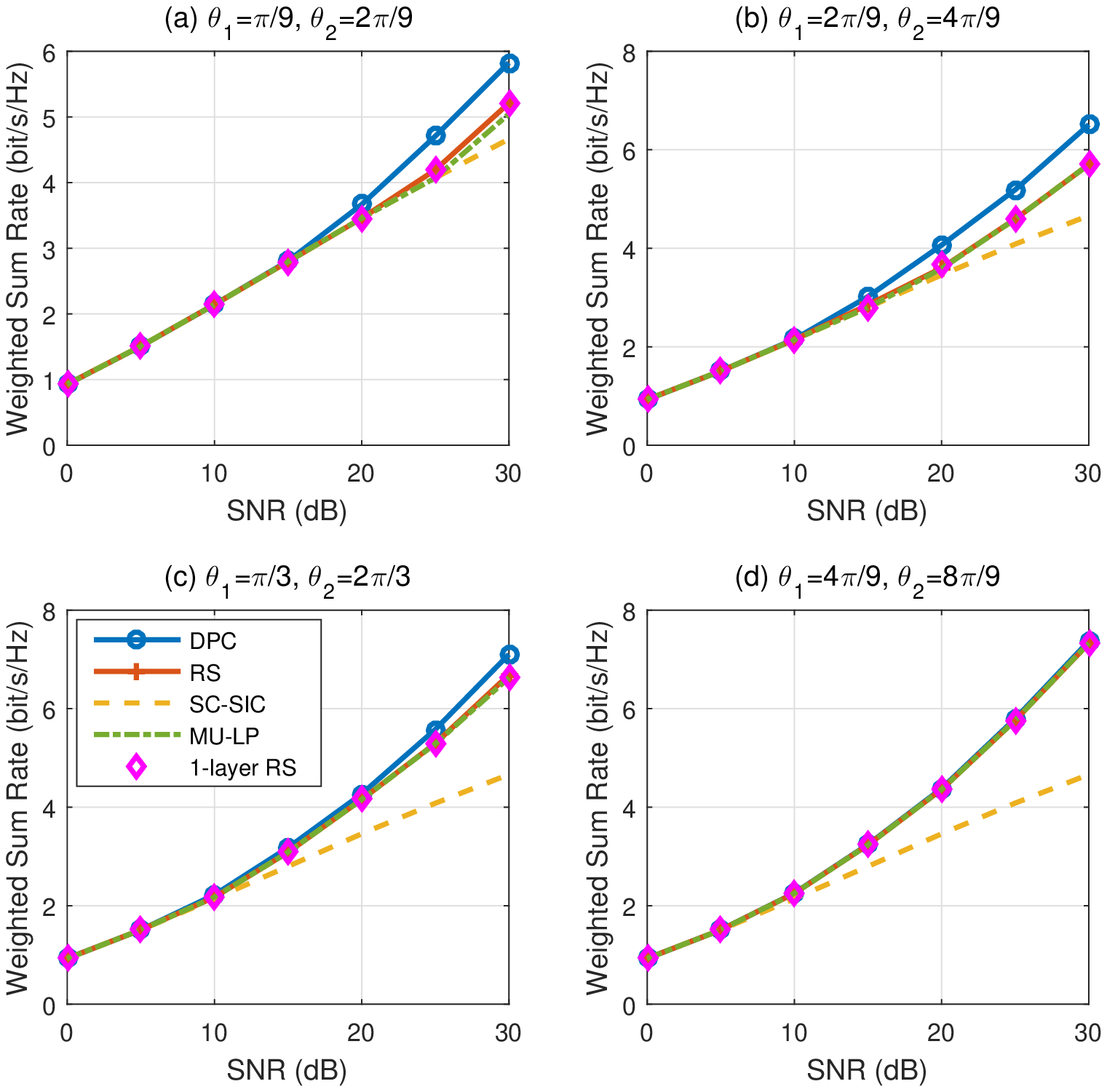}%
	\caption{Weighted sum rate versus SNR comparison of different strategies for underloaded three-user deployment with perfect CSIT, $\gamma_1=0.3, \gamma_2=0.1$, $u_1=0.4, u_2=0.3, u_3=0.3$, $N_t=4$, $R_k^{th}=0, k\in\{1,2,3\}$.}
	\label{fig: three user bias10301 040303 perfect}
\end{figure}

\begin{figure}[t!]
	\centering
	\includegraphics[width=3.4in]{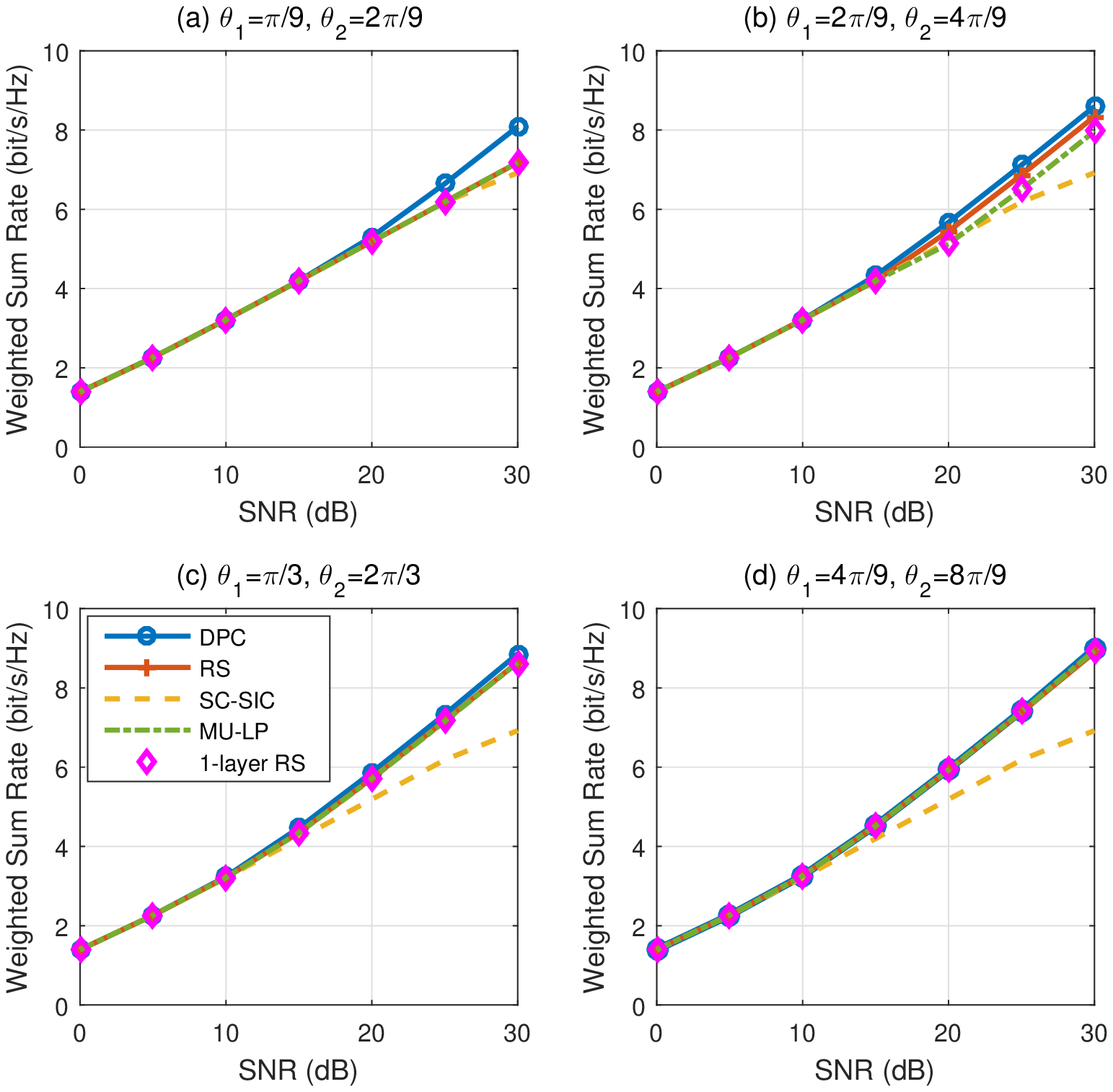}%
	\caption{Weighted sum rate versus SNR comparison of different strategies for underloaded three-user deployment with perfect CSIT, $\gamma_1=0.3, \gamma_2=0.1$, $u_1=0.6, u_2=0.3, u_3=0.1$, $N_t=4$, $R_k^{th}=0, k\in\{1,2,3\}$.}
	\label{fig: three user bias10301 060301 perfect}
\end{figure}

 In all figures (Fig.  \ref{fig: three user bias1 020305 perfect}--\ref{fig: three user bias10301 060301 perfect}), the WSR of RS is equal to or better than that of MU--LP and SC--SIC. Considering a specific scenario where  $\theta_1=\frac{2\pi}{9},\theta_2=\frac{4\pi}{9}$ and $\mathbf{u}=[0.6,0.3,0.1]$, the WSR of RS is better than that of  MU--LP and  SC--SIC as shown in Fig.  \ref{fig: three user bias1 060301 perfect}(b), Fig.  \ref{fig: three user bias1103 060301 perfect}(b) and Fig.  \ref{fig: three user bias10301 060301 perfect}(b). As SNR increases, the WSR improvement of RS is  generally more obvious. For a fixed weight vector, the WSR of SC--SIC becomes closer to that of RS as the channel gain differences among users increase. For example, we compare Fig. \ref{fig: three user bias1 020305 perfect}, Fig. \ref{fig: three user bias1103 020305 perfect} and Fig. \ref{fig: three user bias10301 020305 perfect} for a fixed  $\mathbf{u}=[0.2,0.3,0.5]$.  When $\mathbf{u}=[0.4,0.3,0.3]$, the WSR  of RS and MU--LP are almost identical. In such scenario, RS reduces to MU--LP. In subfigure (d) of each figure,  $\theta_1=\frac{4\pi}{9},\theta_2=\frac{8\pi}{9}$, the channels of user-$1$ and user-$2$, the channels of user-$2$ and user-$3$ are sufficiently orthogonal while the channels of user-$1$ and user-$3$ are almost in opposite directions. In such circumstance, the WSRs of RS and MU--LP strategies overlap with the optimal WSR achieved by DPC.


\subsection{Overloaded three-user deployment with perfect CSIT}
\subsubsection{Two transmit antenna deployment}
\begin{figure}[t!]
	\centering
	\includegraphics[width=3.4in]{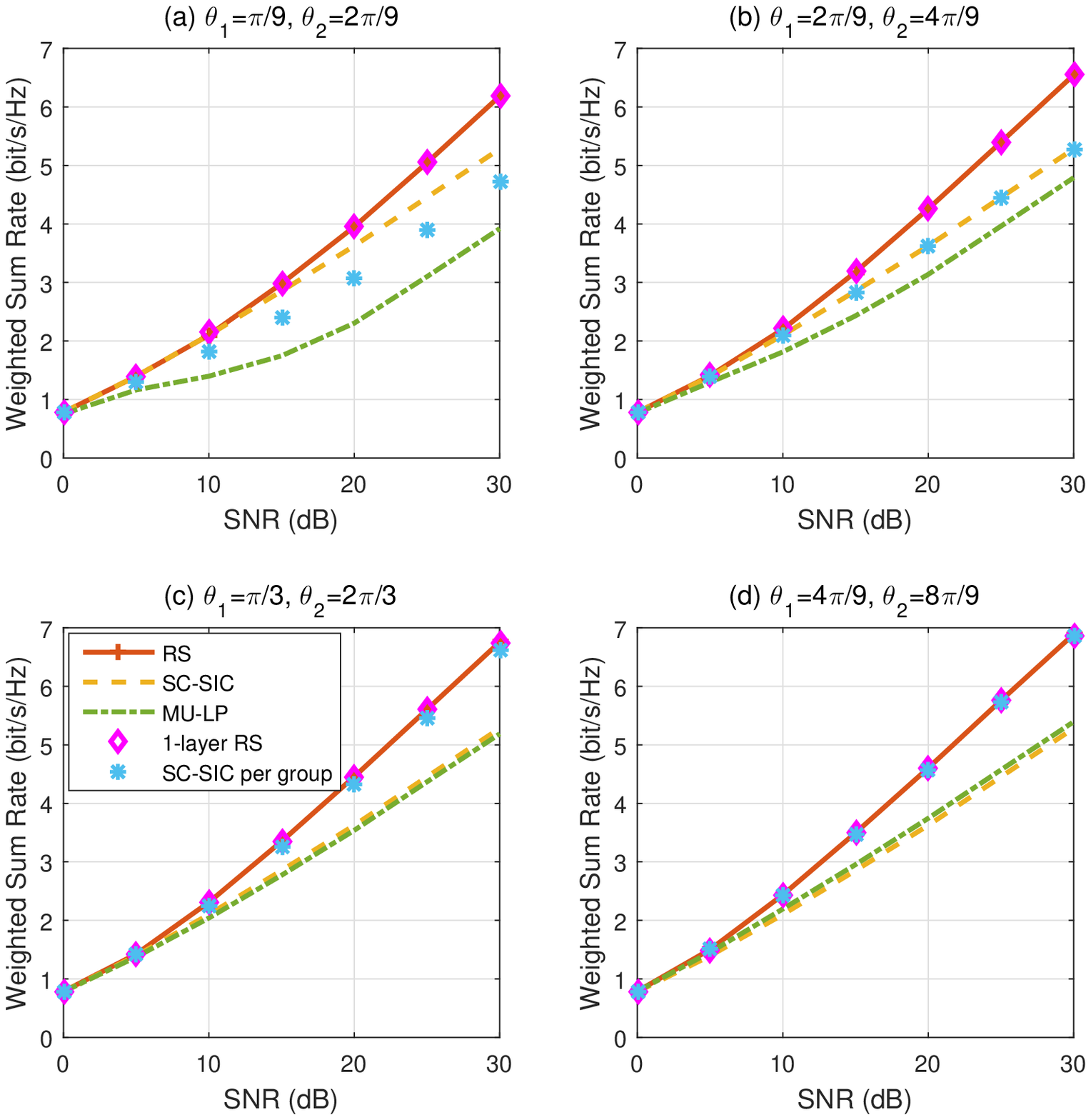}%
	\caption{Weighted sum rate versus SNR comparison of different strategies for overloaded three-user deployment with perfect CSIT, $\gamma_1=\gamma_2=1$, $u_1=0.2, u_2=0.3, u_3=0.5$, $N_t=2$, $\mathbf{r}_{th}=[0.02,0.08,0.19,0.3,0.4,0.4,0.4]$ bit/s/Hz.}
	\label{fig: three user bias1 020305 perfect overloaded}
\end{figure}
\begin{figure}[t!]
	\centering
	\includegraphics[width=3.1in]{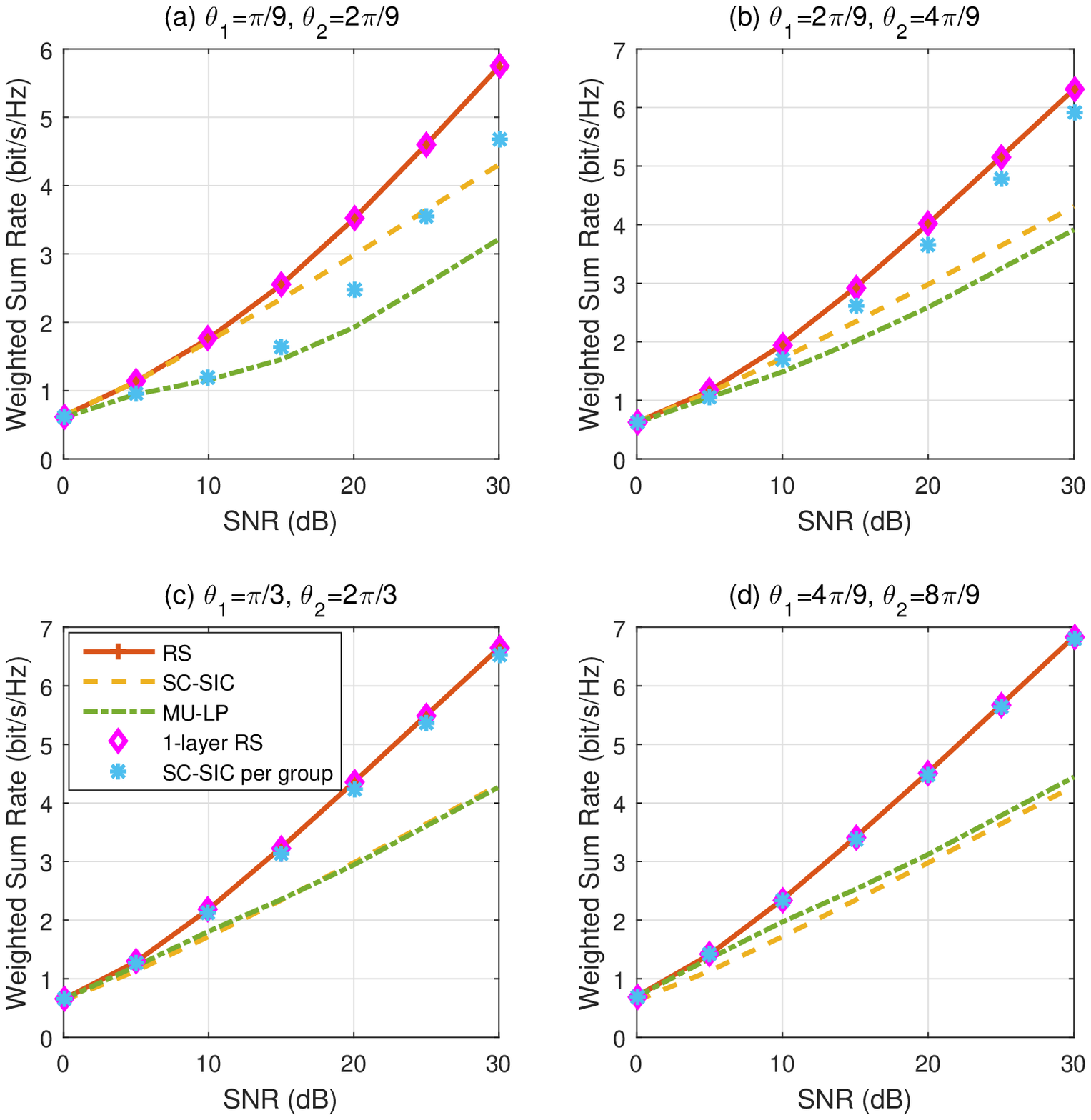}%
	\caption{Weighted sum rate versus SNR comparison of different strategies for overloaded three-user deployment with perfect CSIT, $\gamma_1=\gamma_2=1$, $u_1=0.4, u_2=0.3, u_3=0.3$, $N_t=2$, $\mathbf{r}_{th}=[0.02,0.08,0.19,0.3,0.4,0.4,0.4]$ bit/s/Hz.}
	\label{fig: three user bias1 040303 perfect overloaded}
\end{figure}
\begin{figure}[t!]
	\centering
	\includegraphics[width=3.1in]{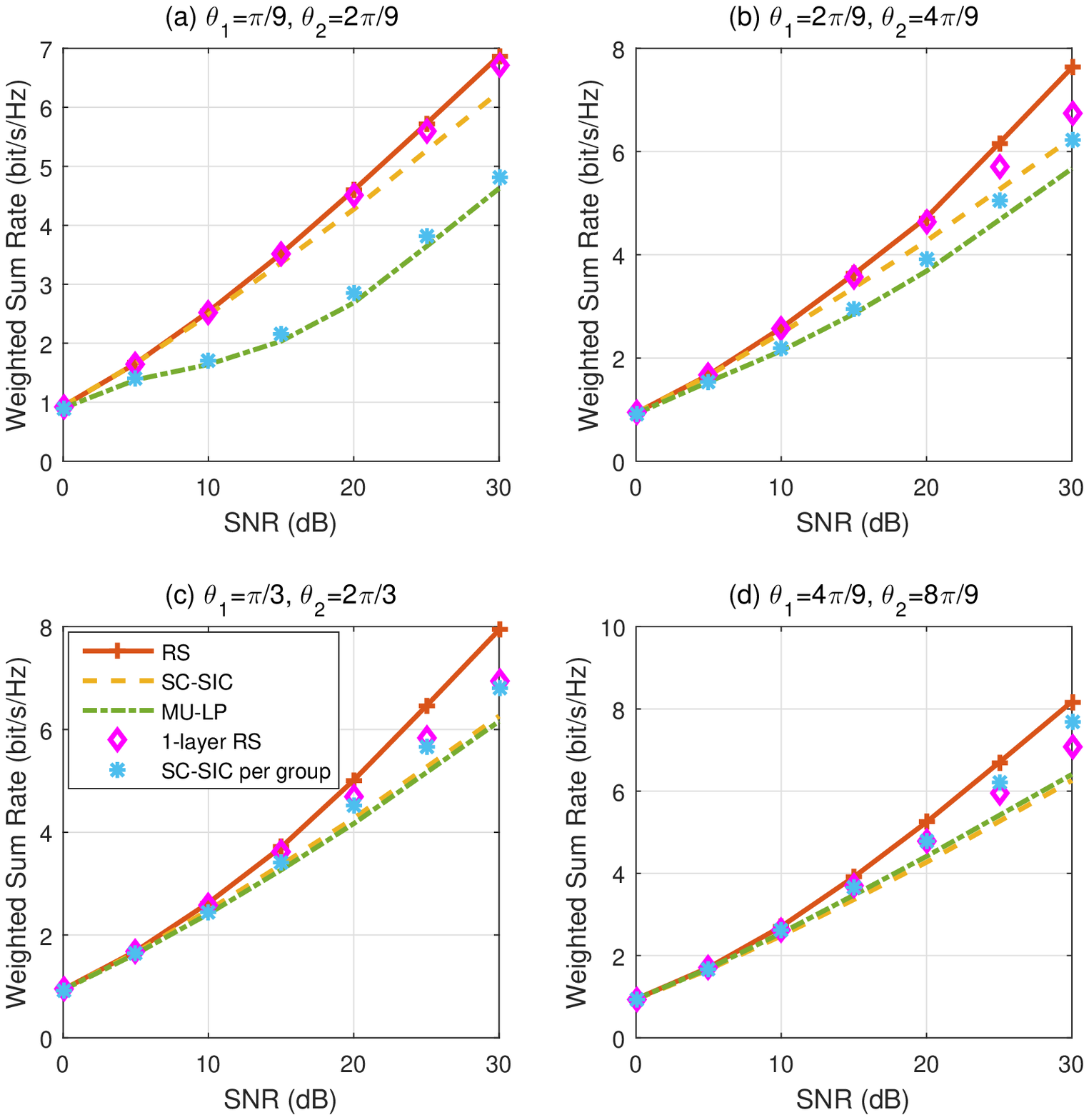}%
	\caption{Weighted sum rate versus SNR comparison of different strategies for overloaded three-user deployment with perfect CSIT, $\gamma_1=\gamma_2=1$, $u_1=0.6, u_2=0.3, u_3=0.1$, $N_t=2$, $\mathbf{r}_{th}=[0.02,0.08,0.19,0.3,0.4,0.4,0.4]$ bit/s/Hz.}
	\label{fig: three user bias1 060301 perfect overloaded}
\end{figure}
\begin{figure}[t!]
	\centering
	\includegraphics[width=3.1in]{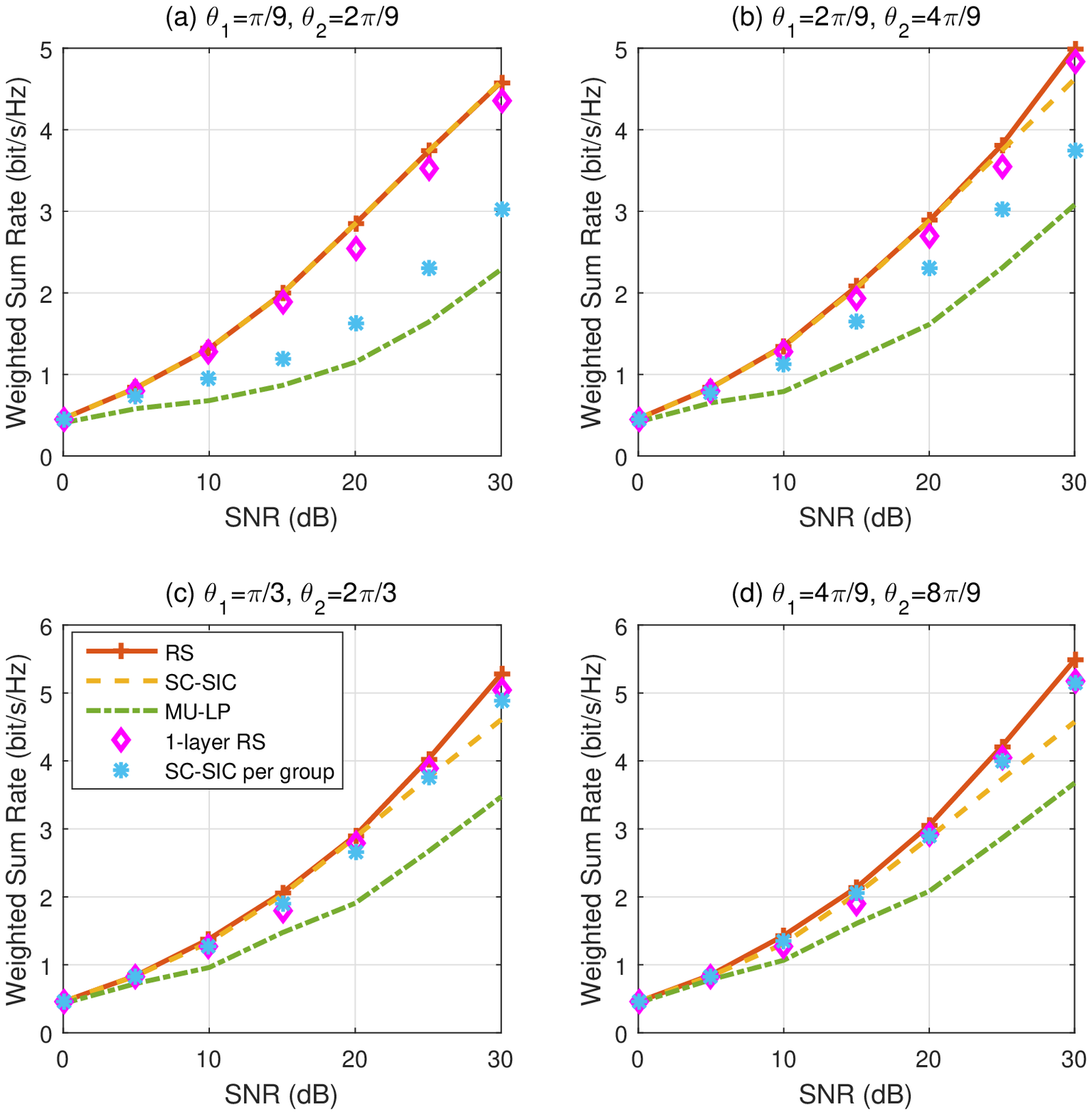}%
	\caption{Weighted sum rate versus SNR comparison of different strategies for overloaded three-user deployment with perfect CSIT, $\gamma_1=1, \gamma_2=0.3$, $u_1=0.2, u_2=0.3, u_3=0.5$, $N_t=2$, $\mathbf{r}_{th}=[0.02,0.08,0.19,0.3,0.4,0.4,0.4]$ bit/s/Hz.}
	\label{fig: three user bias1103 020305 perfect overloaded}
\end{figure}
\begin{figure}[t!]
	\centering
	\includegraphics[width=3.1in]{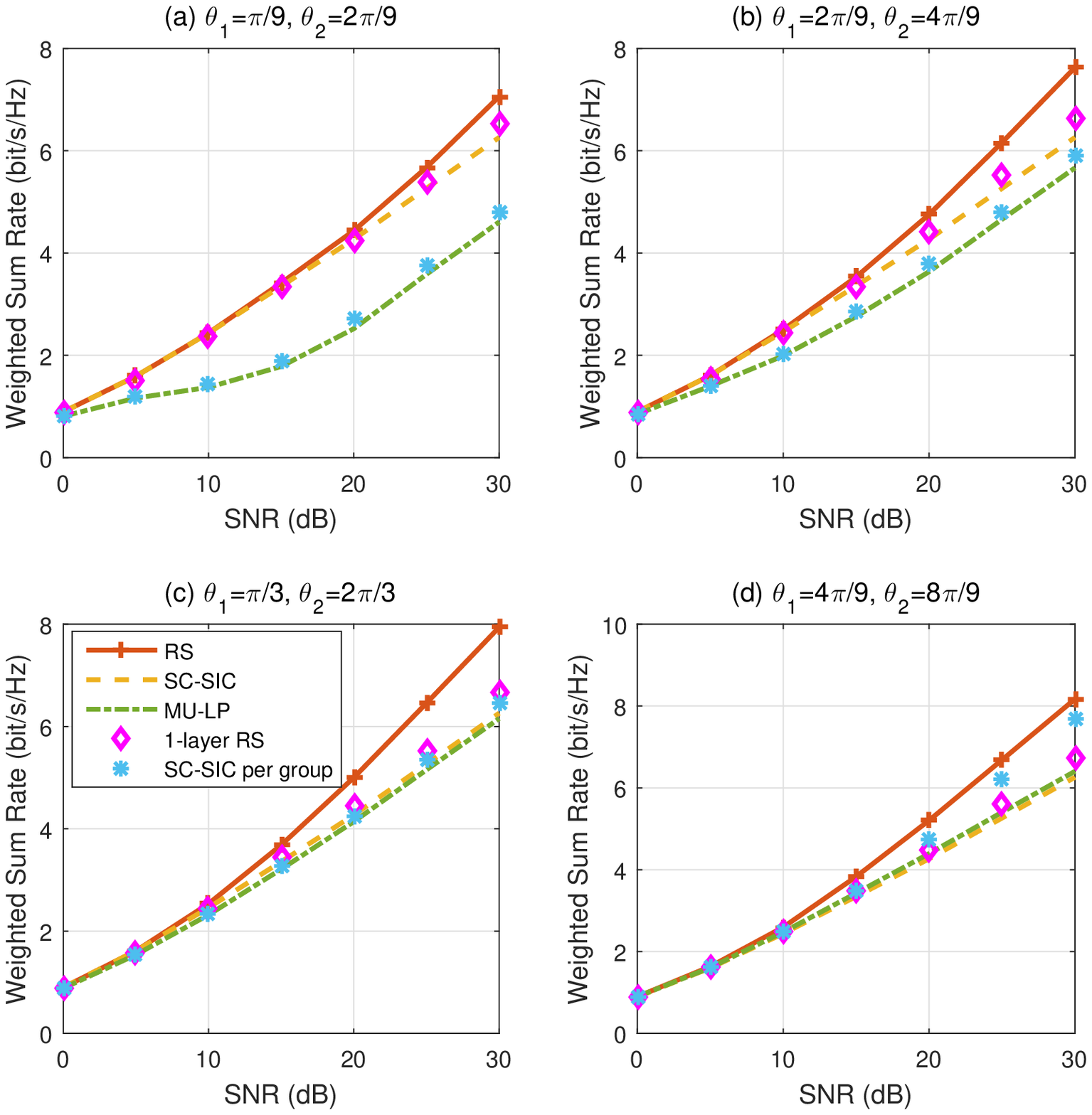}%
	\caption{Weighted sum rate versus SNR comparison of different strategies for overloaded three-user deployment with perfect CSIT, $\gamma_1=1, \gamma_2=0.3$, $u_1=0.6, u_2=0.3, u_3=0.1$, $N_t=2$, $\mathbf{r}_{th}=[0.02,0.08,0.19,0.3,0.4,0.4,0.4]$ bit/s/Hz.}
	\label{fig: three user bias1103 060301 perfect overloaded}
\end{figure}
\par  Fig. \ref{fig: three user bias1 020305 perfect overloaded}--\ref{fig: three user bias1103 060301 perfect overloaded} show the results when $\gamma_1, \gamma_2$ and $\mathbf{u}$ are varied as discussed in Appendix C.

\par RS exhibits a clear WSR gain over SC--SIC, SC--SIC per group and MU--LP in all figures (Fig. \ref{fig: three user bias1 020305 perfect overloaded}--\ref{fig: three user bias1103 060301 perfect overloaded}). 1-layer RS outperforms SC--SIC, SC--SIC per group and MU--LP in most figures. It further shows that 1-layer RS outperforms the joint switching between SC--SIC and SC--SIC per group  in most user deployments while the complexity of 1-layer RS is much reduced. 
In Fig. \ref{fig: three user bias1 020305 perfect overloaded}(a)--(c) and Fig. \ref{fig: three user bias1 040303 perfect overloaded}(a)--(c), 1-layer RS achieves the same WSR as RS. It implies that RS reduces to 1-layer RS in these user deployments. Both of RS and 1-layer RS achieve higher WSRs than all other strategies.

\begin{figure}[t!]
	\centering
	\includegraphics[width=3.0in]{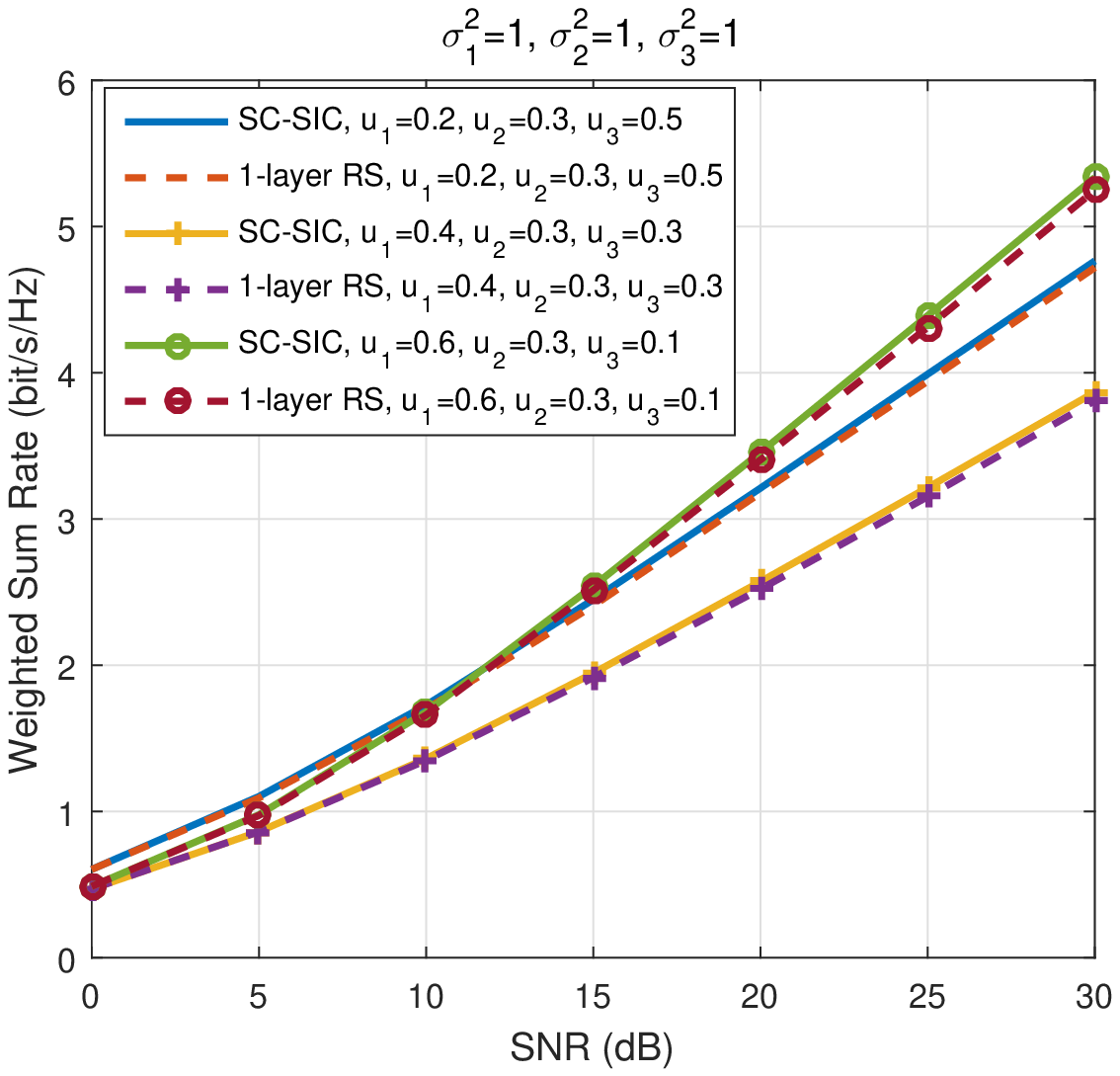}%
	\caption{Weighted sum rate versus SNR comparison of different strategies for overloaded three-user deployment with perfect CSIT, $\sigma_1^2=\sigma_2^2=\sigma_3^2=1$, $N_t=1$, $\mathbf{r}_{th}=[0,0,0.01,0.03,0.1,0.2,0.3]$ bit/s/Hz.}
	\label{fig: three user bias11 siso overloaded}
\end{figure}

\begin{figure}[t!]
	\centering
	\includegraphics[width=3.0in]{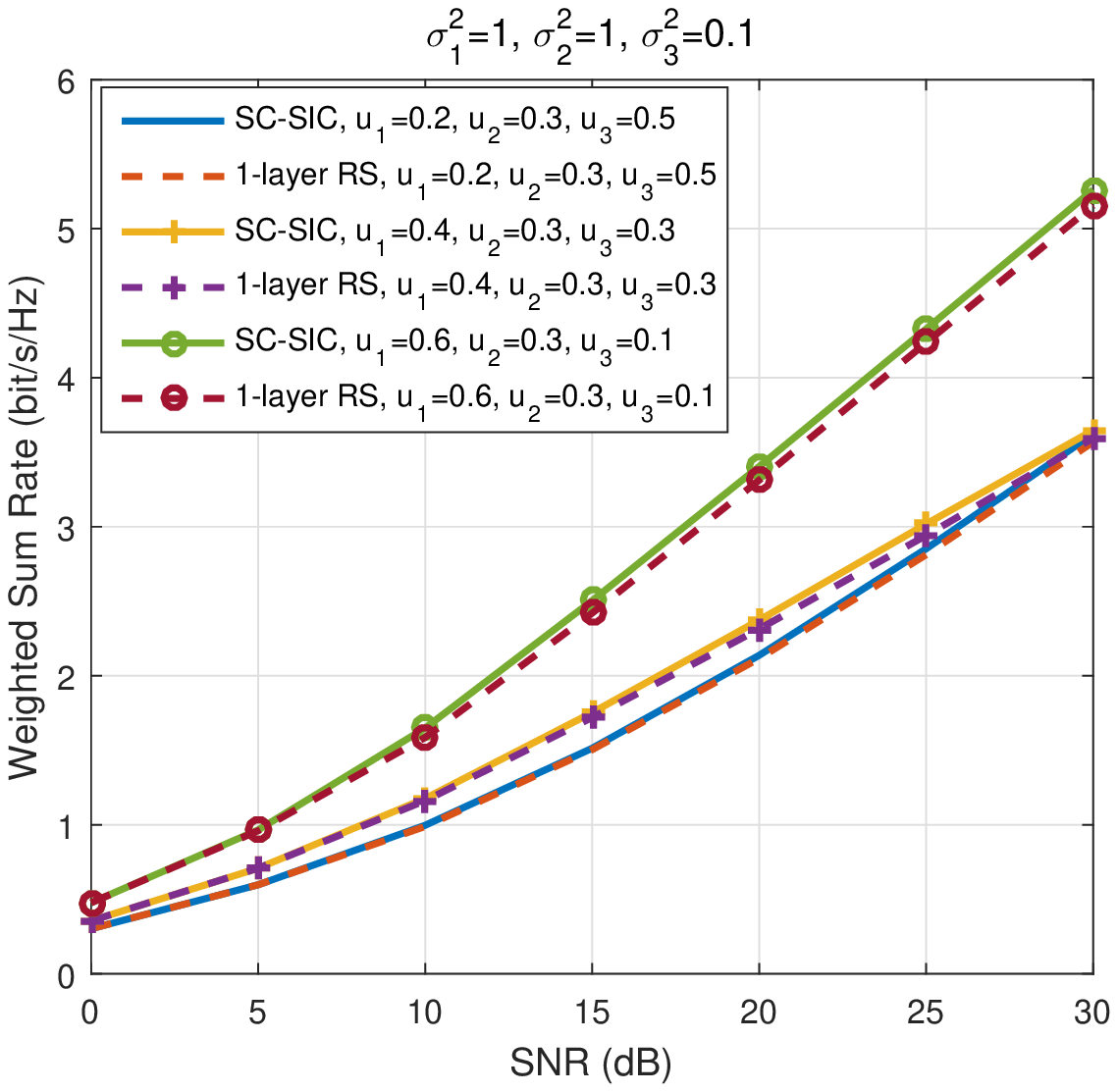}%
	\caption{Weighted sum rate versus SNR comparison of different strategies for overloaded three-user deployment with perfect CSIT, $\sigma_1^2=\sigma_2^2=1, \sigma_3^2=0.3$, $N_t=1$, $\mathbf{r}_{th}=[0,0,0.01,0.03,0.1,0.2,0.3]$ bit/s/Hz.}
	\label{fig: three user bias103 siso overloaded}
\end{figure}

\subsubsection{Single transmit antenna deployment}
 \par Fig. \ref{fig: three user bias11 siso overloaded} and  Fig. \ref{fig: three user bias103 siso overloaded} show the average rate regions of different strategies over 10 random channel realizations when $\sigma_1^2=\sigma_2^2=\sigma_3^2=1$ and $\sigma_1^2=\sigma_2^2=1, \sigma_3^2=0.3$, respectively. We further show that 1-layer RS is an attractive alternative to SC--SIC.

\subsection{Underloaded three-user deployment with imperfect CSIT}
\par We consider the imperfect CSIT scenarios. The channel model in the two-user deployment with imperfect CSIT is extended here. The estimated channel of user-1, user-2 and user-3 are initialized using equation (\ref{eq: channel three users}). 
For the given channel estimate at the BS, the channel realization  is $\mathbf{h}_k=\widehat{\mathbf{h}}_{k}+\widetilde{\mathbf{h}}_{k},\forall k\in\{1,2,3\}$, where $\widetilde{\mathbf{h}}_{k}$ is the estimated error of user-$k$. $\widetilde{\mathbf{h}}_{k}$ has i.i.d. complex Gaussian entries drawn from $\mathcal{CN}(0,\sigma_{e,k}^2)$. The error covariance of user-1, user-2 and user-3 are $\sigma_{e,1}^2=P_t^{-0.6}$, $\sigma_{e,2}^2=\gamma_1 P_t^{-0.6}$ and $\sigma_{e,3}^2=\gamma_2 P_t^{-0.6}$, respectively. The precoders are initialized and designed using the estimated channels $\widehat{\mathbf{h}}_{1},\widehat{\mathbf{h}}_{2},\widehat{\mathbf{h}}_{3}$ and the same methods as stated in perfect CSIT scenarios.  1000 different channel error samples are generated for each user. Each point in the rate region is the average rate over the generated 1000 channels.

\par
\begin{figure}[t!]
	\centering
	\includegraphics[width=3.1in]{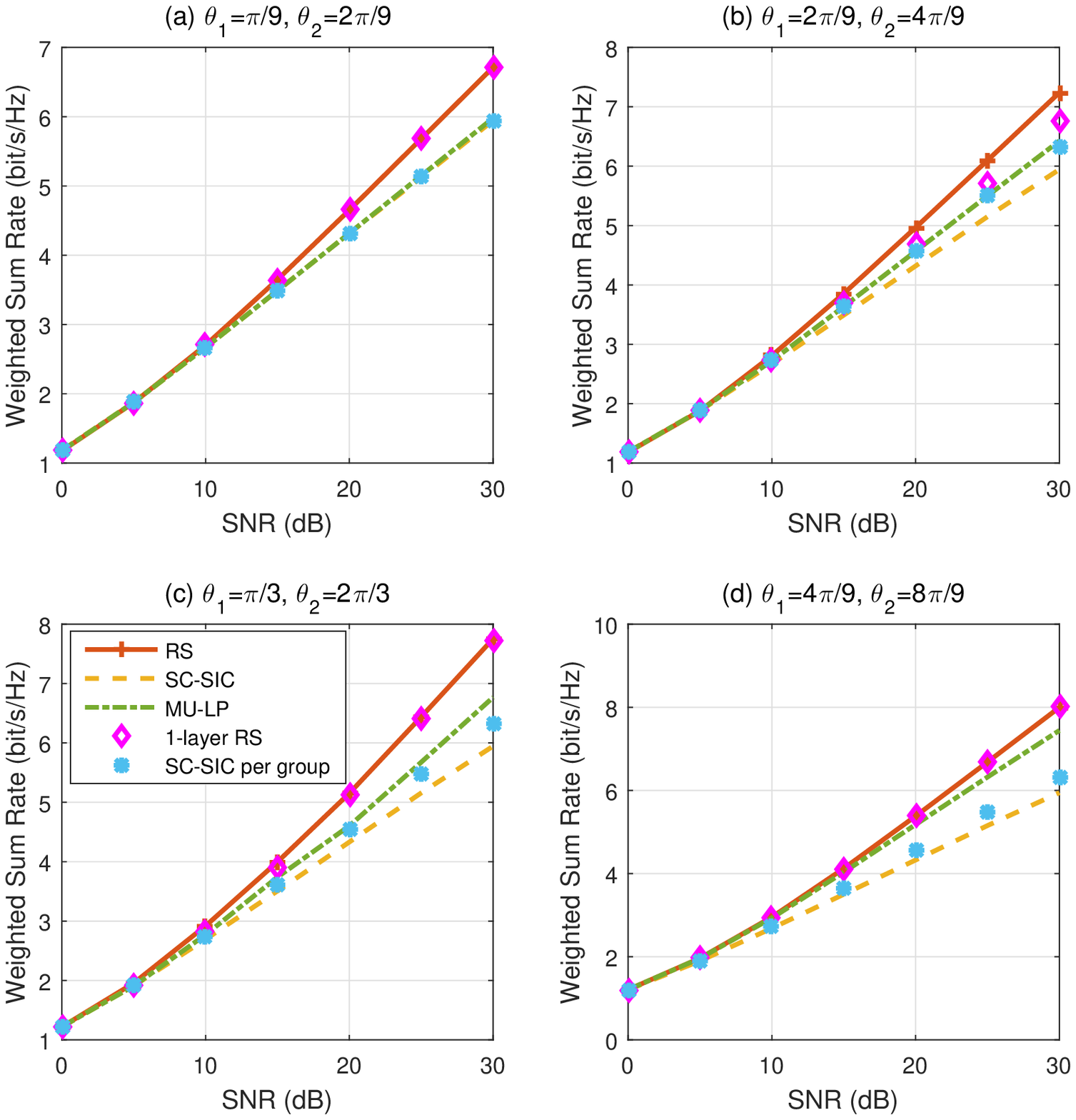}%
	\caption{Weighted sum rate versus SNR comparison of different strategies for underloaded three-user deployment with imperfect CSIT, $\gamma_1=\gamma_2=1$, $u_1=0.2, u_2=0.3, u_3=0.5$, $N_t=4$, $R_k^{th}=0, k\in\{1,2,3\}$.}
	\label{fig: bias11 weights020305 underloaded imperfect}
\end{figure}

\begin{figure}[t!]
	\centering
	\includegraphics[width=3.1in]{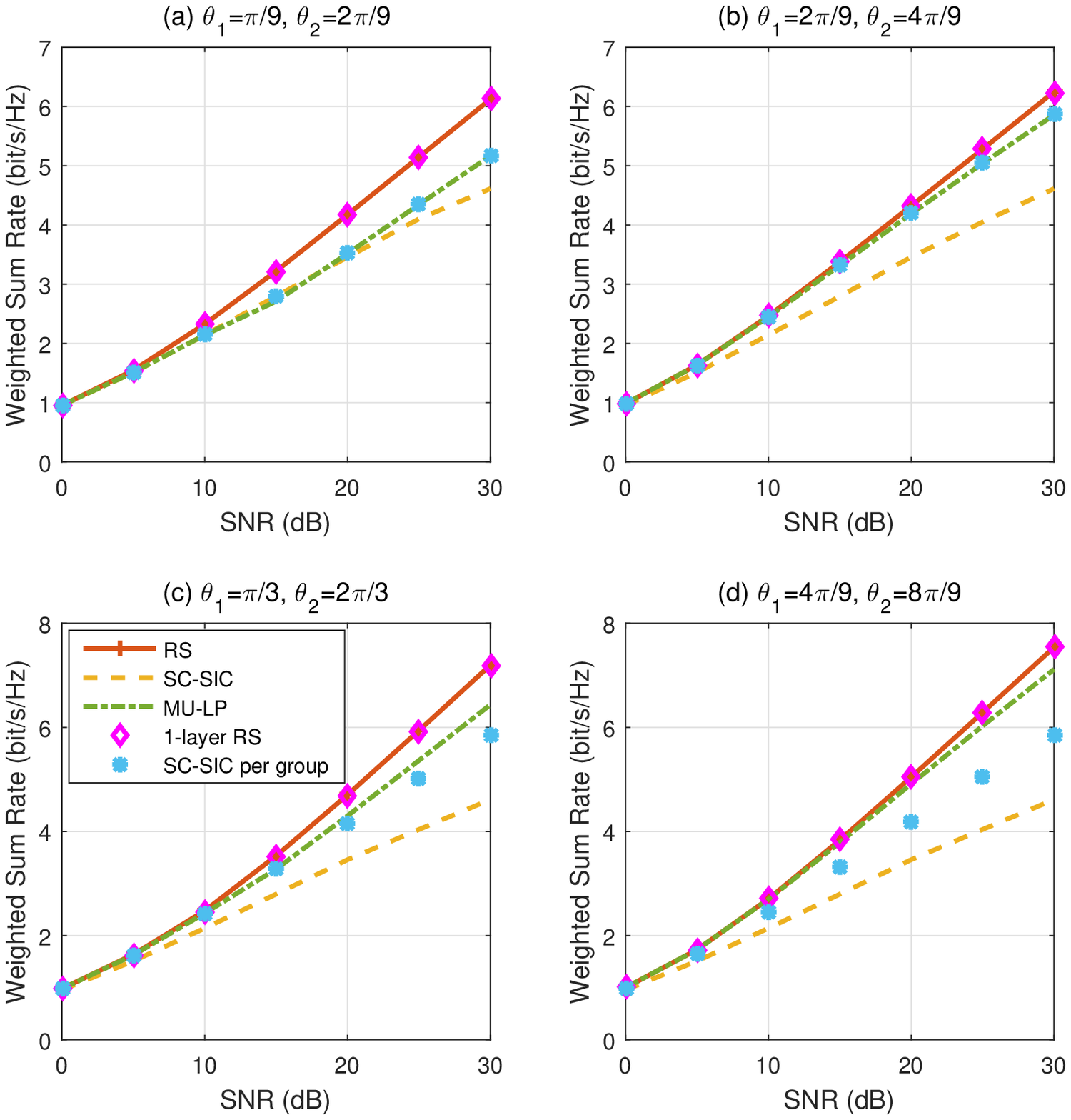}%
	\caption{Weighted sum rate versus SNR comparison of different strategies for underloaded three-user deployment with imperfect CSIT, $\gamma_1=\gamma_2=1$, $u_1=0.4, u_2=0.3, u_3=0.3$, $N_t=4$, $R_k^{th}=0, k\in\{1,2,3\}$.}
	\label{fig: bias11 weights040303 underloaded imperfect}
\end{figure}

\begin{figure}[t!]
	\centering
	\includegraphics[width=3.1in]{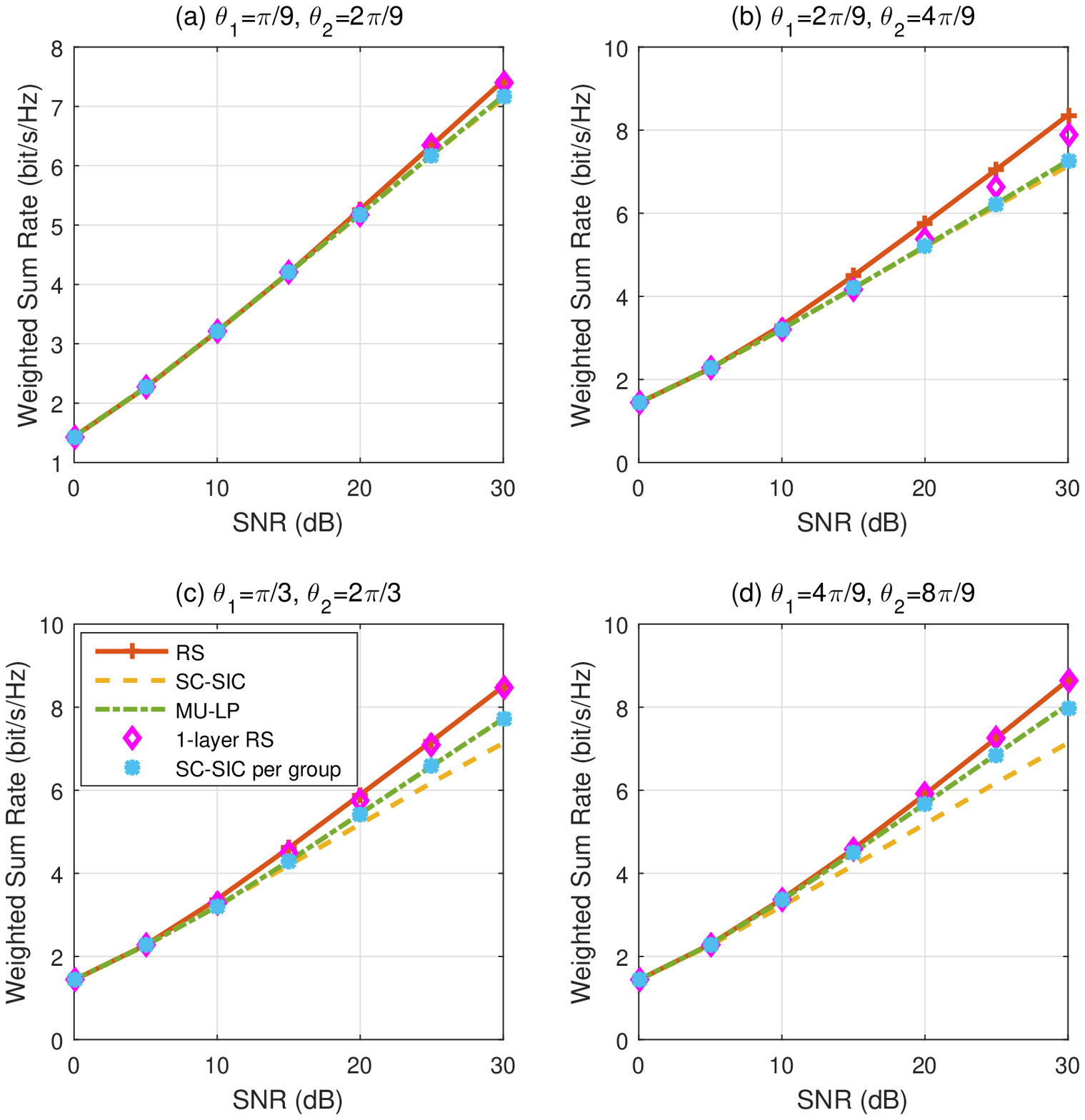}%
	\caption{Weighted sum rate versus SNR comparison of different strategies for underloaded three-user deployment with imperfect CSIT, $\gamma_1=\gamma_2=1$, $u_1=0.6, u_2=0.3, u_3=0.1$, $N_t=4$, $R_k^{th}=0, k\in\{1,2,3\}$.}
	\label{fig: bias11 weights060301 underloaded imperfect}
\end{figure}

\begin{figure}[t!]
	\centering
	\includegraphics[width=3.1in]{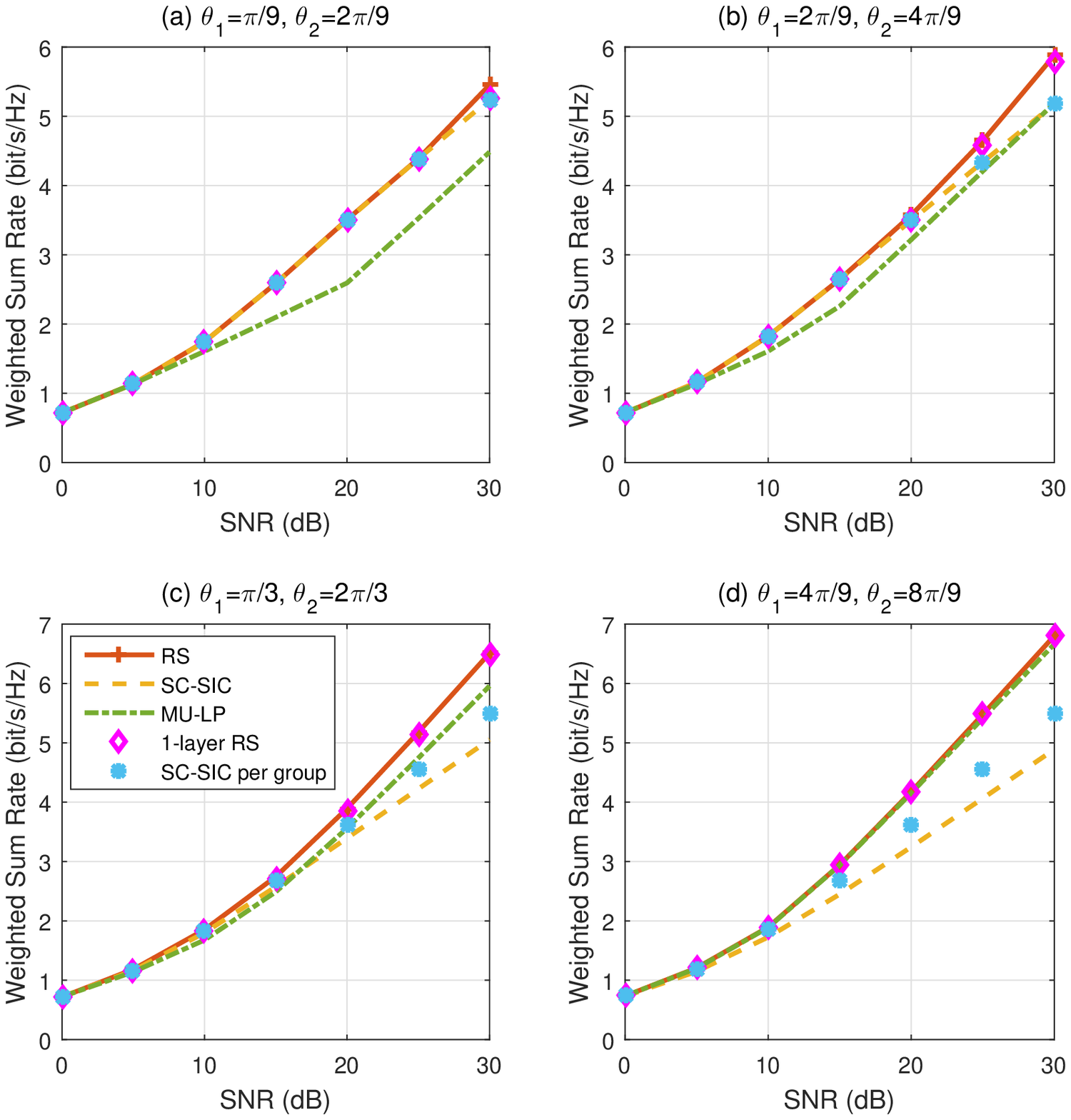}%
	\caption{Weighted sum rate versus SNR comparison of different strategies for underloaded three-user deployment with imperfect CSIT, $\gamma_1=1,\gamma_2=0.3$, $u_1=0.2, u_2=0.3, u_3=0.5$, $N_t=4$, $R_k^{th}=0, k\in\{1,2,3\}$.}
	\label{fig: bias103 weights020305 underloaded imperfect}
\end{figure}

\begin{figure}[t!]
	\centering
	\includegraphics[width=3.1in]{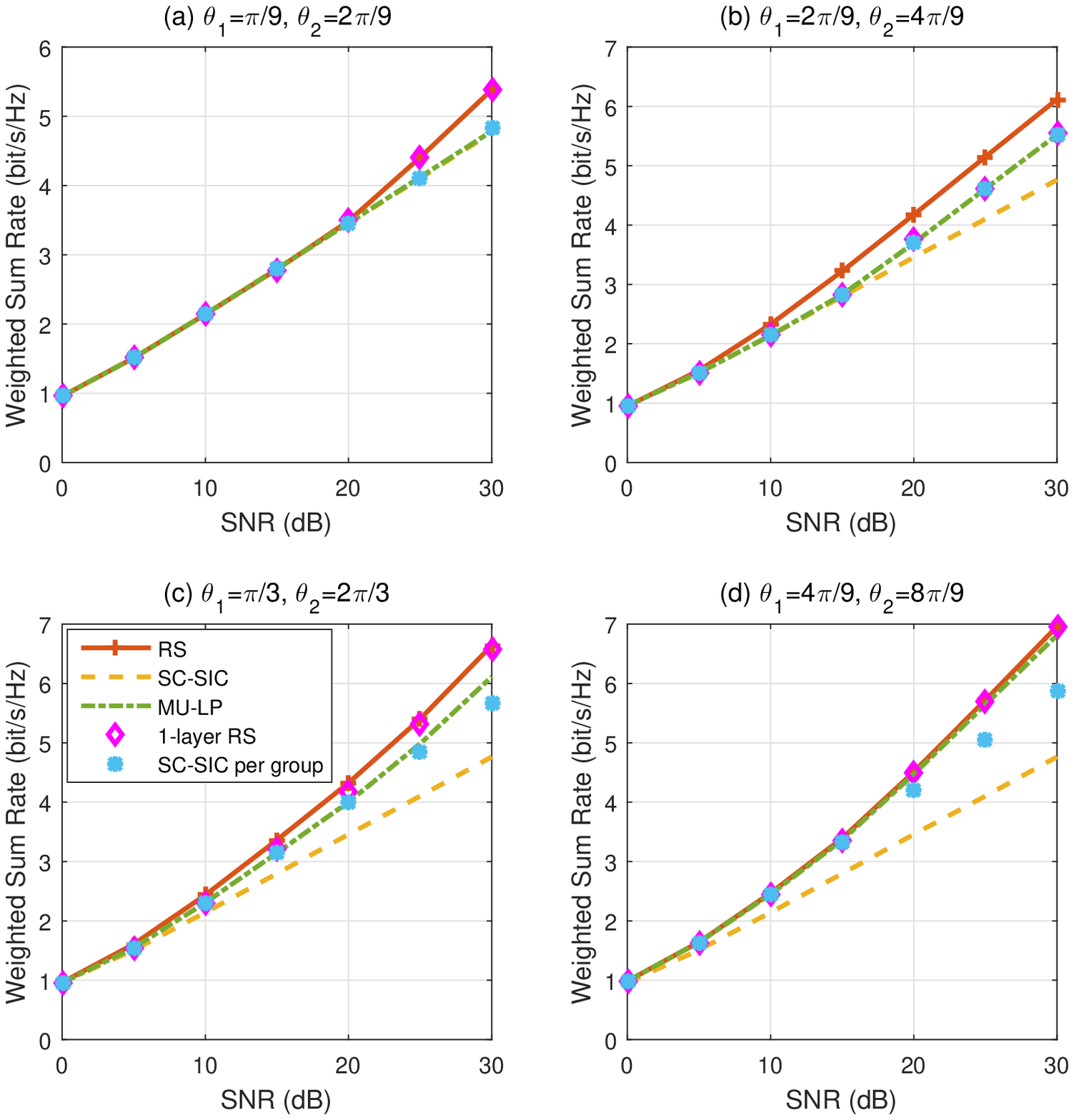}%
	\caption{Weighted sum rate versus SNR comparison of different strategies for underloaded three-user deployment with imperfect CSIT, $\gamma_1=1,\gamma_2=0.3$, $u_1=0.4, u_2=0.3, u_3=0.3$, $N_t=4$, $R_k^{th}=0, k\in\{1,2,3\}$.}
	\label{fig: bias103 weights040303 underloaded imperfect}
\end{figure}

\begin{figure}[t!]
	\centering
	\includegraphics[width=3.1in]{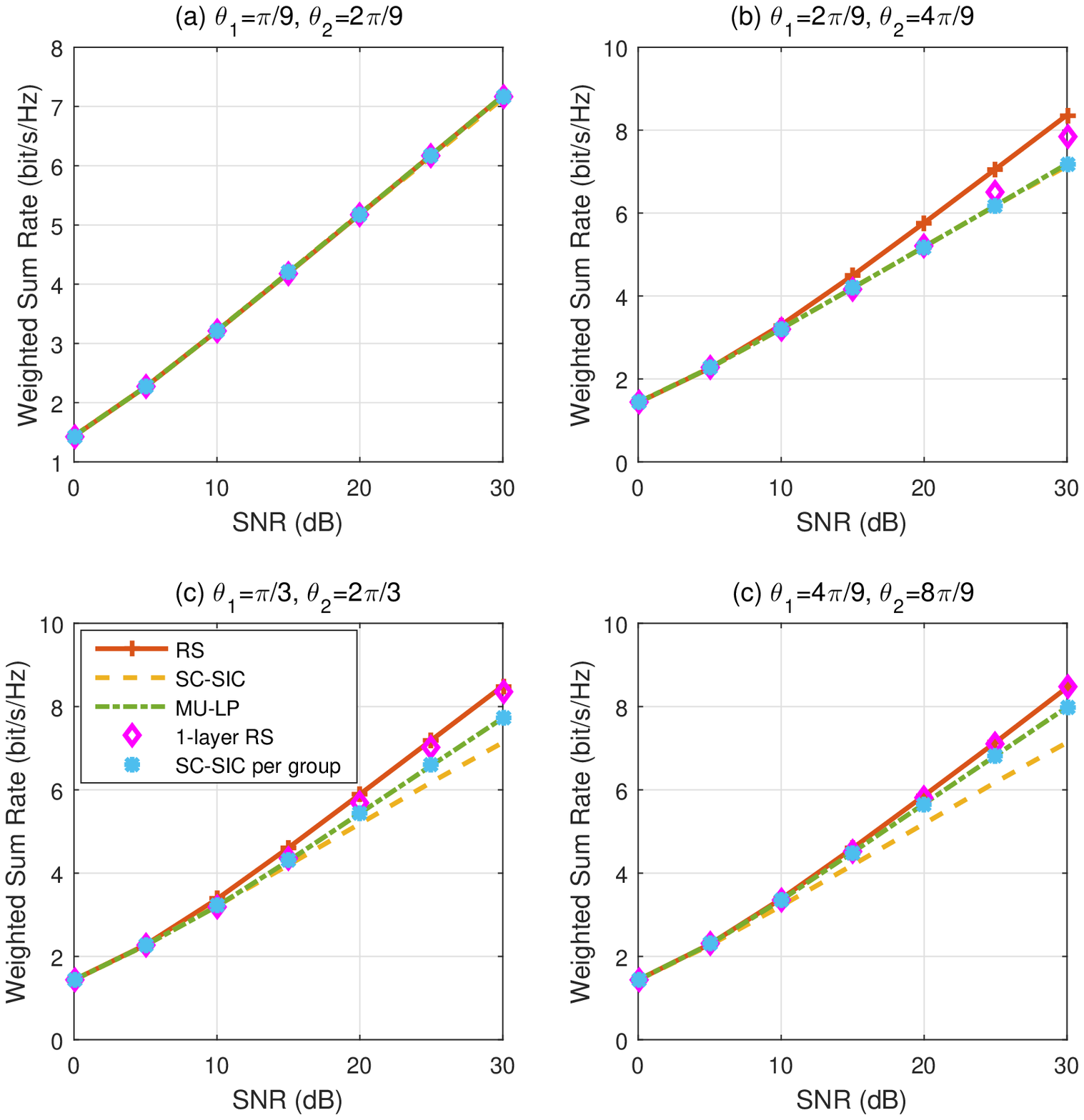}%
	\caption{Weighted sum rate versus SNR comparison of different strategies for underloaded three-user deployment with imperfect CSIT, $\gamma_1=1,\gamma_2=0.3$, $u_1=0.6, u_2=0.3, u_3=0.1$, $N_t=4$, $R_k^{th}=0, k\in\{1,2,3\}$.}
	\label{fig: bias103 weights060301 underloaded imperfect}
\end{figure}

 Comparing with the simulation results in perfect CSIT, the WSR gap between RS and MU--LP increases in imperfect CSIT.  In contrast, the WSR gap between RS and 1-layer RS decreases in imperfect CSIT. 1-layer RS achieves equal or better WSRs than SC--SIC, SC--SIC per group and MU--LP in all figures (Fig. \ref{fig: bias11 weights020305 underloaded imperfect}--\ref{fig: bias103 weights060301 underloaded imperfect}). As mentioned earlier, all forms of RS are suited to any network load and channel circumstances of users. Moreover,  all forms of RS are robust to imperfect CSIT.

\subsection{Overloaded three-user deployment with imperfect CSIT}
\par
\begin{figure}[t!]
	\centering
	\includegraphics[width=3.1in]{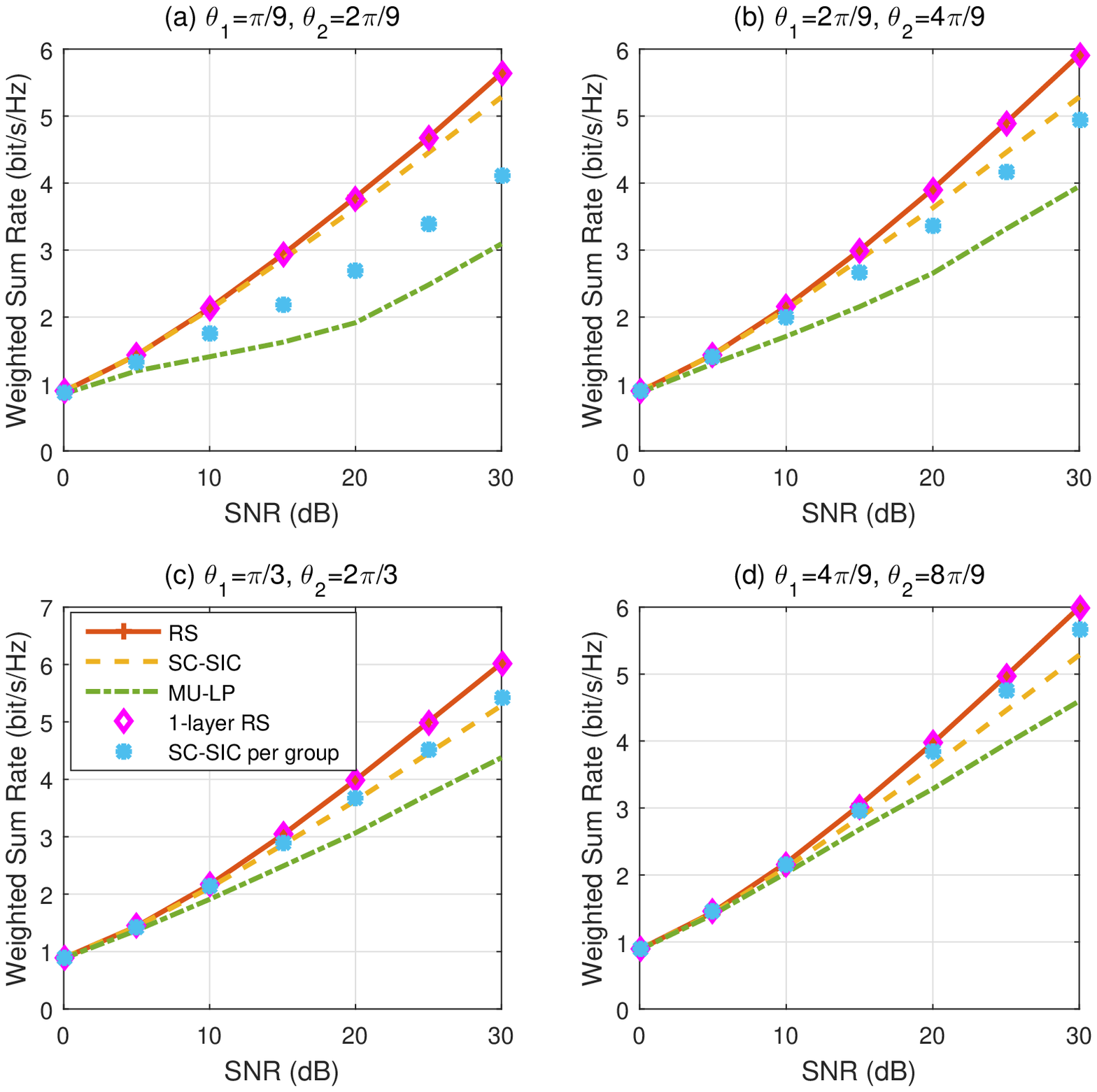}%
	\caption{Weighted sum rate versus SNR comparison of different strategies for overloaded three-user deployment with imperfect CSIT, $\gamma_1=\gamma_2=1$, $u_1=0.2, u_2=0.3, u_3=0.5$, $N_t=2$, $\mathbf{r}_{th}=[0.02,0.08,0.19,0.3,0.4,0.4,0.4]$ bit/s/Hz.}
	\label{fig: bias11 weights020305 overloaded imperfect}
\end{figure}

\begin{figure}[t!]
	\centering
	\includegraphics[width=3.1in]{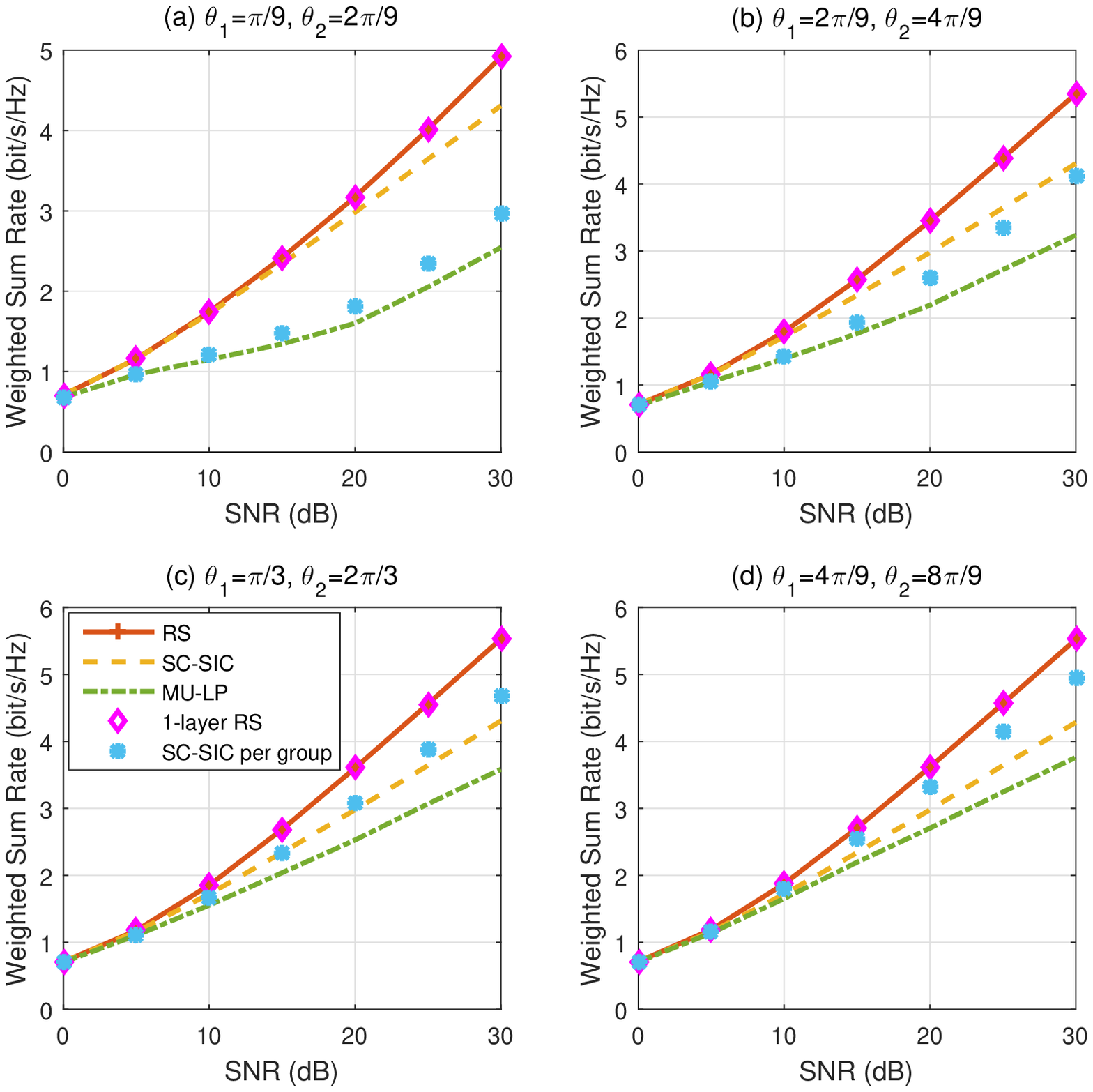}%
	\caption{Weighted sum rate versus SNR comparison of different strategies for overloaded three-user deployment with imperfect CSIT, $\gamma_1=\gamma_2=1$, $u_1=0.4, u_2=0.3, u_3=0.3$, $N_t=2$, $\mathbf{r}_{th}=[0.02,0.08,0.19,0.3,0.4,0.4,0.4]$ bit/s/Hz.}
	\label{fig: bias11 weights040303 overloaded imperfect}
\end{figure}

\begin{figure}[t!]
	\centering
	\includegraphics[width=3.1in]{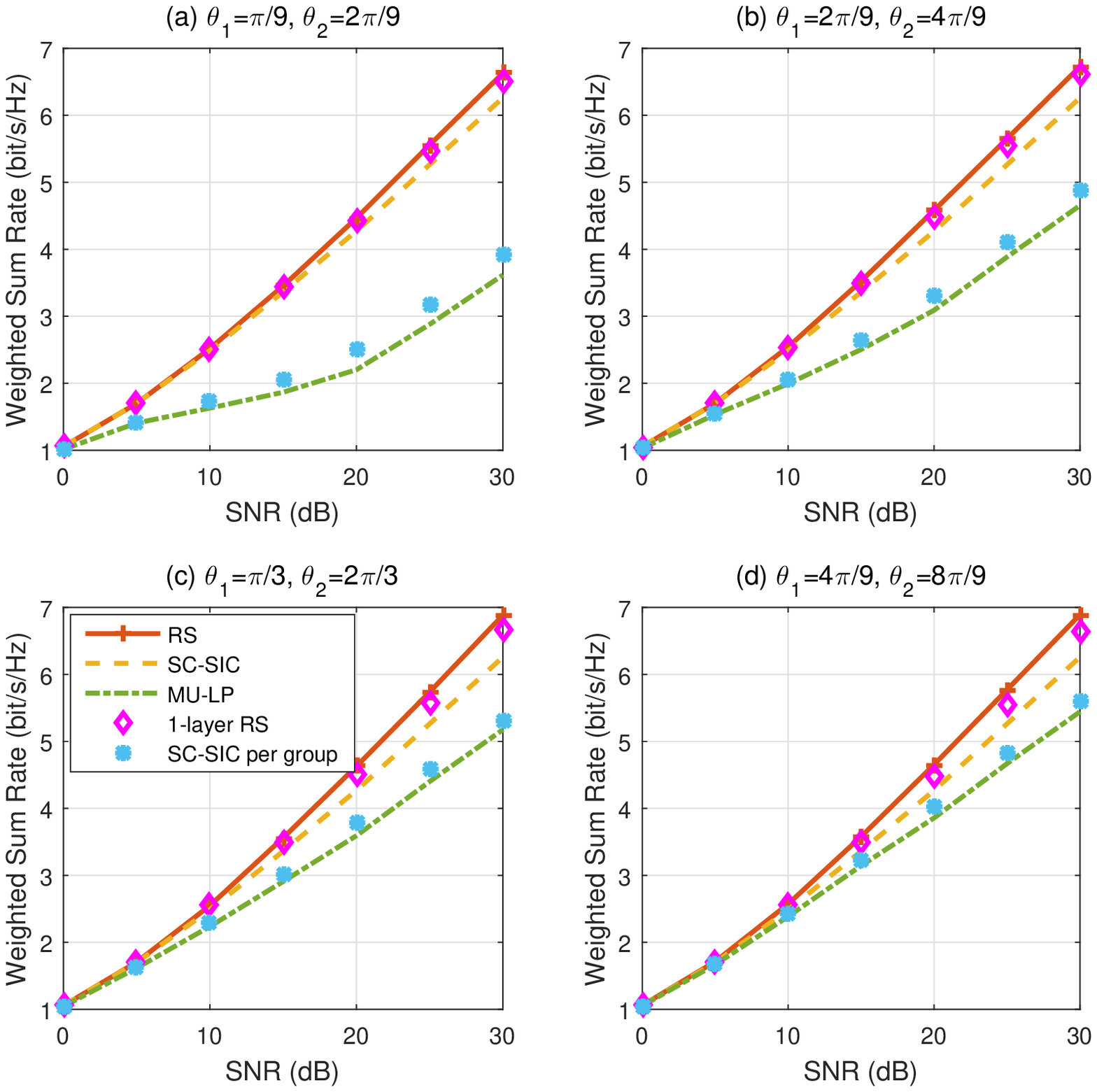}%
	\caption{Weighted sum rate versus SNR comparison of different strategies for overloaded three-user deployment with imperfect CSIT, $\gamma_1=\gamma_2=1$, $u_1=0.6, u_2=0.3, u_3=0.1$, $N_t=2$, $\mathbf{r}_{th}=[0.02,0.08,0.19,0.3,0.4,0.4,0.4]$ bit/s/Hz.}
	\label{fig: bias11 weights060301 overloaded imperfect}
\end{figure}

\begin{figure}[t!]
	\centering
	\includegraphics[width=3.1in]{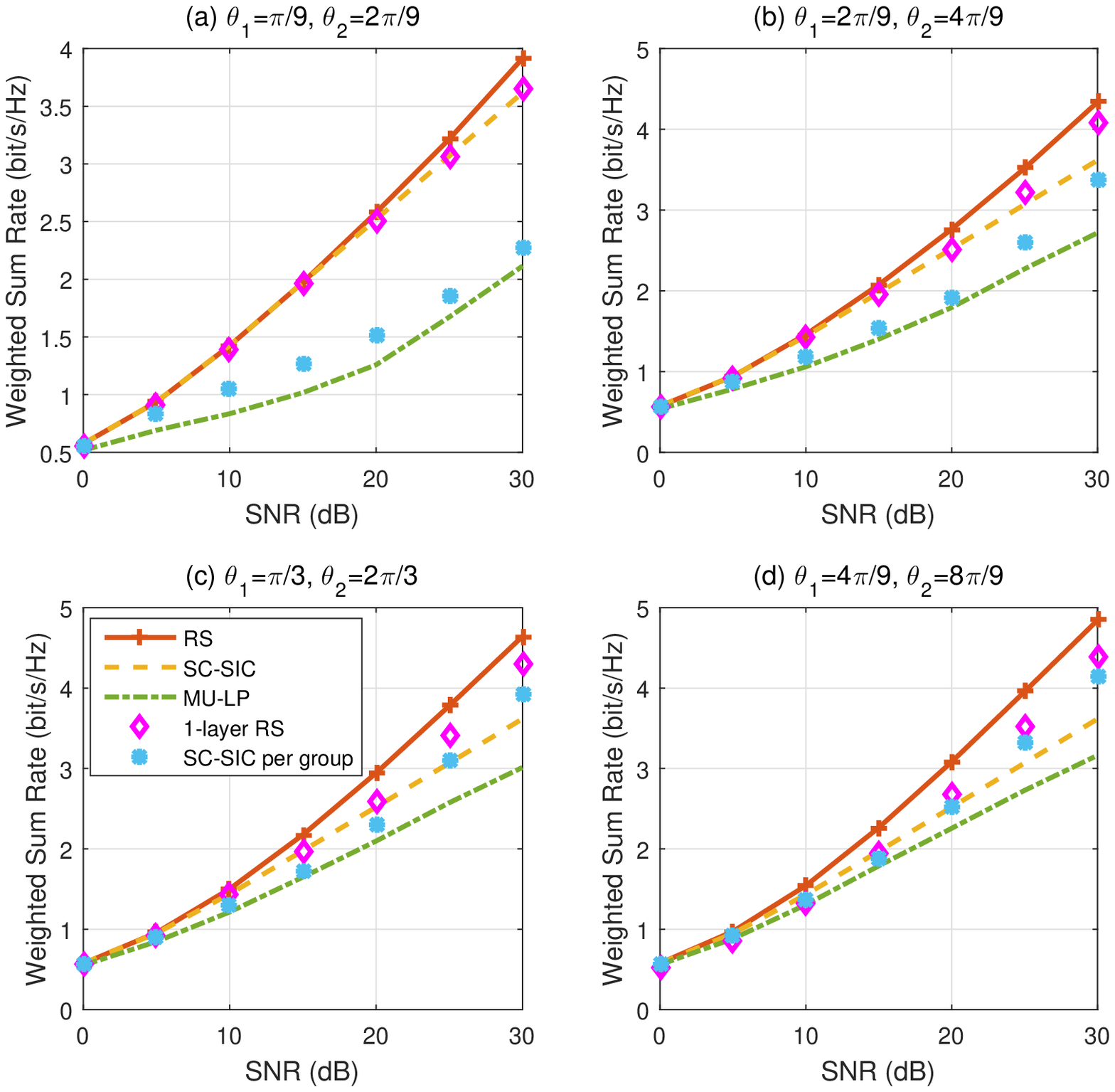}%
	\caption{Weighted sum rate versus SNR comparison of different strategies for overloaded three-user deployment with imperfect CSIT, $\gamma_1=1,\gamma_2=0.3$, $u_1=0.2, u_2=0.3, u_3=0.5$, $N_t=2$, $\mathbf{r}_{th}=[0.02,0.08,0.19,0.3,0.4,0.4,0.4]$ bit/s/Hz.}
	\label{fig: bias103 weights020305 overloaded imperfect}
\end{figure}

\begin{figure}[t!]
	\centering
	\includegraphics[width=3.1in]{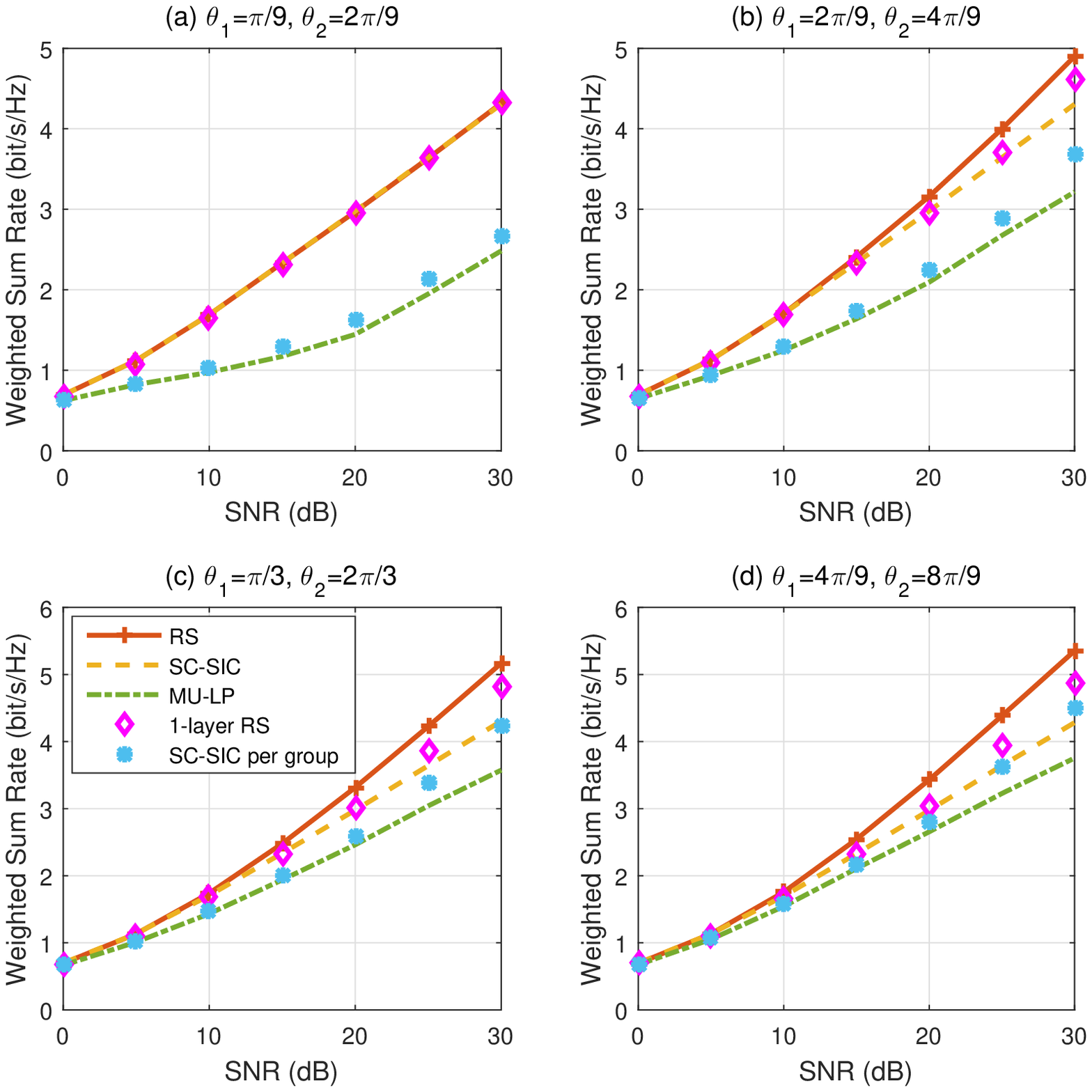}%
	\caption{Weighted sum rate versus SNR comparison of different strategies for overloaded three-user deployment with imperfect CSIT, $\gamma_1=1,\gamma_2=0.3$, $u_1=0.4, u_2=0.3, u_3=0.3$, $N_t=2$, $\mathbf{r}_{th}=[0.02,0.08,0.19,0.3,0.4,0.4,0.4]$ bit/s/Hz.}
	\label{fig: bias103 weights040303 overloaded imperfect}
\end{figure}

\begin{figure}[t!]
	\centering
	\includegraphics[width=3.1in]{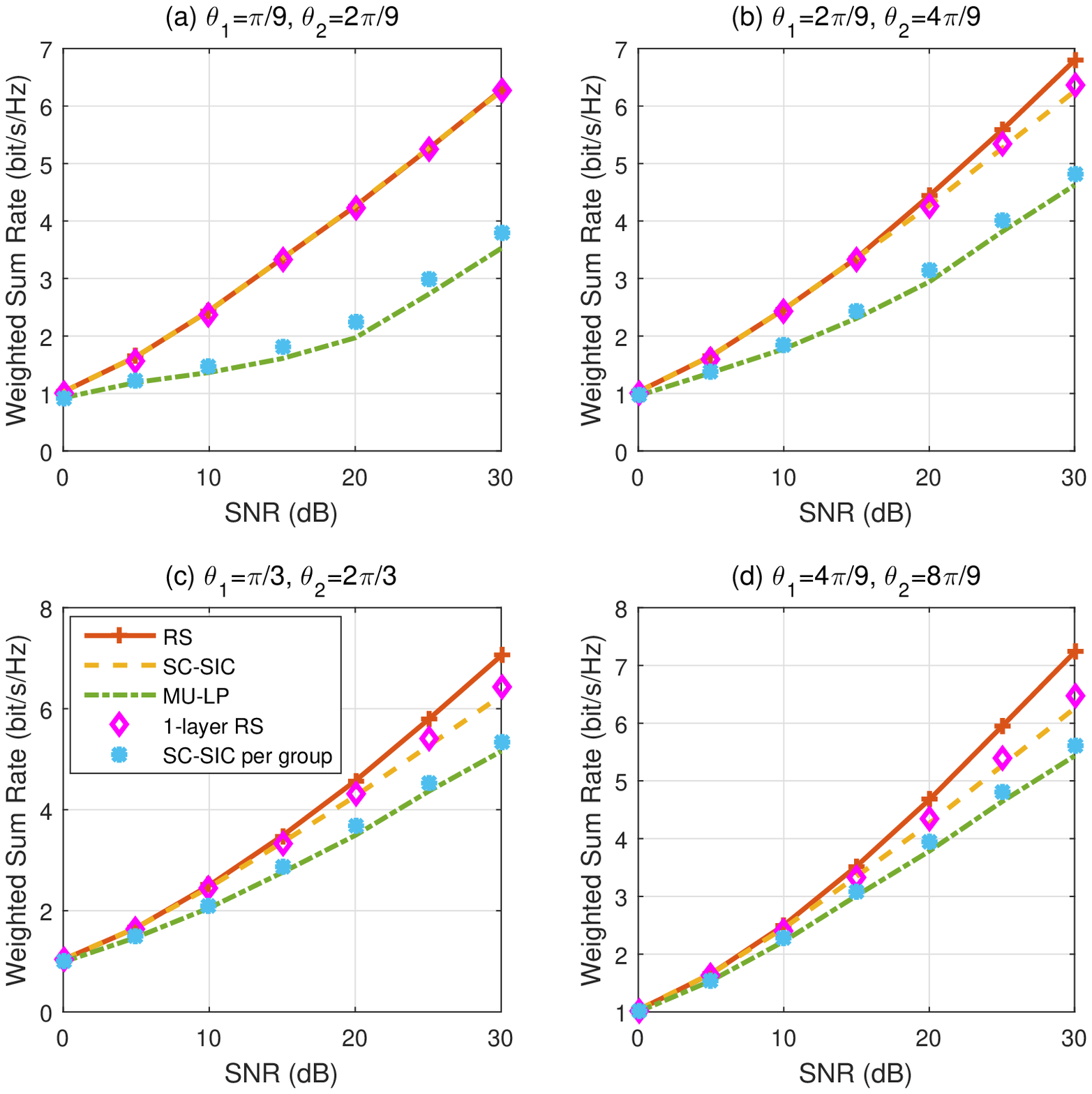}%
	\caption{Weighted sum rate versus SNR comparison of different strategies for overloaded three-user deployment with imperfect CSIT, $\gamma_1=1,\gamma_2=0.3$, $u_1=0.6, u_2=0.3, u_3=0.1$, $N_t=2$, $\mathbf{r}_{th}=[0.02,0.08,0.19,0.3,0.4,0.4,0.4]$ bit/s/Hz.}
	\label{fig: bias103 weights060301 overloaded imperfect}
\end{figure}

 We further investigate the overloaded three-user deployment with imperfect CSIT. The BS is equipped with two antennas ($N_t=2$). Fig. \ref{fig: bias11 weights020305 overloaded imperfect}--\ref{fig: bias103 weights060301 overloaded imperfect} shows the simulation results when the rate threshold is $\mathbf{r}_{th}=[0.02,0.08,0.19,0.3,0.4,0.4,0.4]$ bit/s/Hz. Comparing Fig. \ref{fig: bias11 weights020305 overloaded imperfect} with Fig. \ref{fig: three user bias1 020305 perfect overloaded}, the WSR gaps between RS and  SC--SIC per group, RS and  MU--LP are increasing dramatically while the WSR gap between  RS and  SC--SIC is decreasing. The inter-group interference of SC--SIC per group becomes difficult to coordinate due to the limited number of transmit antenna and imperfect CSIT. RS is able to overcome the limitations of SC--SIC per group and MU--LP by dynamically determining the level of multi-user interference to decode and treat as noise.


\subsection{Overloaded four-user deployment with perfect CSIT}

\par
\begin{figure}[t!]
	\centering
	\includegraphics[width=3.2in]{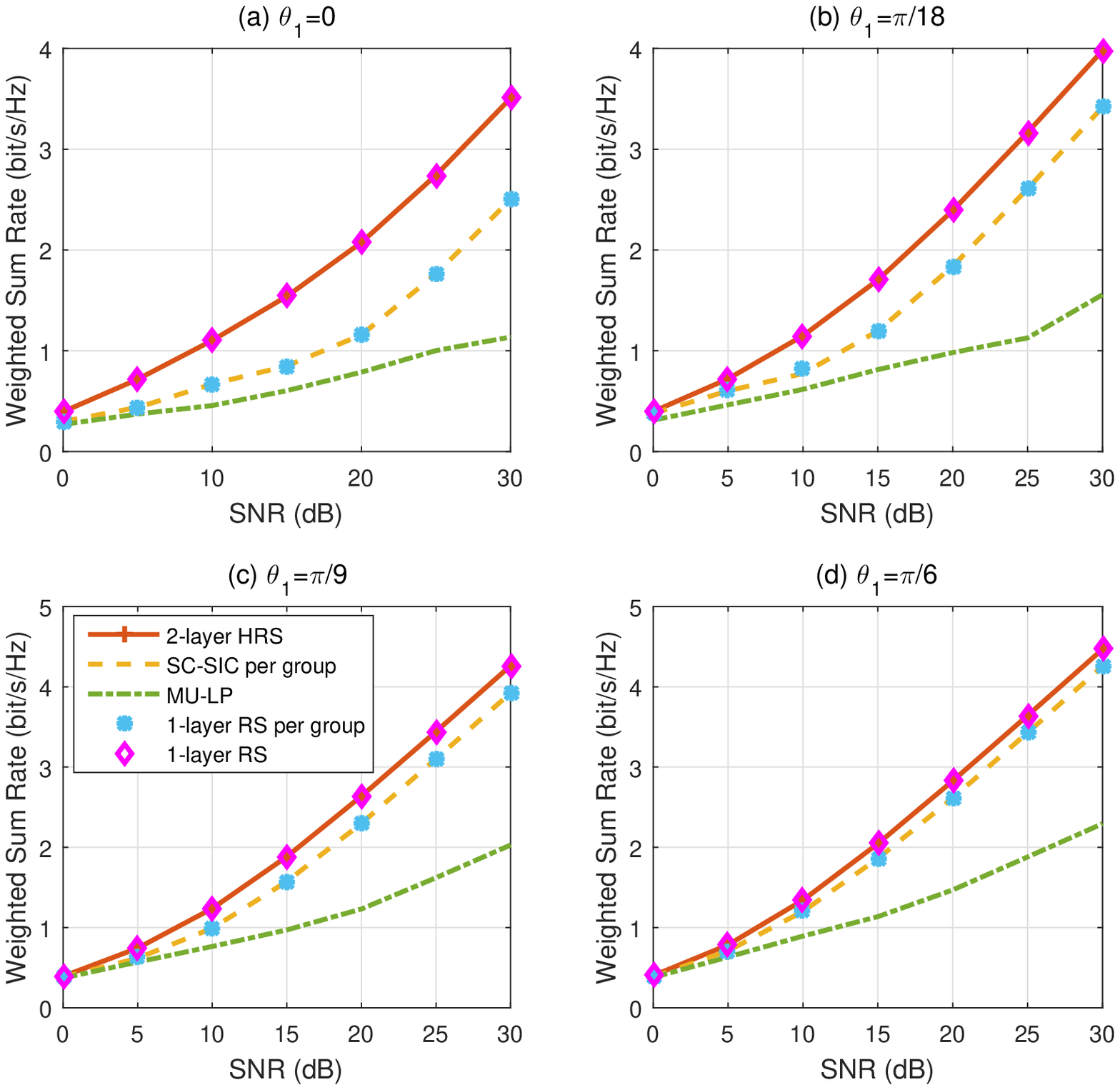}%
	\caption{Weighted sum rate versus SNR comparison of different strategies for overloaded four-user deployment with perfect CSIT, $\gamma_1=1$,  $\theta_2=\theta_1+\frac{\pi}{9}$, $\mathbf{r}_{th}=[0.03,  0.1,  0.2,  0.3,  0.4,  0.4,  0.4]$ bit/s/Hz.}
	\label{fig: bias1 weights025025 angle20 overloaded}
\end{figure}

\begin{figure}[t!]
	\centering
	\includegraphics[width=3.2in]{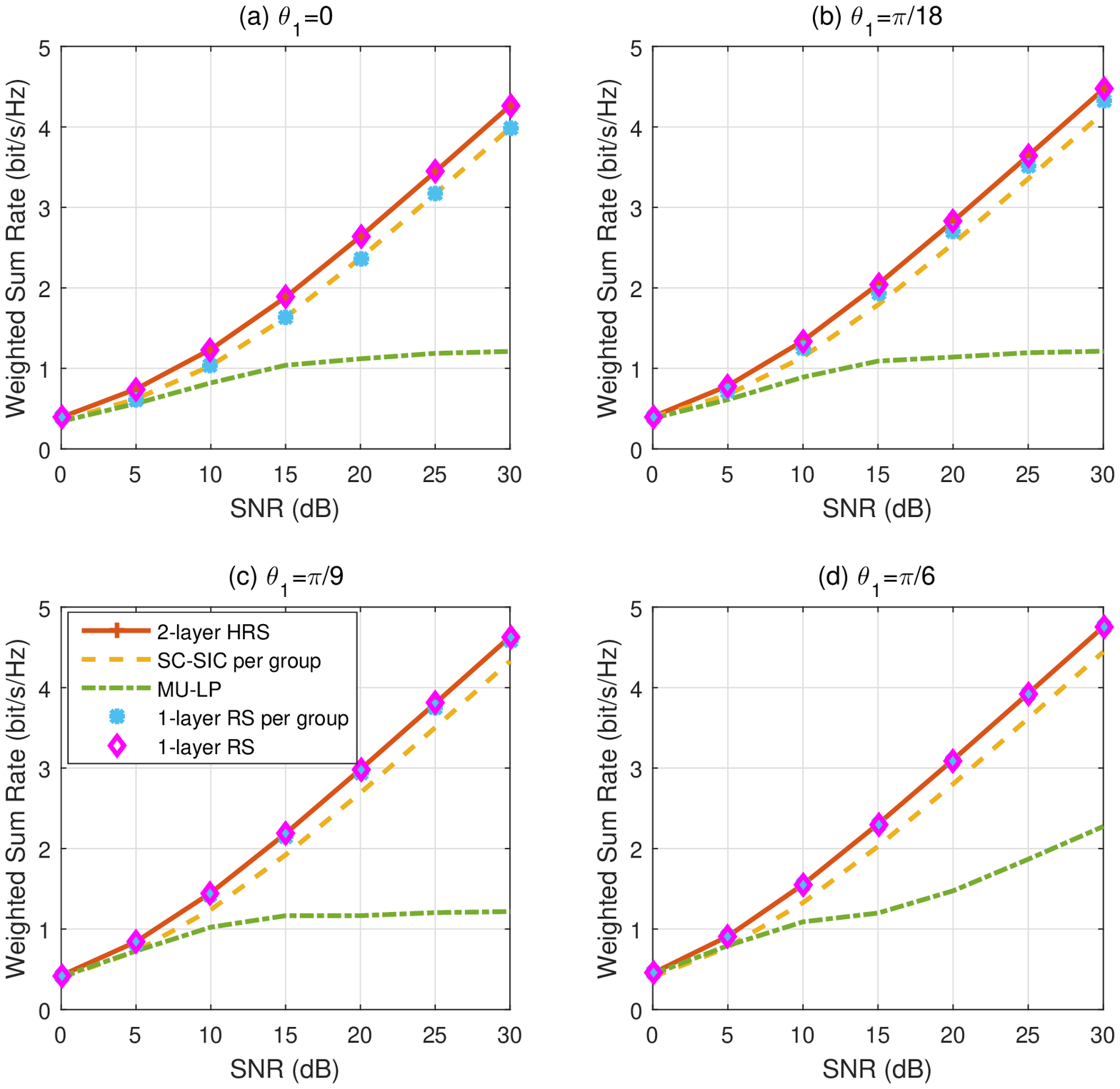}%
	\caption{Weighted sum rate versus SNR comparison of different strategies for overloaded four-user deployment with perfect CSIT, $\gamma_1=1$,  $\theta_2=\theta_1+\frac{\pi}{3}$, $\mathbf{r}_{th}=[0.03,  0.1,  0.2,  0.3,  0.4,  0.4,  0.4]$ bit/s/Hz.}
	\label{fig: bias1 weights025025 angle60 overloaded}
\end{figure}

 Fig. \ref{fig: bias1 weights025025 angle20 overloaded} and Fig. \ref{fig: bias1 weights025025 angle60 overloaded} show the results when $\gamma_1=1$. Comparing with SC--SIC per group, 1-layer RS per group always achieves equal or better WSR. 1-layer RS per group is more general than SC--SIC per group. It enables the capability of partially decoding interference and partially treating interference as noise in each user group. When there is a sufficient channel gain difference between users within each group and a sufficient inter-group angle, the WSR of SC--SIC per group becomes closer to the WSR of RS comparing Fig.  \ref{fig: bias1 weights025025 angle60 overloaded}  and Fig. \ref{fig: bias103 weights025025 angle60}. 

\subsection{Overloaded ten-user deployment with perfect CSIT}

\par
\begin{figure}[t!]
	\centering
	\includegraphics[width=3.0in]{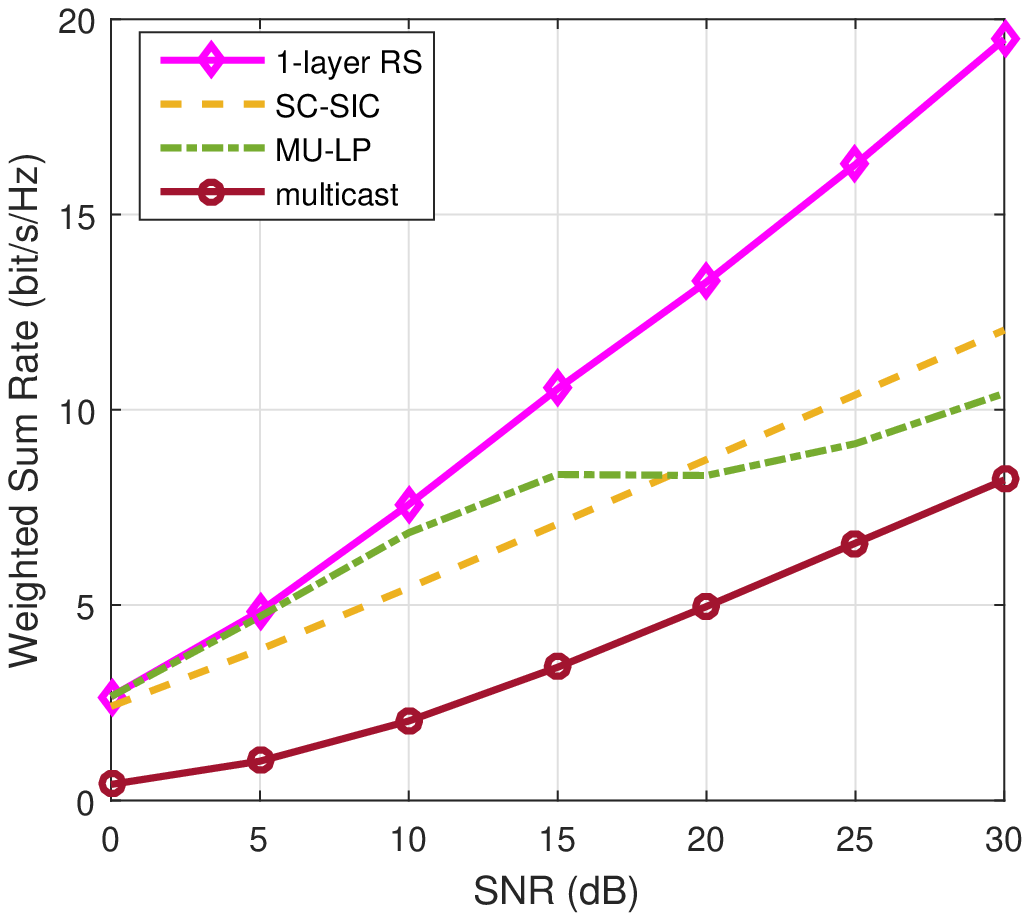}%
	\caption{Weighted sum rate versus SNR comparison of different strategies for overloaded ten-user deployment with perfect CSIT, $\sigma_1^2=\sigma_2^2=\ldots=\sigma_{10}^2=1$, $N_t=2$, SNR=30 dB, $\mathbf{r}_{th}=[0, 0.001, 0.004, 0.01,  0.03,  0.06,  0.1]$ bit/s/Hz.}
	\label{fig: ten user equal covariance low threshold}
\end{figure}

 Fig. \ref{fig: ten user equal covariance low threshold} shows the simulation results when $\sigma_1^2=\sigma_2^2=\ldots=\sigma_{10}^2=1$, $\mathbf{r}_{th}=[0, 0.001, 0.004, 0.01,  0.03,  0.06,  0.1]$ bit/s/Hz.   Comparing with  Fig. \ref{fig: ten user equal covariance individual rate}, the rate threshold of each SNR is reduced in Fig. \ref{fig: ten user equal covariance low threshold}.  The WSR achieved by MU--LP is approaching RS when SNR is 0 dB or 5 dB in Fig. \ref{fig: ten user equal covariance low threshold}. This is because the rate threshold is set to 0 when SNR is 0 dB or 5 dB.   When the rate threshold is 0, MU--LP could deliver 2 interference free streams since there are 2 transmit antennas. It  achieves a DoF of 2 while SC--SIC is always limited by a DoF of 1.  

\par
\begin{figure}[t!]
	\centering
	\includegraphics[width=3.0in]{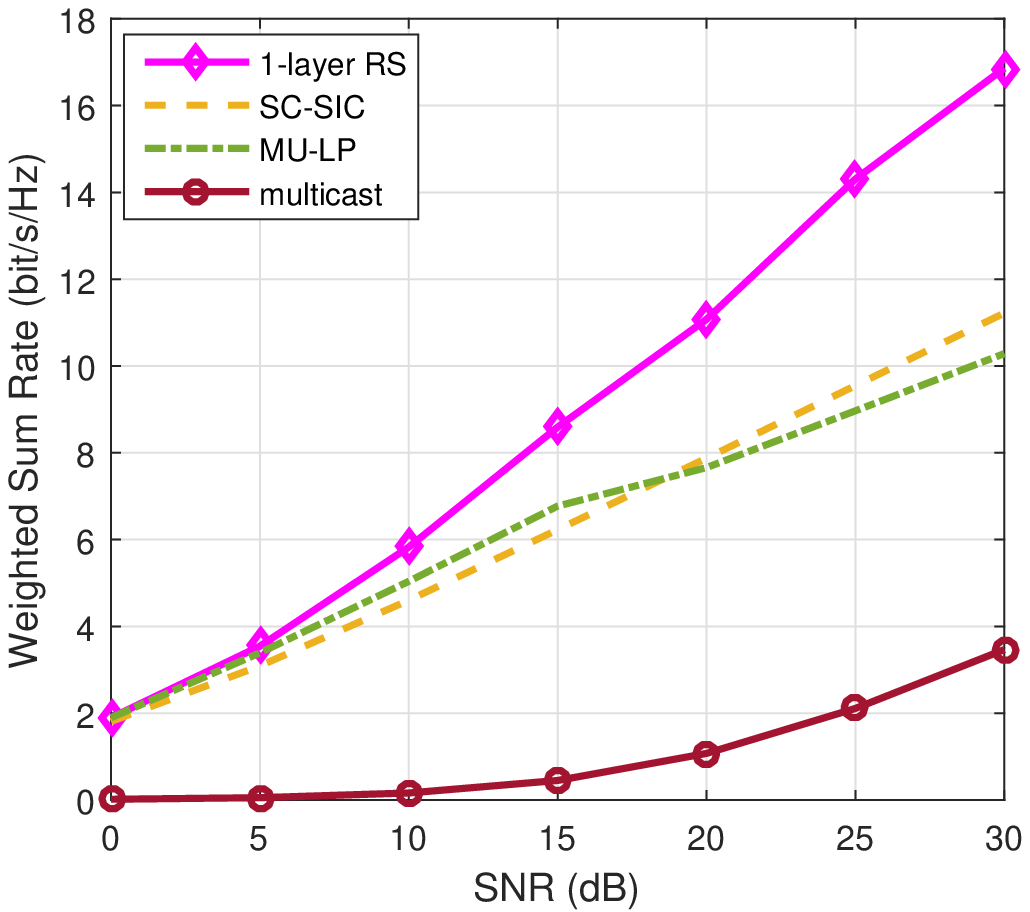}%
	\caption{Weighted sum rate versus SNR comparison of different strategies for overloaded ten-user deployment with perfect CSIT, $\sigma_1^2=1, \sigma_2^2=0.9, \ldots \sigma_{10}^2=0.1$, $N_t=2$, SNR=30 dB, $\mathbf{r}_{th}=[0, 0.001, 0.004, 0.01,  0.03,  0.06,  0.1]$ bit/s/Hz. }
	\label{fig: ten user covariance varied low threshold}
\end{figure}

 Fig. \ref{fig: ten user covariance varied low threshold} shows the simulation results when $\sigma_1^2=1, \sigma_2^2=0.9, \ldots \sigma_{10}^2=0.1$. The rate threshold is the same as in Fig. \ref{fig: ten user equal covariance low threshold}.  In the extremely overloaded scenario, the WSR gap between RS  and SC--SIC is still large despite the diversity in channel strengths. Here again, SC--SIC makes an inefficient use of the transmit antennas and achieves a DoF of 1. In contrast, 1-layer RS, with a low scheduler and receiver complexity, achieves a good performance in all network loads.

\bibliographystyle{IEEEtran}
\bibliography{reference}

\end{document}